\documentclass[sigconf]{acmart}

\usepackage{algorithm}
\usepackage[noend]{algpseudocode}
\usepackage{amsmath, amsfonts}
\usepackage{paralist}
\usepackage{multirow}
\usepackage{array}
\usepackage{thm-restate}
\usepackage{amsbsy}
\usepackage{adjustbox}
\usepackage{subfig}
\usepackage{siunitx}
\usepackage{multirow}
\usepackage{comment}
\usepackage{rotating}
\usepackage{enumitem}
\usepackage{adjustbox}
\usepackage{tikz}
\usetikzlibrary{arrows,shapes,snakes,automata,backgrounds,petri}
\usetikzlibrary{calc,positioning}
\usepackage[table]{colortbl}
\usepackage{color}
\usepackage{array}

\usepackage{macros}

\begin{document}
\title{Probabilistic Reasoning at Scale: Trigger Graphs to the Rescue}

\author{Efthymia Tsamoura}
\affiliation{%
  \institution{Samsung AI Research}
  \city{Cambridge}
  \country{United Kingdom}
}
\email{efi.tsamoura@samsung.com}

\author{Jaehun Lee}
\affiliation{%
  \institution{Samsung}
  \city{Seoul}
  \country{South Korea}
}
\email{jaehun20.lee@samsung.com}

\author{Jacopo Urbani}
\affiliation{%
  \institution{Vrije Universiteit Amsterdam}
  \city{Amsterdam}
  \country{The Netherlands}
}
\email{jacopo@cs.vu.nl}

\begin{abstract}

The role of uncertainty in data management has become more prominent than ever before, especially because of the growing importance of machine learning-driven applications that produce large uncertain databases.
A well-known approach to querying such databases is to blend rule-based reasoning with uncertainty. However, techniques proposed so far struggle with large databases.
In this paper, we address this problem by presenting a new technique for probabilistic reasoning that exploits Trigger Graphs (TGs) -- a notion recently introduced for the non-probabilistic setting. The intuition is that TGs can effectively store a probabilistic model by avoiding an explicit materialization of the lineage and by grouping together similar derivations of the same fact. Firstly, we show how TGs can be adapted to support the possible world semantics. Then, we describe techniques for efficiently computing a probabilistic model, and formally establish the correctness of our approach. We also present an extensive empirical evaluation using a prototype called \sys{}. Our comparison against other leading engines shows that \sys{} is not only faster, even against approximate reasoning techniques, but can also reason over probabilistic databases that existing engines cannot scale to.

\end{abstract}

\maketitle

\section{Introduction}\label{section:introduction}

\leanparagraph{Motivation} Uncertainty is inherent to modern data management.
Traditionally, the roots of uncertainty are traced back to mining knowledge from
unstructured data sources \cite{comet,probase,deepdive,kvault} and to querying
sensor measurements \cite{peex}, but now its presence is even more predominant
due to the widespread usage of neural architectures. The database community has
extensively studied the problem of efficiently querying uncertain data, with many seminal results having conflated into Probabilistic databases
(PDBs)~\cite{pdbs}. {PDBs enable querying uncertain data under an
elegant semantics known as \textit{possible world semantics} \cite{pdbs}.}

A second notion with a rich history in the data management community is that of rule-based languages. In particular, Datalog,
a language that allows expressing recursive queries in a declarative fashion,
finds multiple applications both in academia and in industry~\cite{AbiteboulHV95,10.5555/2415092,Datalography}.
One application is querying
\emph{Knowledge Graphs} (KGs), a graph-like type of \textit{Knowledge Base} (KB). Beyond querying KGs, Datalog also finds applications in AI and machine learning, giving rise to a paradigm known as neurosymbolic
AI \cite{DBLP:journals/corr/abs-1907-08194,neurolog}.
For instance, \citeauthor{DBLP:conf/eccv/ZhuFF14} in \cite{DBLP:conf/eccv/ZhuFF14} and \citeauthor{scallop} in \cite{scallop} use rules for visual question answering.

\noindent\textbf{Problem.} Adopting rule-based reasoning for querying uncertain data requires integrating reasoning with uncertainty in a principled fashion.
For instance, to query the predictions of a neural network~\cite{scallop}, or a Web-mined KG  (e.g., Google’s Knowledge Vault \cite{kvault}) using Datalog rules, we need to extend the Datalog semantics with uncertainty. Despite that
blending logic with uncertainty has a long tradition in databases and AI \cite{sato:iclp95,DBLP:journals/ml/RichardsonD06,DBLP:journals/jmlr/BachBHG17,BaranyCKOV17}, current approaches for probabilistic rule-based reasoning either face scalability limitations or impose several syntactic restrictions. For instance,  Markov Logic Networks (MLNs) \cite{DBLP:journals/ml/RichardsonD06}, Probabilistic Soft Logic (PSL) \cite{DBLP:journals/jmlr/BachBHG17} and ICL~\cite{poole:pilp08} require either the rules to be ground or to satisfy several syntactic restrictions ensuring non-recursion. In the context of PDBs, although
there are techniques to efficiently query them under certain cases as shown by
\citeauthor{dichotomy2} in \cite{dichotomy2},
those techniques only support non-recursive queries.

The state of affairs remains the same for rule-based languages that allow
non-ground Datalog rules under the possible world semantics
\cite{sato:iclp95,deraedt:ijcai07,fierens:tplp15}. Since the problem is
untractable in the worst case, reasoning can become
prohibitively expensive even for PDBs of a few thousands of facts (see
\cite{sota1,DBLP:conf/eccv/ZhuFF14} for a discussion). To tackle the above
limitations, several approximation techniques~\cite{scallop,k-best,k-optimal} that
 have been proposed.
However, beyond being impractical if high confidence is needed, e.g.,
autonomous driving or health care, approximate techniques still
cannot scale beyond a certain level.

As an alternative to approximation techniques, the authors in
\cite{DBLP:journals/corr/abs-1911-07750} have recently proposed $\Delta
\Tcp$, a new technique that builds upon work on provenance semirings
    \cite{provenance-semirings} and the well-known $\Tcp$ operator from the logic
    programming community~\cite{vlasselaer:ijar16}.
$\Delta \Tcp$ improves the scalability of exact probabilistic reasoning by reducing it to non-probabilistic reasoning~\cite{DBLP:journals/corr/abs-1911-07750}.
Despite outperforming prior art in terms of runtime, $\Delta \Tcp$ still faces several performance bottlenecks. Firstly,
$\Delta \Tcp$ is required to perform
Boolean formula comparisons after reasoning with the rules to
ensure termination (\textbf{L1}). However, Boolean comparisons can be very
expensive when querying graphs
\cite{fierens:tplp15,renkens:aaai14,vlasselaer:ijar16}. Secondly, {$\Delta\Tcp$}
may keep multiple copies of the same formula (\textbf{L2})
increasing the memory consumption. Beyond these two limitations, $\Delta \Tcp$
introduced an additional one: rewriting the rules into more complex ones and
maintaining additional structures (\textbf{L3}).
These limitations can introduce performance bottlenecks in some cases. For
instance, in our experiments with $\Delta\Tcp$ on the well-known benchmark
LUBM~\cite{lubm}, we measured that the overhead introduced by \textbf{L1} and \textbf{L3} can take up to $96\%$ of the total runtime.

\noindent\textbf{Our approach.} In this paper, we introduce a new technique for
performing probabilistic rule-based reasoning under the possible world semantics
that overcomes the three limitations mentioned above. In doing so, we show that
we can reason over PDBs in a much more scalable way than it is
currently possible.

Our technique is based on \emph{Trigger Graphs} (TGs), a structure that was
recently introduced for non-probabilistic rule-based reasoning
~\cite{tsamoura2021materializing}. A TG is an acyclic directed graph that
captures all the operations that should be performed to compute the model of a
non-probabilistic database using a set of rules, i.e., to compute
an extension of the database for which all the rules are logically satisfied.
It has been shown in \cite{tsamoura2021materializing} that reasoning with TGs is much more efficient than reasoning using prior techniques~\cite{chaserevisited,chasebench} due to the ability of TGs to avoid redundant derivations.

A limitation of TGs is that they cannot be used as-is for probabilistic
reasoning. In this paper, we show that with the right modifications, TGs can
maintain the provenance of the derivations, which can be directly used to
compute their {probability}
\cite{kimmig_demoen_de_raedt_costa_rocha_2011}. In doing so, probabilistic reasoning can be implemented in a way that overcomes limitations
\textbf{L1}--\textbf{L3} from above. Regarding \textbf{L1}, TGs eliminate the
requirement to perform Boolean formula comparisons.
Regarding \textbf{L2}, storing the derivation provenance in the TG removes the
need to store the same formula multiple times. Moreover, we show how we can
collapse multiple derivation trees into one to save space and hence improving
the runtime. Finally, regarding \textbf{L3}, TGs allow us to maintain the provenance natively
overcoming the requirement to rewrite the rules into more complex forms or to
maintain additional structures.

We implemented our technique in a new engine called \textit{Lineage TGs} (\sys{}) and compared its
performance against leading engines, namely \problog{}
\cite{schoenfisch:ijar17,vBDJ-CIKM19} and \vp{}
\cite{DBLP:journals/corr/abs-1911-07750}. Our empirical evaluation considers
scenarios from the Web ($\lubm$~\cite{lubm}, $\dbpedia$~\cite{dbpedia} and
$\claros$ \cite{claros}) and probabilistic logic programming communities
($\smokers$ \cite{smokers}). We additionally ran experiments using popular
real-world KGs ($\yago$ and $\wnrr$ \cite{wn18rr}) and rules mined with
state-of-the-art techniques (AnyBurl~\cite{anyburl}).  Finally, we also
considered a recent benchmark called $\vqar$ \cite{scallop}. In $\vqar$, the probabilistic facts
are derived by neural networks, while the rules are used to answer queries over
images~\cite{scallop}. This benchmark is challenging for prior art because reasoning leads to an explosion of derivations.
Indeed, the benchmark has motivated \scallop~\cite{scallop}, a
recent approximate probabilistic reasoning engine with state-of-the-art
performance. %
We compared \sys{} against \scallop{} and observed that $\ours$ often outperforms
Scallop even though Scallop does not search for all the explanations. Noticeably,
\sys{} is the only engine that can compute the full probabilistic model of
$\vqar$ due to its ability to maintain compact model representations.

Overall, our experimental results show that our approach outperforms the other
engines, often significantly, both in terms of runtime and RAM consumption.
Moreover, in multiple scenarios, $\ours$ can mean the difference between
answering queries over PDBs using rules and not answering them at all.

To summarize, our contributions are as follows:\\
    $\bullet$ We introduce a new technique for reasoning over
    large PDBs based on the distribution semantics and TGs.

    \noindent$\bullet$ We introduce an extension that avoids the
    combinatorial explosions of derivations via compression.

    \noindent$\bullet$ We show that our approach is correct and that it provides
    anytime bounds like prior art
    (\cite{vlasselaer:ijar16,DBLP:journals/corr/abs-1911-07750}).

    \noindent$\bullet$ We implement our technique in a new engine, called
    $\ours$, and compare its performance against state-of-the-art engines using
    a portfolio of benchmarks from various communities.

\section{Preliminaries}\label{section:preliminaries}

We start our discussion with a short recap of some basic notions related to logic and (probabilistic) rule-based reasoning.

A \textit{term} is either
a constant or a variable. \emph{Atoms} have the form
${p(t_1,\dots,t_n)}$, where $p$ is an $n$-ary predicate, and each $t_i$ is a
term. An atom is \textit{ground} if its terms are all constants. Ground atoms
are also called \textit{facts}.
A \emph{term mapping} $\sigma$ is a (possibly partial) mapping from terms to
terms; we write ${\sigma = \{ t_1 \mapsto s_1, \dots, t_n \mapsto s_n \}}$ to
denote that ${\sigma(t_i) = s_i}$ for ${1 \leq i \leq n}$. Let $\alpha$ be a
term, a formula or a set of terms or formulas.  Then $\sigma(\alpha)$ is
obtained by replacing each occurrence of a term $t$ in $\alpha$ that also occurs
in the domain of $\sigma$ with $\sigma(t)$ (i.e., terms outside the domain of
$\sigma$ remain unchanged). We refer to $\sigma(\alpha)$ as an
\emph{instantiation} of $\alpha$.
Symbol $\models$ denotes logical entailment.
For a set of ground atoms $I$ and an atom
$\alpha$, ${I \models \alpha}$ holds if ${\alpha \in I}$.  Symbol $\equiv$
denotes logical equivalence.

A Datalog \emph{rule} is a universally quantified implication of the form
\begin{align} p(\vec{X}) &\leftarrow \bigwedge \nolimits_{j=1}^n p_j(\vec{X}_j).
\label{eq:rule} \end{align}
Above, $\vec{X}$ and $\vec{X}_j$, $1 \leq j \leq n$
are vectors of variables
and each variable occurring in $\vec{X}$ also occurs in some $\vec{X}_j$.
From now on, we will refer to Datalog rules as rules.
We refer to the right part of a rule as its premise and to the left as its conclusion.

\leanparagraph{Logic programs} A (non-probabilistic) \textit{logic program} $\Pp$ is a pair ${(
\Rp, \Fp )}$, where $\Rp$ is a set of rules and $\Fp$ is a set of facts.
The \textit{Herbrand base} $\Ap$ of a program $\Pp$ denotes the set of all ground
atoms that can be computed using all constants and predicates occurring in $\Pp$.
An \textit{interpretation} of $\Pp$ is an assignment of each atom in the Herbrand base of $\Pp$ to either true or false.
We can equivalently see an interpretation as a subset of $\Ap$ including only the atoms that are assigned to true.
An interpretation is a \textit{model} of $\Pp$ if ${I \models r}$ holds for each rule $r$ in $\Pp$.
The \textit{least} Herbrand model of $\Pp$ is the one with the fewest atoms
among all models of $\Pp$. Every program of Datalog rules admits a finite model.
We use ${\Pp \models \alpha}$ or ${(\Rp, \Fp ) \models \alpha}$ to denote ${\Rp \cup \Fp \models \alpha}$, where $\alpha$ is a ground atom.

\emph{Queries} are defined using a fresh
\emph{predicate} $\predQ$.
A tuple ${\vec a}$ of constants is an \emph{answer} to
$\predQ$ w.r.t. a program ${(\Rp, \Fp )}$
iff ${\Rp \cup \Fp \models \predQ(\vec a)}$.
The above definition allows us to represent \emph{conjunctive queries} (CQs) \cite{Chandra:1977:OIC:800105.803397}
by introducing a rule defining $\predQ$ in its conclusion \cite{BenediktMT18}.

\leanparagraph{PDBs}
A \emph{tuple-independent Probabilistic Database} (PDB) $\mathcal{D}$ is a pair ${(\Fp, \pi)}$, where $\Fp$ is a set of facts. Each fact is viewed
as an independent Bernoulli random variable that becomes true (resp. false) with
probability $\pi(f)$ (resp. $1-\pi(f)$) \cite{pdbs}. Below, we will
    write
${\pi(f) :: f}$ to denote a fact $f$ and its probability of being true.
A PDB induces a distribution on all database instances, which we call \emph{possible worlds}.
Each subset of $\Fp$ is a possible world.
Viewing the database facts as independent random variables allows us to compute the probability
$\Pr(\mathcal{C})$ of a possible world $\mathcal{C}$ in $\mathcal{D}$ as the
product of the probabilities of the facts that are true in $\mathcal{C}$ multiplied by the product of the probabilities of the facts that are false
in $\mathcal{C}$.
The probability of a formula $\varphi$ in $\mathcal{D}$ is then
the sum of the probabilities of all possible worlds in which $\varphi$ holds.         

\noindent \textbf{Probabilistic logic programs} \cite{vlasselaer:ijcai15} extend PDBs with rule-based reasoning.
A {probabilistic logic program}, or probabilistic program for short, is a triple ${\Pp = (\Rp, \Fp, \pi)}$, where $\Rp$, $\Fp$ and $\pi$ are defined as above.
The probability of a formula $\varphi$ in $\Pp$ is defined analogously to
PDBs. However, this time we consider all possible worlds of $(\Fp,\pi)$ which
along with the rules $\Rp$ entail $\varphi$:
\begin{align} \label{eq:p_suc}
	 \sum \limits_{C \subseteq \Fp \mid  C \cup \Rp \models \varphi}\, \Pr(C)
\end{align}
Notice that it is possible to assign probabilities also to the rules by adding extra
``dummy'' facts to the rule premises with probabilities equal to that of the rules~\cite{DeRaedt2015}.

An \emph{explanation} of a ground atom $\alpha$ in $\Pp$ is a minimal subset $C$
of $\Fp$ so that together with $\Rp$ it entails $\alpha$. We
denote explanations using the conjunction of the constituting atoms. The
\emph{lineage} of $\alpha$ in $\Pp$ is the disjunction of its explanations in
$\Pp$.

\begin{example} \label{example:running}
    Consider the set of rules $\Rp$ describing graph reachability
    \begin{align}
    	p(X,Y) &\leftarrow e(X,Y)			\tag{$r_1$}\\
    	p(X,Y) &\leftarrow p(X,Z) \wedge p(Z,Y) \tag{$r_2$}
    \end{align}
    According to $\Rp$, there exists a path from $X$ to $Y$
    if there exists either an edge from $X$ to $Y$, or
    paths from $X$ to $Z$ and $Z$ to $Y$.
    Consider also the set $\Fp$ including the probabilistic facts ${e(a,b)}$, ${e(b,c)}$, ${e(a,c)}$ and ${e(c,b)}$.
    Each fact $f$ is true with a probability denoted as ${\pi(f)}$.

    Consider fact ${p(a,b)}$.
    Its probability in ${\mathcal{P} = (\mathcal{R},\mathcal{F}, \pi)}$
    equals the sum of the probabilities of all possible worlds that include either fact ${e(a,b)}$ or facts ${e(a,c)}$ and ${e(c,b)}$.
    Thus, ${e(a,b)}$ and ${e(a,c) \wedge e(c,b)}$ are the two explanations of ${p(a,b)}$ in $\mathcal{P}$ and
    ${e(a,b) \vee e(a,c) \wedge e(c,b)}$ is its lineage.
\end{example}

For the rest of the section we fix a program ${\Pp=(\Rp, \Fp, \pi)}$.
Probabilistic programs have multiple least Herbrand models:
for each possible world $\mathcal{C}$ of the PDB $(\Fp, \pi)$,
the least Herbrand model of the logic program ${(\Rp, \mathcal{C})}$ is also a (least Herbrand) model of $\Pp$.

\leanparagraph{Probabilistic reasoning}
State-of-the-art probabilistic reasoning techniques, like the ones proposed by ProbLog \cite{DBLP:journals/corr/abs-1911-07750,vlasselaer:ijar16} and Scallop \cite{scallop}
build upon work on provenance semirings \cite{provenance-semirings}.

Let $\mathsf{V}$ denote the set of Boolean variables associated with the facts
from $\Fp$. The idea is to first associate each fact $\alpha$ with a Boolean formula
$\lambda_\alpha$ over $\mathsf{V}$, which represents $\alpha$'s provenance
\cite{provenance-semirings}, and then to compute $\Pr(\alpha)$ via
\textit{weighted model counting} (WMC)~\cite{WMC-hardness} on $\lambda_\alpha$. To compute $\alpha$ and $\lambda_\alpha$, \citeauthor{vlasselaer:ijar16} borrowed ideas from bottom-up Datalog evaluation and introduced the notion of the \emph{least parameterized model} and the $\Tcp$ operator for computing it \cite{vlasselaer:ijar16}. A least parameterized model of $\Pp$ includes for each atom $\alpha$
occurring in any of the least Herbrand models of $\Pp$, a pair of the form ${(\alpha, \lambda_{\alpha})}$. We often refer to a least parameterized model as a \emph{probabilistic model}.

$\Tcp$ proceeds in rounds, where
each round $k$ computes, for each atom $\alpha$, a formula $\lambda^k_{\alpha}$ encoding all derivations of atom $\alpha$ of depth ${\leq k}$.
Each round includes three steps: a derivation step (\textbf{DE}), an aggregation step (\textbf{AG}), and a formula update step (\textbf{FU}).
\textbf{DE} instantiates the rules in $\Rp$ using the atoms derived so far and computes a Boolean formula
out of each rule instantiation. Then, for each atom $\alpha$, \textbf{AG} computes a new formula $\mu^{k}_{\alpha}$
by disjointing \emph{all} formulas computed for $\alpha$ at \textbf{DE}.
Finally, \textbf{FU} computes a new formula
    $\lambda^{k}_{\alpha}=\mu^{k}_{\alpha} \vee  \lambda^{k-1}_{\alpha}$
if $\mu^{k}_{\alpha} \not\equiv \lambda^{k-1}_{\alpha}$ or by setting
$\lambda^{k}_{\alpha}=\lambda^{k-1}_{\alpha}$ otherwise.
The technique terminates at round $k$ when \emph{all} the formulas computed during the $k$-th round
are logically equivalent to the formulas computed during the ${(k-1)}$-th round.

\begin{table}[tb]
\small
\centering
	\caption{Formulas in the first three rounds of $\Tcp$ and $\Delta \Tcp$. $\omega$ denotes the formula
${\lambda^2_{p(a,b)} \wedge \lambda^2_{p(b,b)} \vee \lambda^2_{p(a,c)} \wedge \lambda^1_{p(c,b)}}$.}
	\begin{tabular}{ |c c c c | }
		\hline
		R                  & Atom & ${\mu^i}$        & ${\lambda^i}$ \\

		\multirow{4}{*}{1}     & $p(a,b)$       & ${e(a,b)}$    & ${e(a,b)}$ \\
		                       & $p(b,c)$       & ${e(b,c)}$    & ${e(b,c)}$ \\
		                       & $p(a,c)$       & ${e(a,c)}$    & ${e(a,c)}$ \\
		                       & $p(c,b)$       & ${e(c,b)}$    & ${e(c,b)}$ \\
		\hline
		\multirow{3}{*}{2}     & $p(a,b)$       & ${\lambda^1_{p(a,c)} \wedge \lambda^1_{p(c,b)}}$    & ${e(a,c) \wedge e(c,b) \vee e(a,b)}$ \\
		                       & $p(a,c)$       & ${\lambda^1_{p(a,b)} \wedge \lambda^1_{p(b,c)}}$    & ${e(a,b) \wedge e(b,c) \vee e(a,c)}$ \\
		                       & $p(b,b)$       & ${\lambda^1_{p(b,c)} \wedge \lambda^1_{p(c,b)}}$    & ${e(b,c) \wedge e(c,b)}$ \\

		\hline
        \multirow{4}{*}{3}     & $p(a,b)$       &  $\omega$ &  ${\mu^3_{p(a,b)} \vee \lambda^2_{p(a,b)} \equiv \lambda^2_{p(a,b)}}$     \\

		                       & $p(b,c)$       & ${\lambda^2_{p(b,b)} \wedge \lambda^1_{p(b,c)} }$                                                    &  ${\mu^3_{p(b,c)} \vee \lambda^2_{p(b,c)} \equiv \lambda^2_{p(b,c)}}$     \\

		                       & $p(a,c)$       & ${\lambda^2_{p(a,b)} \wedge \lambda^1_{p(b,c)} }$                                                    &  ${\mu^3_{p(a,c)} \vee \lambda^2_{p(a,c)} \equiv \lambda^2_{p(a,c)}}$     \\

		                       & $p(b,b)$       & ${\lambda^2_{p(b,b)} \wedge \lambda^2_{p(b,b)} }$                                                    &  ${\mu^3_{p(b,b)} \vee \lambda^2_{p(b,b)} \equiv \lambda^2_{p(b,b)}}$     \\
		\hline
	\end{tabular}
	\label{table:running:tcp}
\end{table}

\begin{example} \label{example:running:tcp}
    We demonstrate $\Tcp$ over Example~\ref{example:running}.
    In the first round, $\Tcp$ computes all paths of length one by instantiating $r_1$ using the facts in $\Fp$.
    For instance, the instantiation ${p(a,b) \leftarrow e(a,b)}$
    states that there is a path from $a$ to $b$, since there is an edge from $a$ to $b$.
    Hence, ${\lambda^1_{p(a,b)} = e(a,b)}$.
    In the second round, $\Tcp$ computes all paths of length up to two.
    There, the instantiation  ${p(a,b) \leftarrow p(a,c) \wedge p(c,b)}$ is computed.
    This instantiation states that there is a path from $a$ to $b$ as there is a path from $a$ to $c$ and from $c$ to $b$.
    Since ${\lambda^1_{p(a,c)} = e(a,c)}$ and
    ${\lambda^1_{p(c,b)} = e(c,b)}$, formula ${e(a,c) \wedge e(c,b)}$ is computed out of this rule instantiation
    and \textbf{FU} sets $\lambda^2_{p(a,b)}={e(a,c) \wedge e(c,b) \vee e(a,b)}$.
    Then, $\Tcp$ starts the third round to compute all paths of lengths up to three.
    As all formulas computed in the third round are logically equivalent to the ones computed
in the second round, $\Tcp$ terminates.
Table~\ref{table:running:tcp} reports some of the formulas computed in the first
three rounds. For brevity,
column $\mu^i$ in Table~\ref{table:running:tcp}
does not show formulas for all rule instantiations.
Instead, it shows \textit{only}
formulas computed from rule instantiations that involve at least one ``fresh" fact, i.e.,
a fact that has been either derived or its formula has been updated during
the previous round.
For instance, Table~\ref{table:running:tcp}
does not show formulas for facts $p(b,c)$ and $p(c,b)$ in round two, as those facts can only be derived via $r_1$ and database facts at this point.
As we discuss below, the derivations in Table~\ref{table:running:tcp} reflect those of $\Delta\Tcp$.

\end{example}

Computing a probabilistic model is challenging because we are called to
store, for each atom, all explanations in its lineage, which can be
exponentially many \cite{DeRaedt2015}. Moreover, computing the probability
of a given lineage is \#P-hard~\cite{DBLP:journals/siamcomp/Valiant79}. Due to the above, there can be worst-case inputs for which the computation either of the lineage or of its probability
can either take too long or fill the memory.
Although our approach does not change the worst-complexity of the problem, its
goal, similarly to $\Delta \Tcp$, is to improve the scalability and thus reduce
significantly the number of worst-case inputs.

$\Delta\Tcp$ addresses the problem of $\Tcp$, i.e., re-computing the same rule
instantiations, e.g.,
${p(a,b) \leftarrow e(a,b)}$ is computed both in the second and the third round of $\Tcp$.
To avoid those re-computations, \cite{DBLP:journals/corr/abs-1911-07750} introduced $\Delta \Tcp$ as an extension of $\Tcp$ inspired by \emph{Semi Na\"ive Evaluation} (SNE). SNE is a well-known
Datalog technique that restricts the rule instantiations in round $k$
to the ones involving at least one atom whose lineage was updated in the
${(k-1)}$-th round~\cite{AbiteboulHV95}.
The authors in \cite{DBLP:journals/corr/abs-1911-07750} also proposed a
declarative implementation of $\Delta \Tcp$ that works by rewriting the rules' introducing auxiliary
atoms and by adding new rules for populating them.

\section{Motivation} \label{section:motivation}

Example~\ref{example:running:tcp} reveals several limitations of $\Tcp$ and $\Delta \Tcp$.
Firstly, both $\Tcp$ and $\Delta \Tcp$ perform boolean formula comparisons at the end of each round. For instance, they both logically compare at the end of the second round formula $\lambda^1_{p(a,b)}$ with formula $\mu^2_{p(a,b)} \vee \lambda^1_{p(a,b)}$ to update the formula of ${p(a,b)}$ (\textbf{L1}). These comparisons may become
the bottleneck in scenarios involving querying paths  \cite{fierens:tplp15,renkens:aaai14,vlasselaer:ijar16}.

Secondly, both $\Tcp$ and $\Delta \Tcp$ may keep multiple copies of the same formula increasing the memory consumption (\textbf{L2}). For instance, formula ${e(a,b)}$ is kept in both copies of the formulas associated with ${p(a,b)}$ in the first and the second round of $\Tcp$ and $\Delta \Tcp$.
In general, if a formula for a fact $\alpha$ is updated $k$ time in total, then
each formula $\mu^{\ell}_{\alpha}$ that is computed for $\alpha$ at the
\textbf{AG} step of round $\ell$, is kept ${k-\ell}$ times.

Thirdly, regarding $\Delta \Tcp$, the runtime overhead to instantiate the rewritten rules can be substantial, as the execution of each rule involves multiple additional semi-joins and outer-joins, see \cite{chasebench,tsamoura2021materializing}; furthermore, maintaining additional structures introduces extra memory overhead. The above two limitations are referred to as \textbf{L3}. Our objective is to compute the
probability of each fact $\alpha$ in $\Pp$ in a way that overcomes limitations \textbf{L1}--\textbf{L3}.

\begin{figure}[tb]
\begin{center}
    \subfloat[]{
\scalebox{0.6}{%
    \definecolor{lavander}{cmyk}{0,0.48,0,0}
    \definecolor{violet}{cmyk}{0.79,0.88,0,0}
    \definecolor{burntorange}{cmyk}{0,0.52,1,0}
    \def \lav{lavander!90}
    \def \oran{orange!30}
    \tikzstyle{superpeers}=[font=\fontsize{12}{0}\selectfont,fill=red!20]
    \tikzstyle{peers}=[font=\fontsize{12}{0}\selectfont, draw=none, circle, violet, bottom color=white,
                      top color= white, text=violet, minimum size=20pt,inner sep=0pt]
    \tikzstyle{edge} = [draw,thick,->,black!30]
    \tikzstyle{ignored edge} = [draw,dashed,line width=1pt,->,black!30]
    \begin{tikzpicture}[node distance=1.9cm,>=stealth',bend angle=45,auto]

        \node[superpeers] (a) {$e(a,b)$};
        \node[superpeers] (b) [right of=a] {$e(b,c)$};
        \node[superpeers] (c) [right of=b] {$e(a,c)$};
        \node[superpeers] (d) [right of=c] {$e(c,b)$};

        \node[superpeers] (e) [above of=a] [label=left:$\tau_1$] {$p(a,b)$};
        \node[superpeers] (f) [above of=b] [label=left:$\tau_2$] {$p(b,c)$};
        \node[superpeers] (g) [above of=c] [label=left:$\tau_3$] {$p(a,c)$};
        \node[superpeers] (h) [above of=d] [label=left:$\tau_4$] {$p(c,b)$};

        \node[superpeers] (n2) [above of=e,yshift=18mm,xshift=0mm] [label=left:$\tau_8$] {$p(a,b)$};
        \node[superpeers] (n1) [above of=f,yshift=18mm,xshift=0mm] [label=left:$\tau_9$] {$p(a,b)$};
        \node[superpeers] (n3) [above of=g,yshift=18mm,xshift=0mm] [label=left:$\tau_{10}$] {$p(a,c)$};
        \node[superpeers] (n4) [above of=h,yshift=18mm,xshift=0mm] [label=left:$\tau_{11}$] {$p(b,c)$};

        \path[edge] (e) -- (n2);
        \path[edge] (f) -- (n3);
        \path[edge] (f) -- (n4);
        \path[edge] (h) -- (n1);

        \node[superpeers] (i) [above of=e,xshift=6mm] [label=left:$\tau_5$] {$p(a,c)$};
        \node[superpeers] (j) [above of=f,xshift=8mm] [label=left:$\tau_6$] {$p(b,b)$};
        \node[superpeers] (k) [above of=g,xshift=12mm] [label=left:$\tau_7$] {$p(a,b)$};

        \path[edge] (a) -- (e);
        \path[edge] (b) -- (f);
        \path[edge] (c) -- (g);
        \path[edge] (d) -- (h);

        \path[edge] (e) -- (i);
        \path[edge] (f) -- (i);
        \path[edge] (f) -- (j);
        \path[edge] (h) -- (j);
        \path[edge] (g) -- (k);
        \path[edge] (h) -- (k);

        \path[edge] (i) -- (n1);
        \path[edge] (j) -- (n2);
        \path[edge] (k) -- (n3);
        \path[edge] (j) -- (n4);
    \end{tikzpicture}
    \label{figure:derivations}
    }
}\hspace{0.3cm}
\subfloat[]{
    \hfill
    \scalebox{0.5}{
    	\begin{tikzpicture}[->,>=stealth',shorten >=1pt,auto,node distance=2cm, semithick]
	      \tikzstyle{every state}=[font=\fontsize{16}{0}\selectfont,draw=orange!75,fill=orange!20]
	      \tikzstyle{edge} = [font=\fontsize{16}{0}\selectfont,draw,thick,->,black!50]
	      \tikzstyle{edge1} = [font=\fontsize{16}{0}\selectfont,draw,thick,->,blue]
	      \tikzstyle{edge2} = [font=\fontsize{16}{0}\selectfont,draw,thick,->,green]
	      \tikzstyle{edge3} = [font=\fontsize{16}{0}\selectfont,draw,thick,->,red]

	      \node[state]         (A)                              {$v_1/r_1$};
	      \node[state]         (B) [above of=A,yshift=11mm]         {$v_2/r_2$};
	      \node[state]         (D) [above of=B,yshift=11mm]         {$v_4/r_2$};
	      \node[state]         (C) [left of=D,yshift=-5mm]             {$v_3/r_2$};
	      \node[state]         (E) [right of=D,yshift=-5mm]           {$v_5/r_2$};

	      \path[edge] (A) edge [bend right] node {2} (B)
	                  (A) edge [bend left]  node {1} (B);

	      \path[edge1] (A) edge [bend left]  node {1} (C)
	                   (B) edge [bend left] node {2} (C);

	      \path[edge2] (A) edge [bend right] node {2} (D)
	                   (B) edge              node {1} (D);

	      \path[edge3] (B) edge [bend left]  node {1} (E)
	                   (B) edge [bend right] node {2} (E);

	\end{tikzpicture}
    \label{figure:EG}

}}
\end{center}
\caption{(a) Derivations in $(\Delta) \Tcp$ and TG $G$ from Example~\ref{example:running}.}
\end{figure}
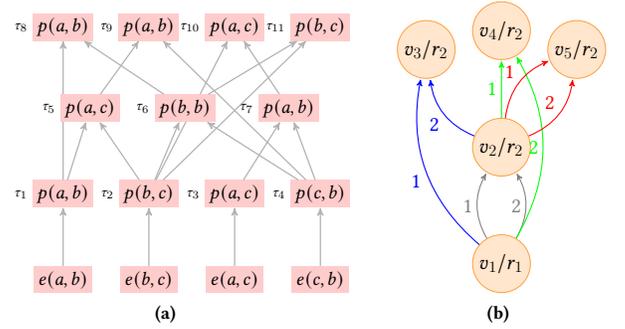

Figure~\ref{figure:derivations} organizes the derivations in the first three
rounds of Example~\ref{example:running} into a graph $\Gamma$ including an edge from fact
${\alpha_1, \dots, \alpha_n}$ to fact $\alpha$, for each rule instantiation
${\alpha \leftarrow \alpha_1 \land \dots \land \alpha_n}$. This figure reveals
the close correspondence between each derivation tree and the formulas computed
at each round of $\Tcp$ and $\Delta \Tcp$. Consider, for instance, fact
${p(a,b)}$. There are four occurrences of ${p(a,b)}$ in $\Gamma$, each one
defining a different derivation tree. We use $\tau_1$ and $\tau_7$ to denote the
derivation trees of ${p(a,b)}$ of depth one and two. Tree $\tau_1$ has a single
leaf node, ${e(a,b)}$ which coincides with the formula of ${p(a,b)}$ in the
first round of $\Tcp$ and $\Delta \Tcp$. Tree $\tau_7$ has two leaf nodes,
${e(a,c)}$ and ${e(c,b)}$. The conjunction of these two nodes results in the
intermediate formula $\mu^2_{p(a,b)}$. Formula ${\lambda^2_{p(a,b)}}$ is
computed by aggregating $\tau_1$ and $\tau_7$.

The above suggests an alternative approach to $\Tcp$ and $\Delta\Tcp$, that is
to maintain all derivation trees of a fact $\alpha$ and aggregate them to
compute its lineage. Computing the probability of the lineage gives us then the
probabilty of  $\alpha$ in $\Pp$~\cite{kimmig_demoen_de_raedt_costa_rocha_2011}.
Computing and maintaining the derivations in an efficient fashion is where
Trigger Graphs (TGs) come to the rescue.

A TG is an acyclic graph where each node is associated with a rule.
Figure~\ref{figure:EG} shows a TG computed out of the rules in
Example~\ref{example:running}. In the figure, we write, next to each node $v_i$,
the rule it is
associated with. For instance, node $v_1$ is associated with $r_1$, while all
the remaining nodes are associated with rule $r_2$. A TG can be seen
as a ``blueprint'' that tells us how to compute least Herbrand models. The
instructions are contained in the edges because they indicate the sets of facts over
which rules will be instantiated. For instance, the two edges from $v_1$ to
$v_2$ indicate that both facts in the premise of $r_2$ will be instantiated over
the facts derived by $r_1$ over $\Fp$.

Our work exploits TGs to efficiently compute all derivation trees.
Probabilistic reasoning in a TG-guided fashion allows us to overcome
\textbf{L1}--\textbf{L3}. Regarding \textbf{L1}, we show that to ensure
termination, we simply need to check whether a fact has been derived multiple
times in the same derivation tree, without performing any comparison of boolean
formulas. Regarding \textbf{L2}, we can exploit the topology of the TG to avoid
storing full copies of the derivation trees on each node, resulting in
significant memory savings. Finally, in contrast
to~\cite{DBLP:journals/corr/abs-1911-07750}, our approach does not need to
rewrite the rules into more complex ones or to introduce additional rules. This
addresses \textbf{L3}.

Notice that we cannot use the definition of TGs
from~\cite{tsamoura2021materializing} for our purposes. This is because TGs are
not designed to store the lineage of the inferred facts. To overcome this
limitation, we must perform several modifications that include extending TGs to
maintain the provenance, defining a new termination criterion (as the one
used for TGs is not sufficient), and implementing mechanisms that collapse the
lineage to avoid a memory blow-up. The modifications conflated into a new type
of TG which we call \emph{lineage TG} (LTG).

\section{Probabilistic Reasoning with TGs}\label{section:probabilistic-TGs}

We present our proposed technique.
We start by recapitulating
the notion of Execution Graphs (EGs), the basis of TGs
\cite{tsamoura2021materializing}, and discuss how to compute Herbrand models with
them. Then, we introduce our procedure for computing EGs that are suitable for
probabilistic query answering (Section~\ref{sec:alg1}).
Finally, we show that
our procedure is correct, i.e., it produces an EG that allows us to correctly
compute the lineage, and hence the probabilities, of the query answers
(Section~\ref{sec:correctness}). We refer to such TGs as \emph{lineage TGs}.

\begin{definition}
\label{definition:execution-graph} [From \cite{tsamoura2021materializing}]
	An \emph{execution graph} (EG) for a set of rules $\Rp$ is an acyclic, node- and edge-labeled digraph ${G = ( V, E , \rulname, \ell)}$,
      where $V$ and $E$ are the {sets of nodes and edges of the graph}, respectively, and $\rulname$ and $\ell$ are the node- and edge-labeling functions.
	{Each node $v$
	(i) is labeled with some rule, denoted by $\rul{v}$, from $\Rp$; and
	(ii) there can be a labeled edge of the form ${u \rightarrow_j v}$, from node $u$ to node $v$, only
	if the $j$-th predicate in the body of $\rul{v}$ equals the head
      predicate of $\rul{u}$.}
\end{definition}

Figure~\ref{figure:EG} shows an EG $G$ for the rules from
Example~\ref{example:running}. %
An EG $G$ for a set of rules $\Rp$ delineates a plan for executing the rules in
$\Rp$ over a set of facts $\Fp$. If $G$ contains a plan that computes a least
Herbrand model, then we say that $G$ is a TG. Indeed, the graph
in Figure~\ref{figure:EG} is a TG. Reasoning over $\Fp$ using $G$ involves
traversing the graph in a bottom-up fashion, instantiating each rule $r$
associated with a node $v$ using the facts associated with its parent nodes and
storing the results within $v$.
If a node $v$ has no parents, then we say it is a \emph{source} node and $r$ is
instantiated using the facts in $\Fp$. The \emph{depth of a node} $v$ in $\EG$ is the number of nodes
in the longest path that ends in $v$.  The \emph{depth} $\Depth{\EG}$ of $\EG$ is 0 if
$\EG$ is the empty graph; otherwise, it is the maximum depth of the nodes in
$\EG$.

Below, we illustrate an example of reasoning over EGs (and TGs).

\begin{example}	\label{example:running:instantiations}
    We demonstrate how reasoning over the facts from Example~\ref{example:running} works using the EG from Figure~\ref{figure:EG}.
    Reasoning starts from $v_1$, then proceeds to $v_2$ and finishes with $v_3$, $v_4$ and $v_5$.
    As $v_1$ has no incoming edges, we instantiate the premise of $r_1$ (the rule associated with rule $v_1$)
    over all facts in $\Fp$. The term mappings\footnote{The notation ${h_i = (c_1,c_2)}$ is short for ${h_i = \{X \mapsto c_1, Y \mapsto c_2\}}$.}
    ${h_1 =(a,b)}$, ${h_2 = (b,c)}$, ${h_3 = (a,c)}$ and ${h_4 = (c,b)}$ will be computed
    when instantiating $r_1$ associated with $v_1$.
    All the facts that result after instatiating the conclusion of $r_1$, using
    each term mapping $h_i$, will be stored in $v_1$.

    After reasoning over $v_1$, the next node to consider is $v_2$, which is
    associated with rule $r_2$.  The edges ${v_1 \rightarrow_1 v_2}$ and ${v_1
    \rightarrow_2 v_2}$ dictate that both atoms in the premise of $r_2$ must be
    instantiated using facts stored within $v_1$. The term
    mappings\footnote{The notation ${h_i = (c_1,c_2,c_3)}$ is short for ${h_i =
    \{X \mapsto c_1, Z \mapsto c_2, Y \mapsto c_3\}}$.} ${h_5 =(a,b,c)}$, ${h_6
= (b,c,b)}$ and ${h_7 = (a,c,b)}$ will instantiate $r_2$ in the context of $v_2$
and $\Fp$ and the derived facts ${p(a,c)}$, ${p(b,b)}$ and ${p(a,b)}$ will be
stored in $v_2$.
\end{example}

\subsection{EGs for probabilistic reasoning}
\label{sec:alg1}

The structure of an EG (or TG) maps to a series of steps to infer the facts in the Herbrand model. By tracing back the rule instantiations, we can compute all derivation trees for the facts and extract their lineage. From now on, we assume without loss of generality that EGs are \emph{canonical}: non-leaf nodes
$v$ are associated with rules of the form ${p(\vec{X}) \leftarrow \bigwedge \nolimits_{j=1}^n p_j(\vec{X}_j)}$
and the EGs include an edge of the form ${u_j \rightarrow_j v}$, for each ${1
\leq j \leq n}$.
As shown in \cite{tsamoura2021materializing}, we can always rewrite the rules into a form leading to canonical EGs.

A major difference against reasoning in a non-probabilistic setting is that now, instead of storing a set of facts inside the nodes, we must store their associated derivation trees.
The trees in the nodes depend on a certain context: the ancestor nodes in the EG.

\begin{definition}\label{def:instrule} Let $(\Rp, \Fp)$ be a program,
    \EG{} be a canonical EG for $\Rp$ and $v$ be a node in \EG{} associated with a rule $r$. The set of trees $\trees(\alpha,v,\Fp)$ is constructed as follows:
    \begin{compactitem}
	    \item if $v$ is a source node,
	    for each instantiation $\alpha \leftarrow \alpha_1\land \ldots \land \alpha_n$
        of $r$ so that each $\alpha_i$ is in $\Fp$,
        $\trees(\alpha,v,\Fp)$ includes a tree with {root $\alpha$} and edges $\alpha_i \rightarrow \alpha$; otherwise,

	   \item for each instantiation $\alpha \leftarrow \alpha_1 \land \ldots \land \alpha_n$ of $r$ so that for each $\trees(\alpha_i,u_i,\Fp) \neq \emptyset$,
	   $\trees(\alpha,v,\Fp)$ includes for each combination of trees ${(\tau_1,\dots,\tau_n)}$ from ${\trees(\alpha_1,u_1,\Fp) \times \dots \times \trees(\alpha_n,u_n,\Fp)}$, a tree with {root node $\alpha$} and an edge from the root of each $\tau_i$ to $\alpha$
	   (recall that \EG{} is canonical).
    \end{compactitem}

    We refer to a tree in $\trees(\alpha,v,\Fp)$ as a {
    derivation tree}.
\end{definition} %

Figure~\ref{figure:derivations} shows the derivation trees
$\tau_1$--$\tau_{11}$ computed when reasoning over the facts in Example~\ref{example:running}
using the EG from Figure~\ref{figure:EG}.
Throughout, we write $\roottree{\tau}$ to denote the fact at the root of the derivation tree $\tau$ and $\children{v}$ to denote the subtrees whose root has an edge to node $v$. Moreover, we tag every node with a label that specifies how the fact can be derived from its (possible) ancestors. The default label is $\andpred$, which indicates that all ancestor facts are needed to derive the fact in the node. In the next section, we will introduce and additional label, namely $\orpred$, to specify alternative derivations.

\begin{algorithm}[tb]
    	\caption{$\mat(\Pp)$, where $\Pp=(\Rp,\Fp,\pi)$}\label{alg:online}
	\begin{algorithmic}[1]
		\State ${k \defeq 0}$; \; $G^0$ is an empty EG for $\Rp$;
		\Do 																												\label{algorithm:while:begin}
			\State ${k \defeq k + 1}$;

			\State Compute $G^k$ starting from $G^{k-1}$ in an incremental fashion				\label{algorithm:online:inc}
			\For{\textbf{each} node $v$ of depth $k$}
			            \State $\treeset{v}{\Fp} \defeq \emptyset$                                   				\label{algorithm:node:init}
			            \For{\textbf{each} ${\trees(\alpha,v,\Fp)\neq \emptyset}$ for some $\alpha$} 		\label{alg:online:homomorphism:start}
			                \For{\textbf{each} derivation tree $\tau\in \trees(\alpha,v,\Fp)$}                  		\label{algorithm:tree:start}
			                    \If{$\tau$ is not redundant w.r.t. $\alpha$}                             				\label{algorithm:filtering}
			                        \State \textbf{add} $\tau$ to $\treeset{v}{\Fp}$                 				\label{algorithm:storetrees}
			                    \EndIf
			                \EndFor                                                                  					\label{algorithm:tree:end}
			            \EndFor

			    \If{$\treeset{v}{\Fp} = \emptyset$}
			        \textbf{remove} $v$ from $G$ \label{algorithm:removenode}
			    \EndIf                                                                       						\label{algorithm:node:end}
			\EndFor

		\doWhile{$\depth{G^k} \neq \depth{G^{k-1}}$}
		\State \textbf{return} $G^{k}(\Fp)$
	\end{algorithmic}
\end{algorithm}

Another major difference against reasoning in a non-probabilistic setting relates to termination, which occurs  when all facts inferred in the current round  are redundant. In a non-probabilistic setting, a fact is redundant if it has been previously derived. In the probabilistic setting though, that condition compromises correctness. To decide whether a derivation is redundant, we must take into account its associated derivation tree. It turns out that it suffices to discard a derivation tree $\tau$ if
$\alpha$ appears in $\tau$ more than once. If that holds, then we say $\tau$ is \emph{redundant} w.r.t. $\alpha$.

We are now ready to present our reasoning procedure that constructs EGs
suitable for probabilistic reasoning. The procedure, called \textit{Probabilistic Reasoning} ($\mat$), is
outlined in Algorithm~\ref{alg:online}. Given a probabilistic program
$\Pp=(\Rp,\Fp,\pi)$ as input, the procedure proceeds in rounds.  At each round
$k$, it first computes an EG of depth $k$ (line 4).
This computation is done incrementally, that is, the procedure considers all
nodes of depth $<k$ and then adds all possible nodes that we can construct by
instantiating the rules over them.
Then, Algorithm~\ref{alg:online}
executes the rules associated with the nodes of depth $k$. Executing a rule
associated with a node $v$ involves computing the corresponding
derivation trees (line~\ref{alg:online:homomorphism:start}) and storing a subset
of them in the set $\treeset{v}{\Fp}$, which contains the trees associated to
$v$ (line \ref{algorithm:storetrees}).
Deciding whether to discard a tree is checked as discussed above (line~\ref{algorithm:filtering}).
Finally, nodes are removed
if $\treeset{v}{\Fp}$ is empty (line \ref{algorithm:removenode}). $\mat$ ends
when all nodes in round $k$ have
been removed.

\begin{example} We demonstrate Algorithm~\ref{alg:online} over the running example.
    In the first iteration, Algorithm~\ref{alg:online} computes an EG including
    only node $v_1$ from Figure~\ref{figure:EG} and stores the trees ${\tau_1}$
    to $\tau_4$ within $\treeset{v_1}{\Fp}$. In the second iteration,
    Algorithm~\ref{alg:online} computes the EG including the nodes $v_1$ and
    $v_2$ from Figure~\ref{figure:EG} and stores the trees ${\tau_5}$, $\tau_6$
    and $\tau_7$ within $\treeset{v_2}{\Fp}$.  In the third iteration,
    Algorithm~\ref{alg:online} adds the nodes $v_3$, $v_4$ and $v_5$ to the
    graph computed in the previous round, see Figure~\ref{figure:EG}. Let us
    focus on $v_3$. Despite that there exists a derivation tree $\tau_8$ in the
    context of $v_3$ and $\Fp$, $\treeset{v_3}{\Fp}$ will be empty. This is due to the
    fact that the fact in the root of $\tau_8$ occurs also in an internal
    node. For similar reasons, no derivation trees are added to
    $\treeset{v_4}{\Fp}$ and $\treeset{v_5}{\Fp}$ and hence, Algorithm~\ref{alg:online} terminates.
\end{example}

Please notice that since the derivations are organized inside the TG, we do not need to
fully store all the trees in $\treeset{v}{\Fp}$. Instead, we can exploit
structure sharing \cite{vlog} and store only the roots of the derivation trees
and pointers to their ancestors. To compute the lineage, we can reconstruct
the derivation trees on-the-fly by traversing the TG.

\subsection{Correctness}
\label{sec:correctness}

The correctness of $\mat(\Pp)$ is shown in a series of steps.
Firstly, we show how derivation trees are used to compute the atoms'
lineage. Essentially, the derivation trees produced when reasoning
over an EG allow us to reconstruct models like the ones from
\cite{DBLP:journals/corr/abs-1911-07750}.

In Lemma~\ref{lemma:tree-formula-correspondence} below, we consider
a simplification of Algorithm~\ref{alg:online} in which the condition in the step in line~\ref{algorithm:filtering}
is ignored so that each tree visited in line~\ref{algorithm:tree:start} is added to node $v$.
We will later revise this assumption. For now, with this simplification in place, we denote by $G^i(\Fp)$ the derivation trees that are stored within the nodes of
depth $i$ in $G$ when reasoning over $\Fp$ and $\Pp$ with Algorithm~\ref{alg:online}. %
For a derivation tree $\tau$, we also denote by
$\phi(\tau)$ the Boolean formula that results after taking the conjunction of
the leaf nodes in $\tau$.
Similarly, we consider a simplification of $\Tcp$ that avoids performing Boolean formulas checks at the end of each round
and denote by $\Ht^i$ the instance computed at the end of the $i$-th iteration, where $\Ht^0 = \{(f,f) \mid p::f \in \Fp \}$.
It turns out that there is a one-to-one correspondence between the
lineage formulas that are computed by these two simplified algorithms.

\begin{restatable}{lemma}{lemmacorrespondence}\label{lemma:tree-formula-correspondence}
	For each ${i \geq 0}$,
		${(\alpha, \lambda^i_{\alpha}) \in \Ht^i}$ if-f $\bigvee \nolimits_{j=1}^m \phi(\tau_j) \equiv \lambda^i_{\alpha}$,
		where ${\tau_1,\dots,\tau_m}$ are all trees in $G^i(\Fp)$ with fact $\alpha$ as root.
\end{restatable}
Lemma~\ref{lemma:tree-formula-correspondence} indicates that to compute the probability of an atom $\alpha$, it suffices to collect all the derivation trees for $\alpha$ stored within $G(\Fp)$, compute the formulas out of each tree and, finally, take the disjunction of those formulas.

Now, let us discuss termination. From Lemma~\ref{lemma:tree-formula-correspondence},
it follows that deciding when to terminate reduces to deciding when the formula
$\phi(\tau)$ of a derivation tree $\tau$ for an atom $\alpha$ is
logically redundant due to the formula $\phi(\tau')$ of another
derivation tree $\tau'$ for $\alpha$, i.e., ${\phi(\tau) \vee \phi(\tau') \equiv
\phi(\tau')}$ holds. It is easy to see that when $\tau$ has $\tau'$ as a
subtree, then $\phi(\tau')$ is a conjunct within formula
$\phi(\tau)$, and hence ${\phi(\tau) \vee \phi(\tau') \equiv \phi(\tau')}$.
When the above holds, we say that the derivation of $\alpha$ under $\tau$ is \emph{superfluous} w.r.t. the derivation of $\alpha$ under $\tau'$.

\begin{restatable}{proposition}{propequiv}\label{proposition:equiv}
    For two derivation trees for $\alpha$, $\tau$ and $\tau'$, if
    $\tau'$ is a subtree of $\tau$, then ${\phi(\tau) \vee \phi(\tau') \equiv
\phi(\tau')}$ holds.
\end{restatable}

To detect superfluous derivations of $\alpha$, it suffices to check whether $\alpha$
occurs in an internal node of its newly computed derivation tree $\tau$, i.e., to check whether $\tau$ is redundant w.r.t. $\alpha$,
as we formalized it in the previous section. Therefore, it is safe to re-enable the check in line~\ref{algorithm:filtering} (which we disabled at the beginning of our discussion) since its task is precisely to discard redundant derivations. In this way, reasoning terminates when all nodes of depth $k$ are empty.

\begin{restatable}{lemma}{thmtermination}\label{theorem:termination}
	$\mat(\Pp)$ terminates for each probabilistic program $\Pp$ admitting a finite Herbrand base.
\end{restatable}

For an atom $\alpha$, we define its lineage in $G(\Fp)$ as the formula that results after taking the disjunction of the formulas of the derivation trees for $\alpha$ in $G(\Fp)$.
We are now ready to introduce the notion of lineage TGs.

\begin{definition}
\label{definition:trigger-graph}
	For a probabilistic program $\Pp=(\Rp,\Fp, \pi)$, an EG $G$ for $\Pp$
    is a \emph{lineage TG} for $\Pp$, if
	for each atom ${\alpha \in \Ap \setminus \Fp}$,
	the lineage of $\alpha$ in $G(\Fp)$ is logically equivalent to the lineage of $\alpha$ in $\Pp$.
\end{definition}

The following results establishes the correctness of Algorithm~\ref{alg:online}, which follows from
Lemma~\ref{lemma:tree-formula-correspondence}, Proposition~\ref{proposition:equiv}
and Lemma~\ref{theorem:termination}.

\begin{restatable}{theorem}{thmparameterizedTG} \label{theorem:parameterized-TG}
    $\mat(\Pp)$ is a lineage TG for any probabilistic program $\Pp$.
\end{restatable}

Moreover, at each round $k$ of reasoning over $\Pp$, the probability of each atom $\alpha$
that is computed based on its lineage in $G^k(\Fp)$ is a lower bound of the actual probability of $\alpha$.
\begin{restatable}{corollary}{thmbounds} \label{theorem:bounds}
	For each probabilistic program $\Pp$ and each atom ${\alpha \in HB(\Pp)}$, the probability of its lineage
	in $G^k(\Fp)$ is less than the probability of $\alpha$ in $\Pp$.
\end{restatable}
The corollary directly follows from
Lemma~\ref{lemma:tree-formula-correspondence} from above and the monotonicity of lineage.

\section{Collapsing the lineage} \label{sec:compression}

A limitation of Algorithm~\ref{alg:online} is that a node may contain multiple derivation trees for the same fact.
The above may lead to an exponential growth in the number of derivations, as each of these derivation trees can be considered in future rule instantiations.
This phenomenon can be   observed in practice. For instance, it can be observed when reasoning under equality rules (\emph{sameAs} \cite{mnph15owl-sameAs-rewriting,BenediktMT18}).
It is also observed in $\vqar$, see Section~\ref{sec:evaluation}.

To counter this problem, we propose an optimization that collapses such trees into a single one to reduce the memory consumption and the runtime.
We provide a demonstrating example.

\begin{figure}[t]
    \center
    \subfloat[All derivation trees are individually stored.]{\includegraphics[scale=0.35]{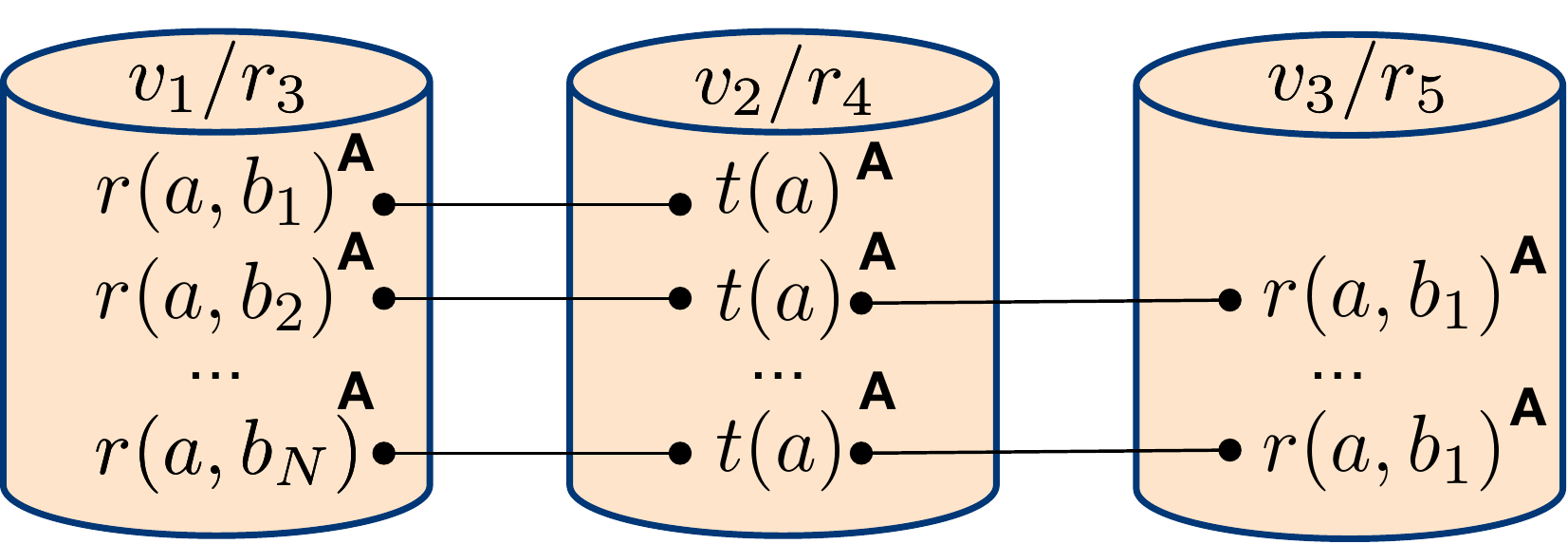}
    \label{fig:compression:default}}\\
    \subfloat[All derivation trees rooted with $t(\alpha)$ are collapsed.]{\includegraphics[scale=0.35]{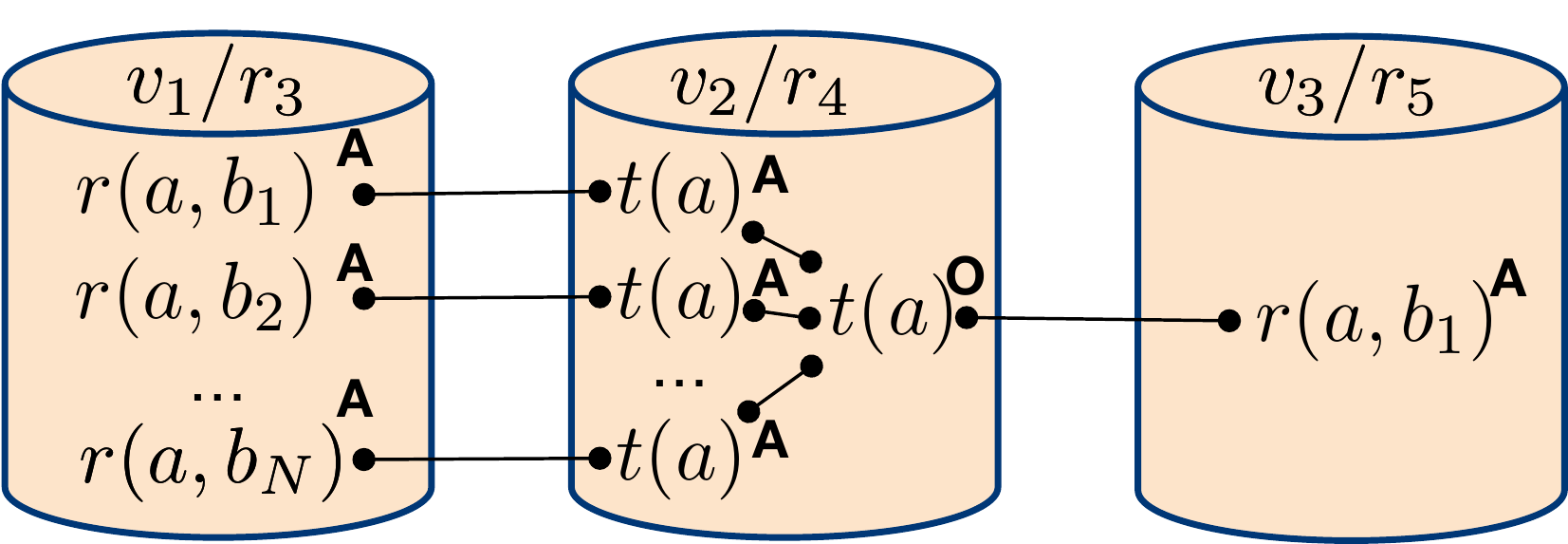}
    \label{fig:compression:better}}
    \caption{Different ways of storing $\Gamma$ in Example~\ref{example:compression}. The superscript \textbf{A} ({\textbf{O}}) means that the fact is labeled with $\andpred$ ($\orpred$). For clarity, edges to facts in $\Fp$ are not shown.}
\end{figure}

\begin{example} \label{example:compression}
    Consider a program with the following the three rules
    \begin{align}
        r(X,Y) & \leftarrow q(X,Y)  \tag{$r_3$}\label{eq:compression-r1}\\
        t(X)    & \leftarrow r(X,Y) \tag{$r_4$}\label{eq:compression-r2}\\
        r(X,Y) & \leftarrow t(X) \wedge s(X,Y) \tag{$r_5$}\label{eq:compression-r3}
    \end{align}
    and let $\Fp$ be a set of facts that include $q(a,b_i)$, for ${1 \leq i \leq N}$, and $s(a,b_1)$.
    Figure~\ref{fig:compression:default} shows the lineage TG $\Gamma$ for the
    corresponding program and the derivation trees stored within the
    nodes in $\Gamma$ (for clarity, edges to facts in $\Fp$ are not shown). Node $v_2$ stores $N$ different trees with
    ${t(a)}$ as root. $N-1$ of such trees, combined with $s(a,b_1)$ by rule $r_5$,
    lead to $N-1$ trees with $r(a,b_1)$ as root, to be stored in $v_3$.
\end{example}

Example~\ref{example:compression} shows that a more space-efficient
technique would be to keep only the derivation tree for ${t(a)}$ in node $v_2$ so that only one
tree with root ${r(a,b_1)}$ is added to $v_3$. The challenge in doing so is
to remember that $t(a)$ can be inferred in $N$ different ways.

To address this issue, we first introduce an additional label called $\orpred$.
Recall that in Section~\ref{section:probabilistic-TGs}, all nodes in
derivation trees are labeled with $\andpred$. Intuitively, facts labeled with $\orpred$ differ from them because they hold if only one ancestor holds. Then, we define the process of collapsing multiple derivation trees into one as follows. %
\begin{definition}\label{definition:collapse}
    Let $\tau_1, \tau_2, \dots, \tau_m$ be a collection of derivation trees that have the same root fact
    $\alpha$. Then, $\collapse{\tau_1, \tau_2, \dots, \tau_m}$, where $m>1$, is the derivation tree $\tau$ defined as follows:

    \begin{enumerate}
        \item $\roottree{\tau}$ is fact $\alpha$ and it is labeled with $\orpred$; and

        \item there exists an edge from $\roottree{\tau}$ to each $\roottree{\tau_i}$.
    \end{enumerate}

\end{definition}

By collapsing the derivation trees, we can reduce the number of future rule
instantiations. For instance, Figure~\ref{fig:compression:better} shows the
effect of this operation on Example~\ref{example:compression}. Here,
all trees in $v_2$ with $t(a)$ as root are collapsed into a single entry. When
rule $r_5$ is applied on $v_2$, then $t(a)$ is considered only once, which leads
to the derivation of one tree instead of $N-1$ ones.

If there are no $\orpred$-labeled nodes, then collecting the lineage of a derivation tree simply consists of collecting the leaves. Otherwise, the process has to be amended so that the branches introduced by $\orpred$-labeled nodes are \textit{unfolded} into multiple trees.
We formalize the notion of unfolding derivation trees as follows.

\begin{definition}
	\label{def:unfolding}
    The unfolding $\unfold{\tau}$ of a derivation tree $\tau$ with $\children{\roottree{\tau}}=\{\tau_1, \dots, \tau_m\}$ is:
	    \begin{itemize}
	        \item[$\star$] $\{\tau\}$, if no node in $\tau$ has label $\orpred$; or
	        \item[$\dagger$] $\bigcup \nolimits_{\tau_i} \unfold{\tau_i}$, if $\roottree{\tau}$ has label $\orpred$; or,
	        \item[$\ddagger$] $\Delta$, where,
	        for each combination of trees ${(\delta_1,\dots,\delta_m)}$ from ${\unfold{\tau_1} \times \dots \times \unfold{\tau_m}}$,
	        $\Delta$ includes a derivation tree $\varepsilon$, such that $\varepsilon$ has the same root fact as $\tau$, label $\andpred$, and there is an edge from each $\roottree{\delta_i}$ to $\roottree{\varepsilon}$.
	    \end{itemize}
\end{definition}
In Definition~\ref{def:unfolding}, $\star$ regards the case where no collapsing took place; $\dagger$ regards the case where multiple derivation trees have been collapsed into one via Definition~\ref{definition:collapse} so that the root of the new tree is an $\orpred$-labelled node; finally, $\ddagger$ regards the case where $\orpred$ appears in an ancestor of $\tau$.
We illustrate Definition~\ref{def:unfolding} over Example~\ref{example:compression}.

\begin{example}\label{example:unfolding}
    Suppose that all trees in node $v_2$ have been collapsed into a single tree $\epsilon$ with root $t(a)$, see Figure~\ref{fig:compression:better}, and that
   node $v_3$ stores a single derivation tree $\tau$
    with the $\andpred$-labeled fact $r(a,b_1)$ as root.
    We discuss the process of unfolding $\tau$.
    Due to the presence of the $\orpred$-labelled node $t(a)$ in a non-root node of $\tau$,
    $\unfold{\tau}$ will fall into case $\ddagger$ and be defined as the set of trees constructed by computing the Cartesian product of the unfoldings of the two children of $\roottree{\tau}$, i.e., $\epsilon$ and the tree with single node $s(a,b_1)$ (the latter is not shown in Figure~\ref{fig:compression:better}). 
    Since $\epsilon$ is an $\orpred$-labeled node,
    $\unfold{\epsilon}$ is defined as the union of the unfoldings of its children, i.e., the $N$ trees with root $t(a)$ (case $\dagger$). Since none of them has an $\orpred$-labelled node, their unfoldings are defined by the base case ($\star$), which are the trees themselves.
\end{example}

As a collapsed tree $\tau$ encapsulates multiple ways to derive the same fact, we say that {$\tau$ is \emph{redundant} w.r.t. fact $\alpha$ if $\alpha$ occurs at least twice in \emph{every} derivation tree in $\unfold{\tau}$.} This means that if we need to check whether $\tau$ is redundant, then we do not always need to fully compute $\unfold{\tau}$ because we can we stop as soon as we find one non-redundant derivation tree. Returning to Example~\ref{example:unfolding}, $r(a,b_1)$ occurs twice in one derivation tree in $\unfold{\tau}$. However, $\tau$ is not redundant w.r.t. $r(a,b_1)$ as it contains other trees in which $r(a,b_1)$ occurs only once.

\begin{algorithm}[tb]
    \caption{$\matcompr(\Pp)$, where $\Pp=(\Rp,\Fp,\pi)$}\label{alg:online:compressed}
    \begin{algorithmic}[1]
        \State ${k \defeq 0}$; \; $G^0$ is an empty EG for $\Rp$;
        \Do
            \State ${k \defeq k + 1}$;

            \State Compute $G^k$ starting from $G^{k-1}$ in an incremental fashion				\label{algorithm:compressed:incremental}
            \For{\textbf{each} node $v$ of depth $k$}
                \State $\mathcal{T} \defeq \{ \trees(\alpha,v,\Fp)\neq \emptyset \mid        \alpha$ is a fact$\}$\;\; $\treeset{v}{\Fp} \defeq \emptyset$    \label{algorithm:alltrees}
                \For{\textbf{each} ${\{\tau_1,\dots,\tau_n\} \in \mathcal{T}}$}
                    \If{the average size of sets in $\mathcal{T}$ is $\geq t$}                                               \label{algorithm:shouldcompr}
                        \State $Z \defeq \{\collapse{\tau_1,\dots,\tau_n}\}$                          \label{algorithm:collapse}
                    \Else ~$Z \defeq \{\tau_1,\dots,\tau_n\}$  						\label{algorithm:donotcompr} 

                    \EndIf
                    \For{\textbf{each} tree $\tau \in Z$}       						\label{algorithm:beginuncompr}
                        \If{$\tau$ is not redundant w.r.t. $\alpha$} 					\label{algorithm:redundancycheck}
                            \State \textbf{add} $\tau$ to $\treeset{v}{\Fp}$					\label{algorithm:redundancypass}
                        \EndIf
                    \EndFor                                         									\label{algorithm:enduncompr}
                \EndFor
                \If{$\treeset{v}{\Fp} = \emptyset$}
                    \textbf{remove} $v$ from $G$
                \EndIf
            \EndFor
        \doWhile{$\depth{G^k} \neq \depth{G^{k-1}}$}
        \State \textbf{return} $G^{k}(\Fp)$
    \end{algorithmic}
    \hrule
\noindent\textbf{Note}: $t$ is a given threshold value (default value is 10).
\end{algorithm}

We outline in Algorithm~\ref{alg:online:compressed}, under the name \textit{Probabilistic COllapsed Reasoning}
($\matcompr$), the reasoning when some derivation trees may be collapsed. The procedure proceeds similarly as in Algorithm~\ref{alg:online}. Firstly, it computes all different derivation trees
that can be obtained via rule instantiations (line~\ref{algorithm:alltrees}).
Then, the algorithm processes one by one all the sets of trees that share the
same fact $\alpha$ as root. The condition in line~\ref{algorithm:shouldcompr}
decides whether the trees in $v$ should be collapsed using the threshold value (see discussion below).
If they should be, then every set of trees is collapsed as in Definition~\ref{definition:collapse}.
Otherwise, they are processed one-by-one (lines~\ref{algorithm:beginuncompr}-\ref{algorithm:enduncompr}).

Several strategies can be implemented to decide whether to collapse the derivation trees within a node.
In line~\ref{algorithm:shouldcompr}, we use a simple threshold value and leave more complex strategies for future work. Our strategy takes into account the average number
of derivation trees within a node having the same root fact. If that average is at least
$t=10$ (we chose the value 10, as it sets a reduction of at least one order of magnitude),
then we collapse the derivation trees in that node; otherwise, we store the trees one by one.

The following result establishes the correctness of Algorithm~\ref{alg:online:compressed}.

\begin{restatable}{theorem}{thmparameterizedTGcompressed} \label{theorem:parameterized-TG-compressed}
    For each probabilistic program $\Pp$, $\matcompr(\Pp)$ is a lineage TG for $\Pp$.
\end{restatable}

The above result follows from the close relationship between Algorithm~\ref{alg:online} and Algorithm~\ref{alg:online:compressed} and the correctness of the notion of redundancy with $\orpred$-labeled nodes.

Our technique for collapsing the lineage is similar to the technique from \cite{circuits-provenance} for computing provenance circuits. A \emph{provenance circuit} is a DAG of Boolean operators and facts
which can represent provenance in a compact fashion avoiding the exponential blow-up of
techniques based on provenance semirings \cite{provenance-semirings}.
In~\cite{circuits-provenance}, the authors provided an algorithm for computing provenance via propagating
circuits during the computation of the model. %
    The technique first creates a circuit that includes every fact in the database.
    Then, at each round $k$, it instantiates all rules using at least one
    fact derived in the ${(k-1)}$-th round-- that is the constrained introduced by SNE, see Section~\ref{section:preliminaries}.
    If such an instantiation ${\alpha \leftarrow \alpha_1 \wedge \dots \wedge \alpha_N}$
    is not possible, the computation terminates.
    Otherwise, it takes the following steps.
    If $\alpha$ is derived for first time, then it adds two $\vee$-nodes to the circuit, one annotated with the fresh variable $X_{\alpha}$
    and the second with the fresh variable $Y_{\alpha}$.
    Then, it adds a fresh $\wedge$-node $u$ and an edge from $X_{\alpha}$ to $u$.
    Finally, for each $\alpha_i$, it adds an edge from $u$ to the node annotated
    with $X_{\alpha_i}$, if $\alpha_i$ is an input fact, and, otherwise, to $Y_{\alpha_i}$.

\begin{figure}[tb]
\begin{center}
    \scalebox{0.6}{
        \def \oran{orange!30}
    \tikzstyle{superpeers}=[font=\fontsize{14}{0}\selectfont,draw=orange!75,fill=orange!20]
    \tikzstyle{edge} = [draw,thick,->]
    \begin{tikzpicture}[]

        \node[superpeers] (j) [above=0.5cm of e, label={[font=\fontsize{14}{0}\selectfont]above:$X^1_{r(a,b_1)}$}] {$\vee$};
        \node[superpeers] (k) [right=of j, label={[font=\fontsize{14}{0}\selectfont]above:$X^1_{r(a,b_2)}$}] {$\vee$};
        \node[] (k2) [right=0.5cm of k] [] {\ldots};
        \node[superpeers] (l) [right=0.5cm of k2, label={[font=\fontsize{14}{0}\selectfont]above:$X^1_{r(a,b_N)}$}] {$\vee$};

        \node[superpeers] (m) [right=2cm of l,label={[font=\fontsize{14}{0}\selectfont]above:$X^2_{t(a)}$}] {$\vee$};
        \node[superpeers] (mf) [below=0.5cm of m, label={[font=\fontsize{14}{0}\selectfont]below:$Y^1_{r(a,b_1)}$}] {$\vee$};
        \node[superpeers] (mg) [right=of mf,label={[font=\fontsize{14}{0}\selectfont]below:$Y^1_{r(a,b_2)}$}] {$\vee$};
        \node[] (mh) [right=0.5cm of mg] [] {\ldots};
        \node[superpeers] (mi) [right=0.5cm of mh,
        label={[font=\fontsize{14}{0}\selectfont]below:$Y^1_{r(a,b_N)}$}] {$\vee$};

        \node[superpeers] (e) [below=of j] {$\wedge$};

        \node[superpeers] (a) [] {$s(a,b_1)$};
        \node[superpeers] (b) [right=0.3cm of a] {$q(a,b_1)$};
        \node[superpeers] (c) [right=0.3cm of b] {$q(a,b_2)$};
        \node[] (c2) [right=0.1cm of c] [] {\ldots};
        \node[superpeers] (d) [right=0.1cm of c2] {$q(a,b_N)$};

        \node[superpeers] (i) [left=0.3cm of a,
        label={[font=\fontsize{14}{0}\selectfont]above:$Y^2_{t(a)}\;\;\;\;\;$}] {$\vee$};

        \path[edge] (j) -- (b);
        \path[edge] (k) -- (c);
        \path[edge] (l) -- (d);
        \path[edge] (m) -- (mf);
        \path[edge] (m) -- (mg);
        \path[edge] (m) -- (mi);

        \path[edge] (j) -- (e);
        \path[edge] (e) -- (a);
        \path[edge] (e) -- (i);
    \end{tikzpicture}
}
\end{center}
\caption{The provenance circuit from \cite{circuits-provenance} for Example~\ref{example:compression}.}
\label{fig:provenance-circuit}
\end{figure}
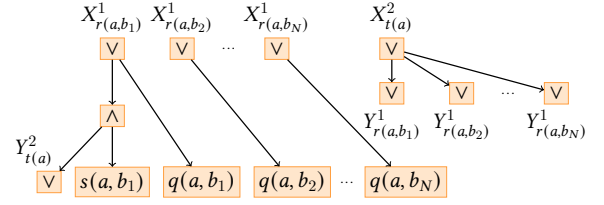

\begin{example}\label{example:provenance-circuits}
    Figure~\ref{fig:provenance-circuit} presents the circuit
    computed out of the probabilistic program from Example~\ref{example:compression}.
    The label of each node is shown at the top of it.
    If a fact has been derived by a rule, then the associated variable has as a superscript the round in which it was created, i.e.,
    variable $X^1_{r(a,b_1)}$ is created during the first round.
\end{example}

The $\vee$ ($\wedge$) nodes in the provenance circuit fulfill the same
function as the $\orpred$ ($\andpred$) labels. However, the way our approach
collapses the lineage is significantly different.
Firstly, our approach collapses only the lineage
stored within a single TG node. Instead, in \cite{circuits-provenance}, the
collapsing considers the entire model
as the technique is based on SNE and not on TGs. Secondly,
our approach is adaptive as the collapsing is activated only if it is
beneficial (see lines 8--9 of Algorithm~\ref{alg:online:compressed}). In contrast,
in~\cite{circuits-provenance}, the operation is always performed, even when not needed.
For instance, in
Example~\ref{example:provenance-circuits} two fresh nodes are created for each $r(a,b_i)$ fact. Instead, the collapsed tree representation in Figure~\ref{alg:online:compressed} stores each $r(a,b_i)$ fact once.

\section{Evaluation}\label{sec:evaluation}

\begin{table}[t]
    \caption{The considered benchmarks.
    In $\smokers$ and $\vqar$, \#DB and \#DR depend on $N$ and on each query, respectively.}
    \vspace{-1em}
\small
\begin{tabular}{p{1cm}
	S[table-format=4,table-column-width=0.7cm]
	S[table-format=4,table-column-width=0.7cm]
	S[table-format=4.1,table-column-width=0.7cm]
	S[table-format=4,table-column-width=0.7cm]
	}
	    & {\#R} & {\#DB} & {\#DR} & {\#Q} \\
	    \hline
	     			   $\lubmM$        	&	127		&	1M		&	1.7M 	&14			\\
	     			   $\lubmL$         	&	127		&	12M		&	18M		&14			\\
	     			   $\dbpedia$      	& 	9k 		&  	29M 	& 	33M 	&50			\\
	     			   $\claros$         	&	2k 		& 	13M 	& 	8M		&50			\\
                       		   $\yagoS$    	&   	221 	& 	1M 		&  	1M		&50   		\\
	     			   $\yagoM$ 		&   	396 	&  	1M 		& 	1M		&76   		\\
	     			   $\yagoL$ 		&   	571 	&    1M 		& 	1M		&50   		\\
	     			   $\wnrrS$  		&   66 		& 86k 		&    87k		&20    		\\
	     			   $\wnrrM$ 		&   116  		& 86k 		&    87k		&20   		\\
                       		   $\wnrrL$  		&   166 		& 86k 		&    87k		&50   		\\
                       		   $\smokers$ 	&   5 		& * 		& 	*		&110 	  	\\
                       		  $\vqar$ 		&   6 		& * 		& 	*		&1000 	  	\\
\end{tabular}
\label{tab:benchmarks}
\end{table}

We implemented our approach in a new engine called \sys{} and evaluated  its
performance against \textbf{\problog{}}
\cite{vlasselaer:ijar16} that implements $\Tcp$,
\textbf{\vp{}} \cite{DBLP:journals/corr/abs-1911-07750}, the
state-of-the-art implementation of $\Delta \Tcp$, and \textbf{\scallop{}}~\cite{scallop}, a recent approximate probabilistic reasoning engine. To our knowledge, these are
the only state-of-the-art engines for reasoning under the possible world semantics.
We ran \sys{} both with and without collapsing the lineage
denoting the cases by ``\sys{}
w/'' and ``\sys{} w/o'', respectively.

To compute the probabilities of the answers given their lineage, we
    considered three state-of-the-art tools: PySDD
    \cite{darwiche:ijcai11}, the d-tree compiler from
    \cite{d-tree} and c2d~\cite{c2d}. PySDD is a well-known WMC solver that can process formulas in Disjunctive Normal Form (DNF), the form of the lineage returned by \sys{}. PySDD will be our default solver, as it is adopted by all our competitors and supports DNF. The d-tree compiler is an alternative technique with
    competitive performance. Finally, c2d is another state-of-the-art solver
    that was ranked among the top three in the 2021 Model Counting Competition
    (\url{https://mccompetition.org/}). This solver requires formulas in
    \textit{Conjunctive Normal Form} (CNF). To convert lineage formulas from DNF to CNF we applied the relaxed Tseitin transformation \cite{relaxed-tseitin}
    that works in polynomial time in the size of the input formula.

All experiments ran on an Ubuntu 16.04 PC with an Intel i7 CPU and
94 GiB RAM.

\subsection{Benchmarks}

We considered benchmarks originating from the database, the probabilistic
programming, and the machine learning communities.

\begin{itemize}[leftmargin=*]
    \item $\lubm$ \cite{lubm} is a popular Datalog benchmark
        that has been used to evaluate \problog{}, \vp{}
        and other engines \cite{rdfox,vlog,tsamoura2021materializing}.
        We considered $\lubmM$ and $\lubmL$ that include 1M and 12M facts, respectively.
        We used the set of same 127 rules with our competitors
        and the 14 available queries.

    \item $\dbpedia$ \cite{dbpedia} is one of the most well-known KGs built from Wikipedia. $\claros$ \cite{claros} is an ontology of cultural heritage.

    \item $\smokers$ \cite{smokers} is a popular KB in the AI community.
        The KB includes PDBs encoding random power-law graphs
        of $N$ nodes and up to $2 \times N$ undirected edges.
        We considered $N$ between 10 and 20 as in \cite{DBLP:journals/corr/abs-1911-07750}
        and the 110 available queries.
        We limit the maximum reasoning depth to four and five steps as in
        \cite{DBLP:journals/corr/abs-1911-07750}.

    \item
        $\vqar$ \cite{scallop} has been proposed
        for rule-based reasoning in the context of visual question answering.
        The benchmark provides over 5000 pairs of queries and probabilistic programs. Each program includes (i) uncertain facts obtained by translating into relational form neural predictions on images and (ii) rules and facts taken from the CRIC ontology~\cite{cric}.
        We considered the 1000 queries requiring the most reasoning steps.
        \vqar{} is challenging because the number of derivations
        explodes combinatorially.

\end{itemize}

\begin{table*}[t]
    \caption{Total time (default is ms) to answer the queries in $\lubmM$
    and $\lubmL$ with $\problog$ (P), $\scallop$ (S), $\vp$ (vP) and $\ours$ (L). Probabilities are computed via PySDD (SDD), d-tree and c2d. Shaded cells contain the best times.}
\begin{adjustbox}{max width= \textwidth}
\begin{tabular}{p{1.5cm} | c  c  c  c  c  c  c  c  c  c  c  c  c c || c  c  c  c  c  c  c  c  c  c  c  c  c c}
	    & $Q_1$ & $Q_2$ & $Q_3$ & $Q_4$ & $Q_5$ & $Q_6$ & $Q_7$ & $Q_8$ & $Q_9$ & $Q_{10}$ & $Q_{11}$ & $Q_{12}$ & $Q_{13}$ & $Q_{14}$ & $Q_1$ & $Q_2$ & $Q_3$ & $Q_4$ & $Q_5$ & $Q_6$ & $Q_7$ & $Q_8$ & $Q_9$ & $Q_{10}$ & $Q_{11}$ & $Q_{12}$ & $Q_{13}$ & $Q_{14}$ \\
	    \hline
	     			   P+SDD     		& 59 & NA & NA & NA & NA & NA & NA & NA & NA & NA & NA & 78 & NA & 150 						& NA & NA & NA & NA & NA & NA & NA & NA & NA & NA & NA & NA & NA & NA\\
	     			   S(30)+SDD      	& 1.3s & NA & 729 & NA & 4.5s & \cellcolor{blue!15}817s & 6s & NA & NA & NA & 63 & 165s & 30s & 326 			& 15.5 & NA & 8.9s & NA & NA & NA & NA & NA & NA & NA & 372 & NA & NA & \cellcolor{blue!15}3.3s\\
	     			   vP+SDD   	     	& 587 & 7.2s & 306 & 5.6s & 13.6s & NA & 6.3s & NA & NA & 1.3s & 2s & 17.3s & 12.4s & 3.1s	& 7.3s & NA & 2.5s & NA & NA & NA & NA & NA & NA & NA & 2s & NA & NA & 38.7s\\
					   L w/o+SDD    	& 57 & 420 & \cellcolor{blue!15}38 & 1.1s & 1.3s & NA & 353 & 35.1s & 348s & 187 & 7 & 10.6s & 541 & 337		& 647 & 52s & 455 & 2.4s & 4.7s & NA & 2s & \cellcolor{blue!15}51.8s & NA & 1.7s & 31 & 12.7s & 6.1s & 4.9s \\
					   L w/+SDD     	& \cellcolor{blue!15}49 & \cellcolor{blue!15}383 & \cellcolor{blue!15}38 & \cellcolor{blue!15}175 & \cellcolor{blue!15}365 & NA & \cellcolor{blue!15}315 & \cellcolor{blue!15}21.8s & 174s & \cellcolor{blue!15}162 & \cellcolor{blue!15}5 & \cellcolor{blue!15}387 & \cellcolor{blue!15}176 & \cellcolor{blue!15}273			& \cellcolor{blue!15}617 & 46.1s & 444 & \cellcolor{blue!15}1.5s & 3.7s & NA & \cellcolor{blue!15}1.9s & 71.4s & NA & \cellcolor{blue!15}1.6s & \cellcolor{blue!15}21 & \cellcolor{blue!15}1.6s & 2.8s & 6s\\ \hline \hline

					   L w/+d-tree    	& \cellcolor{blue!15}49 & 676 & 40 & 461 & 595 & NA & 4.9s & 668s & \cellcolor{blue!15}108s & 1.5s & 6 & 1s & 206 & \cellcolor{blue!15}273		    & \cellcolor{blue!15}617 & \cellcolor{blue!15}42s & \cellcolor{blue!15}411 & 1.7s & \cellcolor{blue!15}2.7s & NA & 6.3s & 658s & NA & 2.9s & 21 & 2s & \cellcolor{blue!15}2.4s & 6s  \\
					   L w/+c2d     	& \cellcolor{blue!15}49 & 41s & 316 & 3.9s  & 62s  & NA & 27s & NA & NA & 2.4s & 6 & 13s & 6.2s & \cellcolor{blue!15}273 			& \cellcolor{blue!15}617  & NA & 1s & 7.4s & 113s & NA & 32s & NA & NA & 4.2s & 21 & 16s & 16s & 6s\\

\end{tabular}
\end{adjustbox} \label{tab:total-lubm10}
\end{table*}

\noindent\textbf{Rule mining benchmarks.} %
We also considered scenarios in which the rules are mined using AnyBurl~\cite{anyburl}, a state-of-the-art KG completion technique that outperforms both prior KG embedding techniques, e.g., ComplEx~\cite{complex}, and other rule mining techniques.
Each rule that is mined by AnyBurl is assigned a confidence value based on
its support in the data. To mine rules, we considered two KGs frequently used by the
machine-learning community: $\yago$3~\cite{mahdisoltani2014yago3} (called
$\yago$ thereafter) and $\wnrr$ \cite{wn18rr}. For each KG, we created three
different benchmarks by choosing for each predicate the top 5, 10, and 15 rules
with the highest confidence.
Both $\yago$ and $\wnrr$ come with sets of training, validation, and testing KG
triples. The training and validation triples are used to mine rules, while
the testing triples are used at reasoning time.

Each benchmark ${B}$ forms the basis to create different scenarios. Each
scenario is constructed from the databases, the rules, and the queries in ${B}$. We denote
scenarios by writing the name of the benchmark followed by possible parameters, e.g.,
$\smokers4$ uses the $\smokers$ KB and sets
the max reasoning depth to four.
Only $\smokers$ and $\vqar$ define the probability function $\pi$.
For the remaining benchmarks, we implemented $\pi$ by assigning
to each fact a random number within ${(0,1]}$. This is the same approach used
in \problog{} and $\vp{}$ \cite{DBLP:journals/corr/abs-1911-07750}.
We created queries of 1, 2, 3, and 4 atoms for benchmarks not providing queries,
using the method from
\cite{DBLP:conf/ecai/JoshiJU20}. The resulting queries require a variable number
of reasoning steps to be answered. Table~\ref{tab:benchmarks} reports statistics for all scenarios.
\#R, \#DB, \#DR and \#Q denote the number of rules, facts, distinct fact derivations and total number of queries.

\begin{figure*}[tb]
    \begin{adjustbox}{max width=\textwidth}
        \begin{tabular}{cccc}
             & \textbf{Reasoning} &  \textbf{Lineage} &  \textbf{Probability}\\
            \rotatebox{90}{\hspace{8mm}$\boldsymbol{\lubmM}$} &
            \x {\includegraphics[width = 0.4\textwidth]{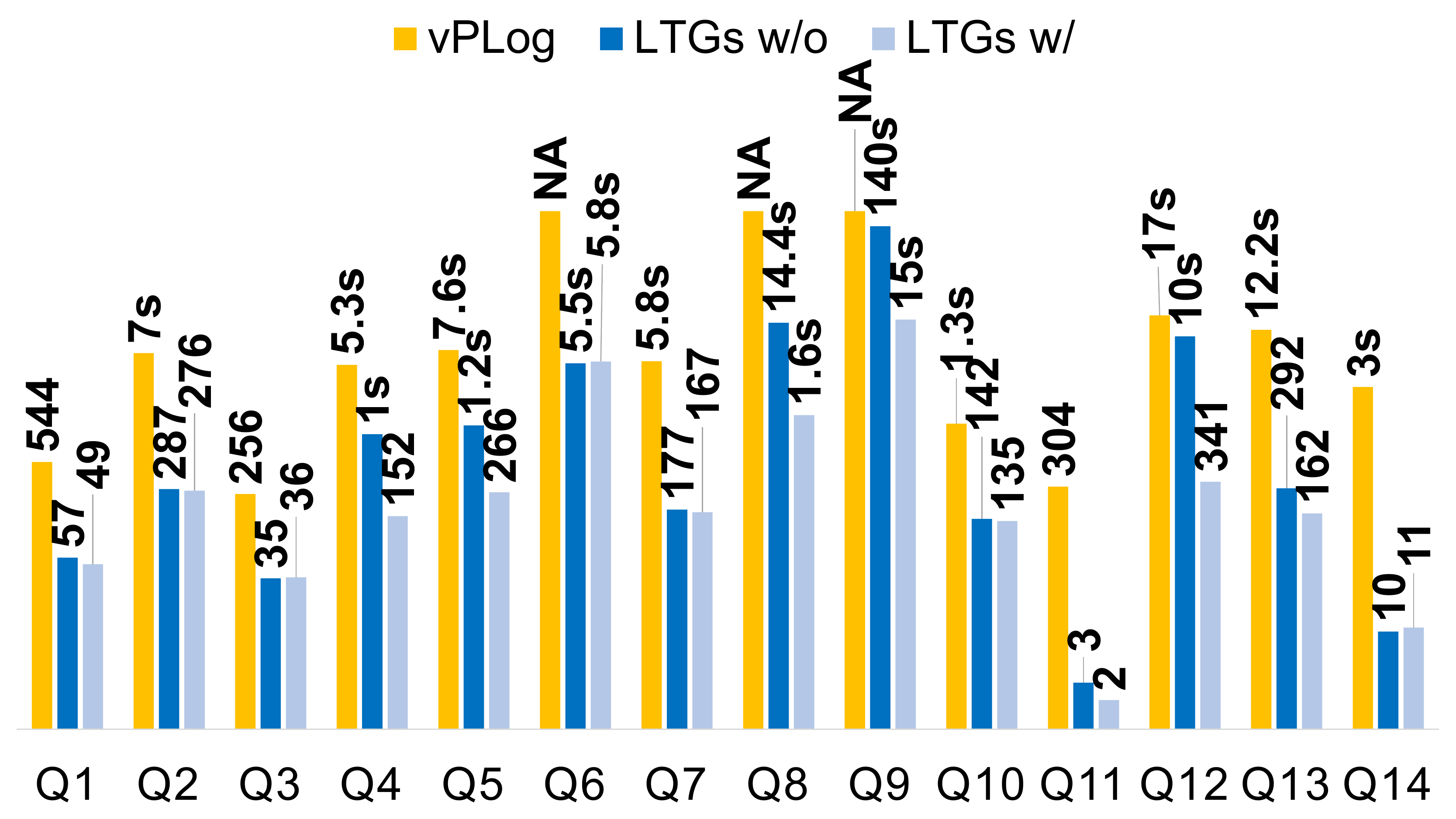}} &
            \x {\includegraphics[width = 0.4\textwidth]{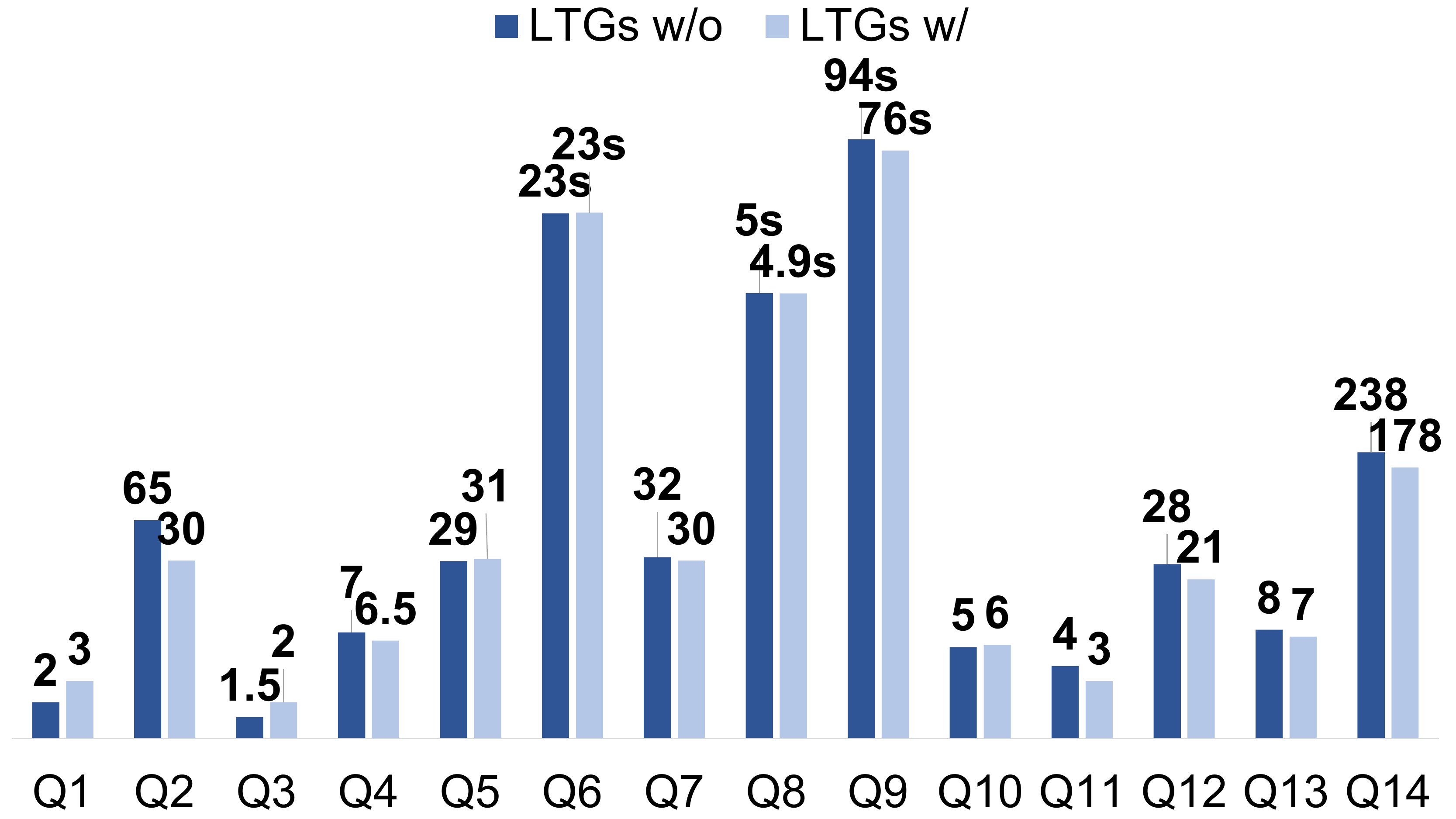}} &
            \x {\includegraphics[width = 0.4\textwidth]{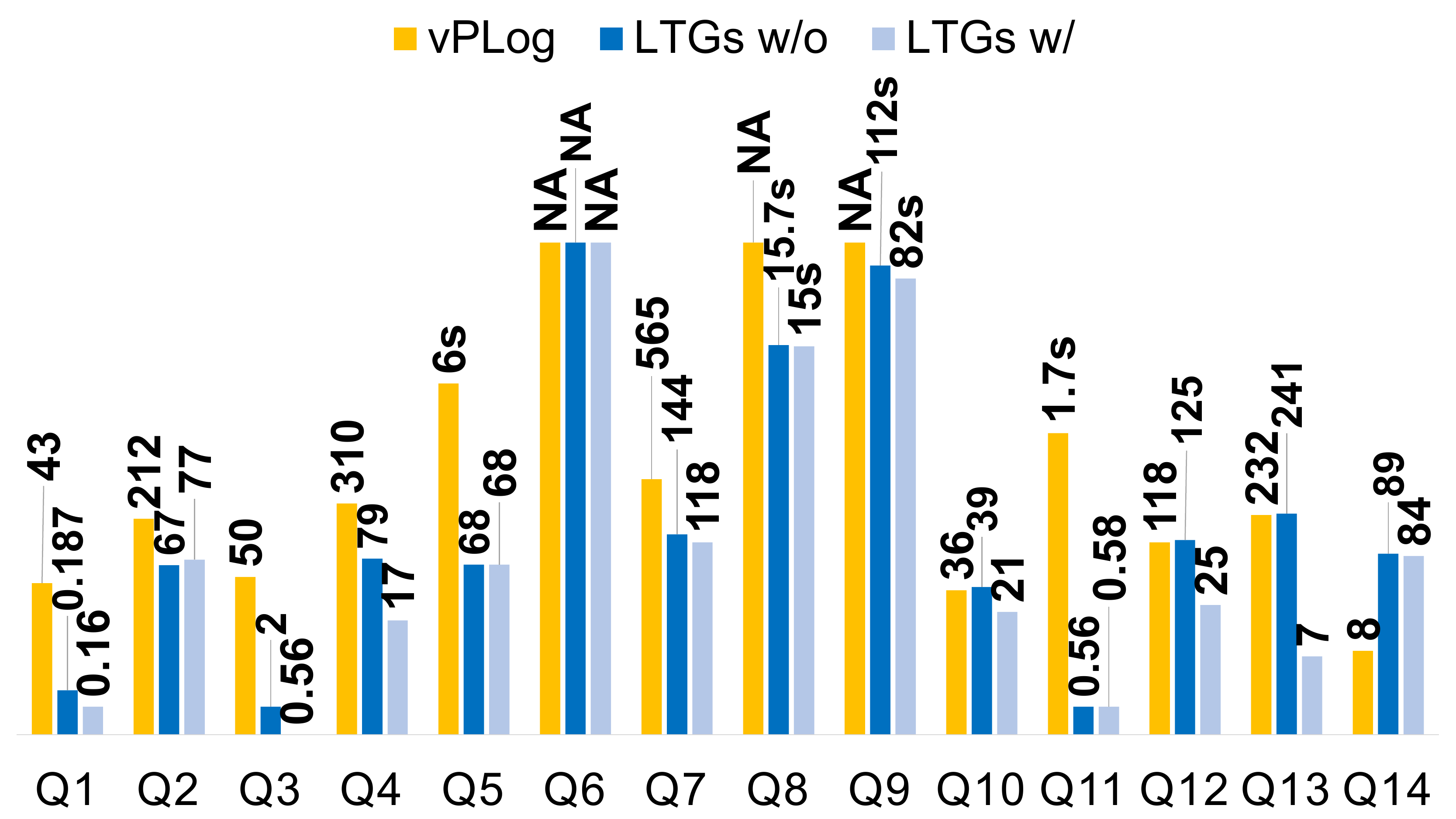}} \\

            \rotatebox{90}{\hspace{0.8cm}{$\boldsymbol{\lubmL}$}} &
            \x {\includegraphics[width = 0.4\textwidth]{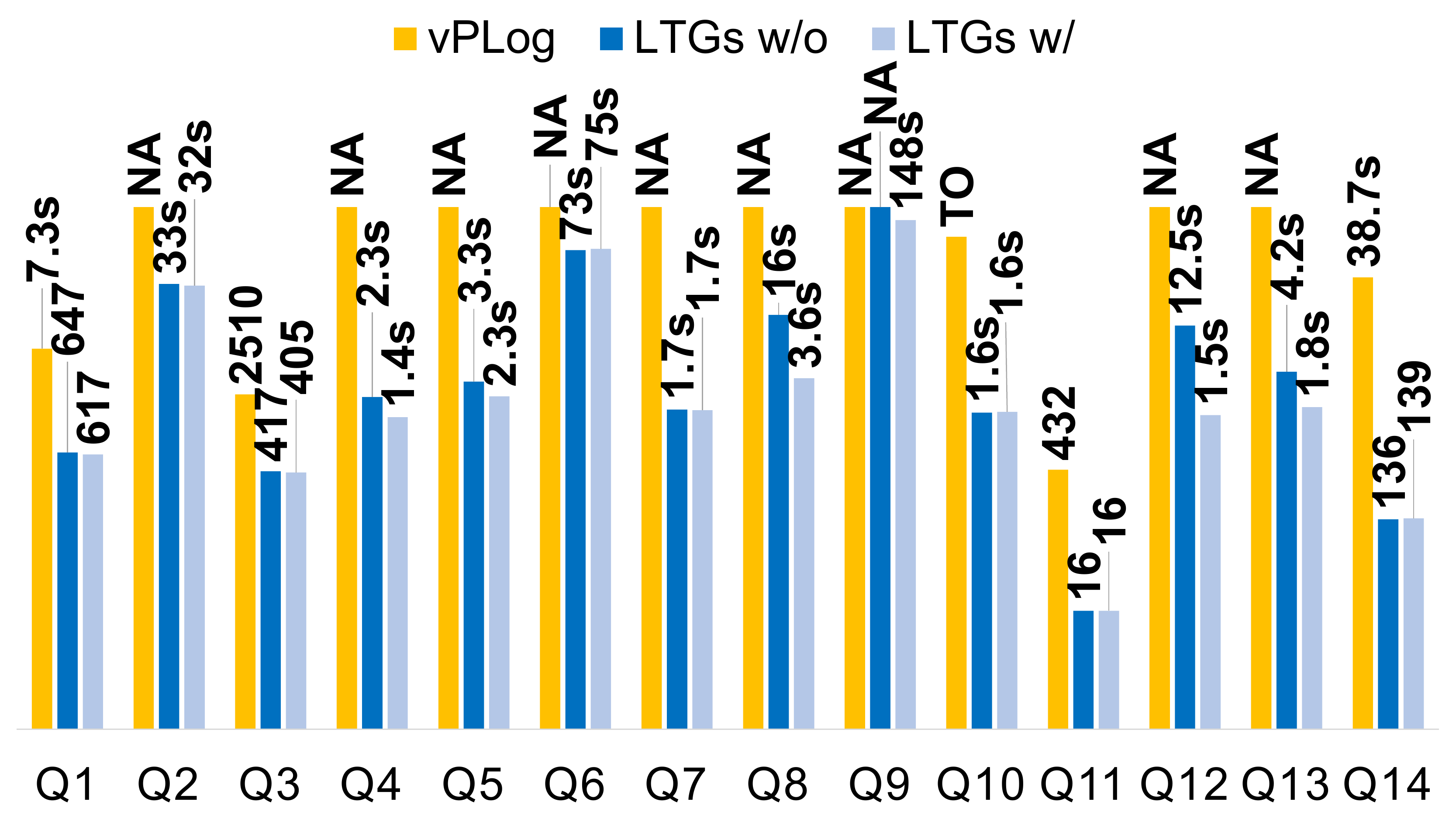}} &
            \x {\includegraphics[width = 0.4\textwidth]{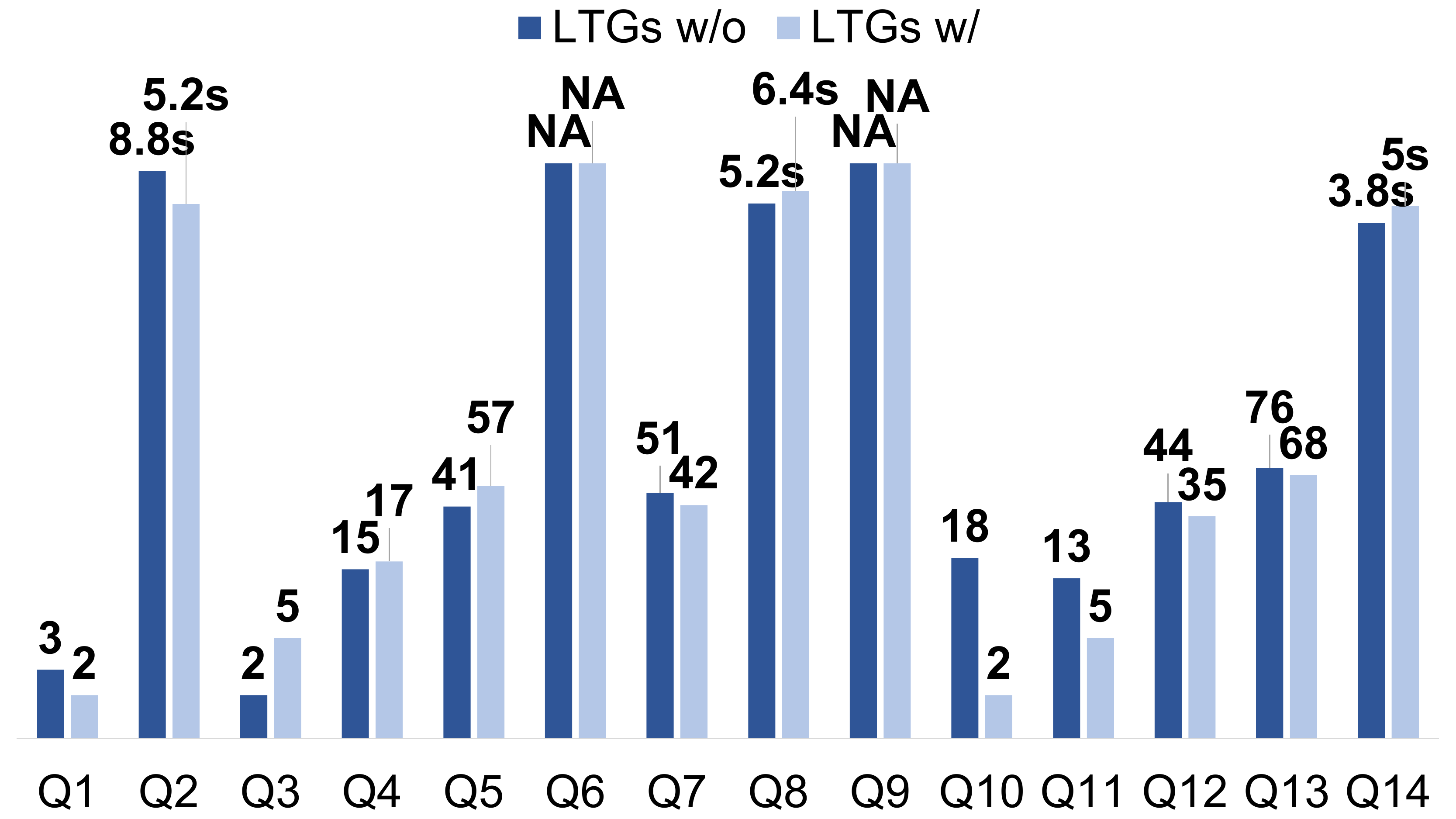}} &
            \x {\includegraphics[width = 0.4\textwidth]{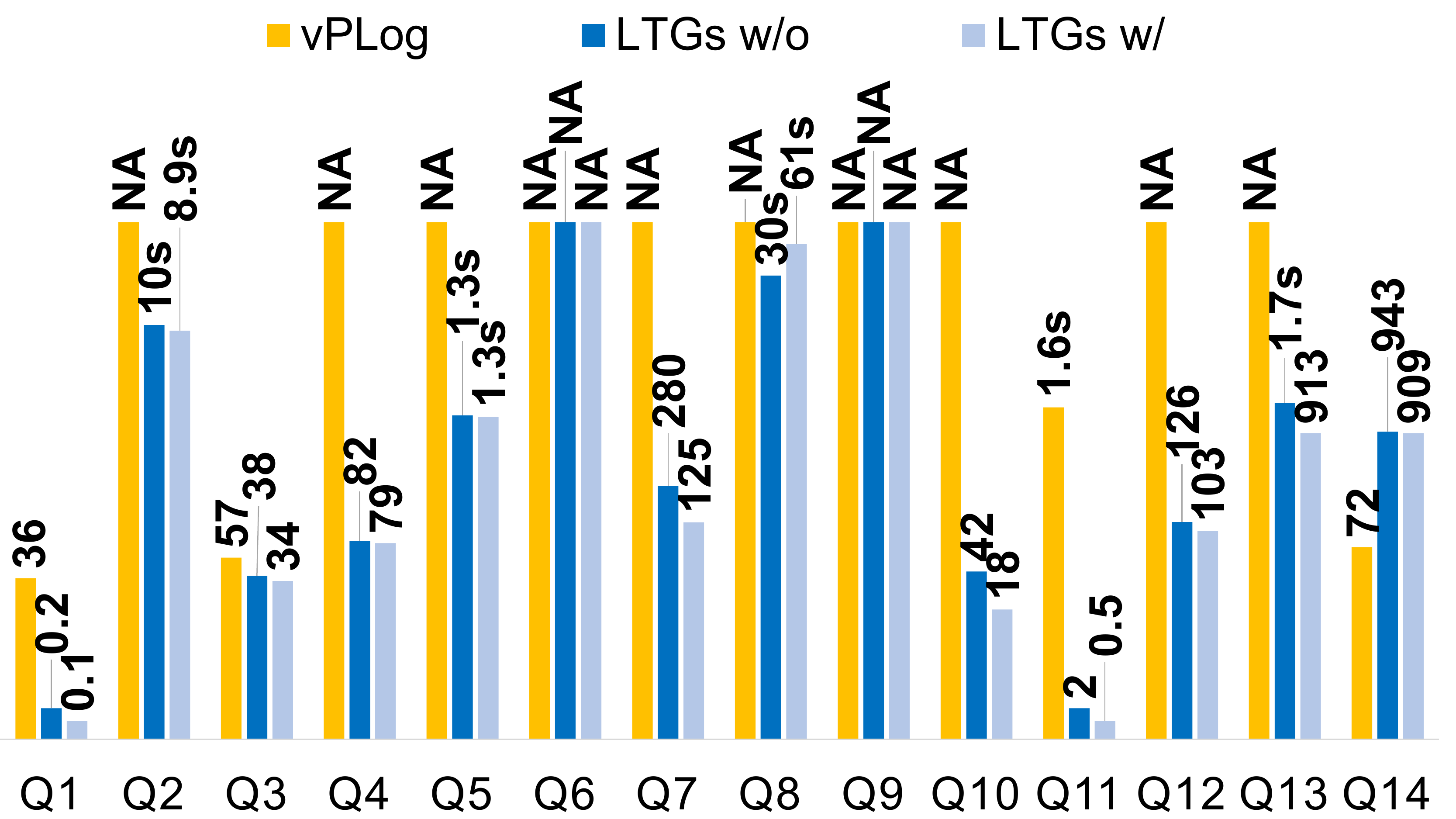}} \\
        \end{tabular}
    \end{adjustbox}
    \caption{Runtime breakdown in ms to answer the $\lubm$ queries.}
    \label{fig:performance:lubm}
\end{figure*}

\begin{figure}[tb]
    \begin{adjustbox}{max width=\columnwidth}
        \begin{tabular}{cc}
            \rotatebox{90}{\hspace{0.8cm}{$\boldsymbol{\lubmL}$}} &
            \x {\includegraphics[width = 0.35\textwidth]{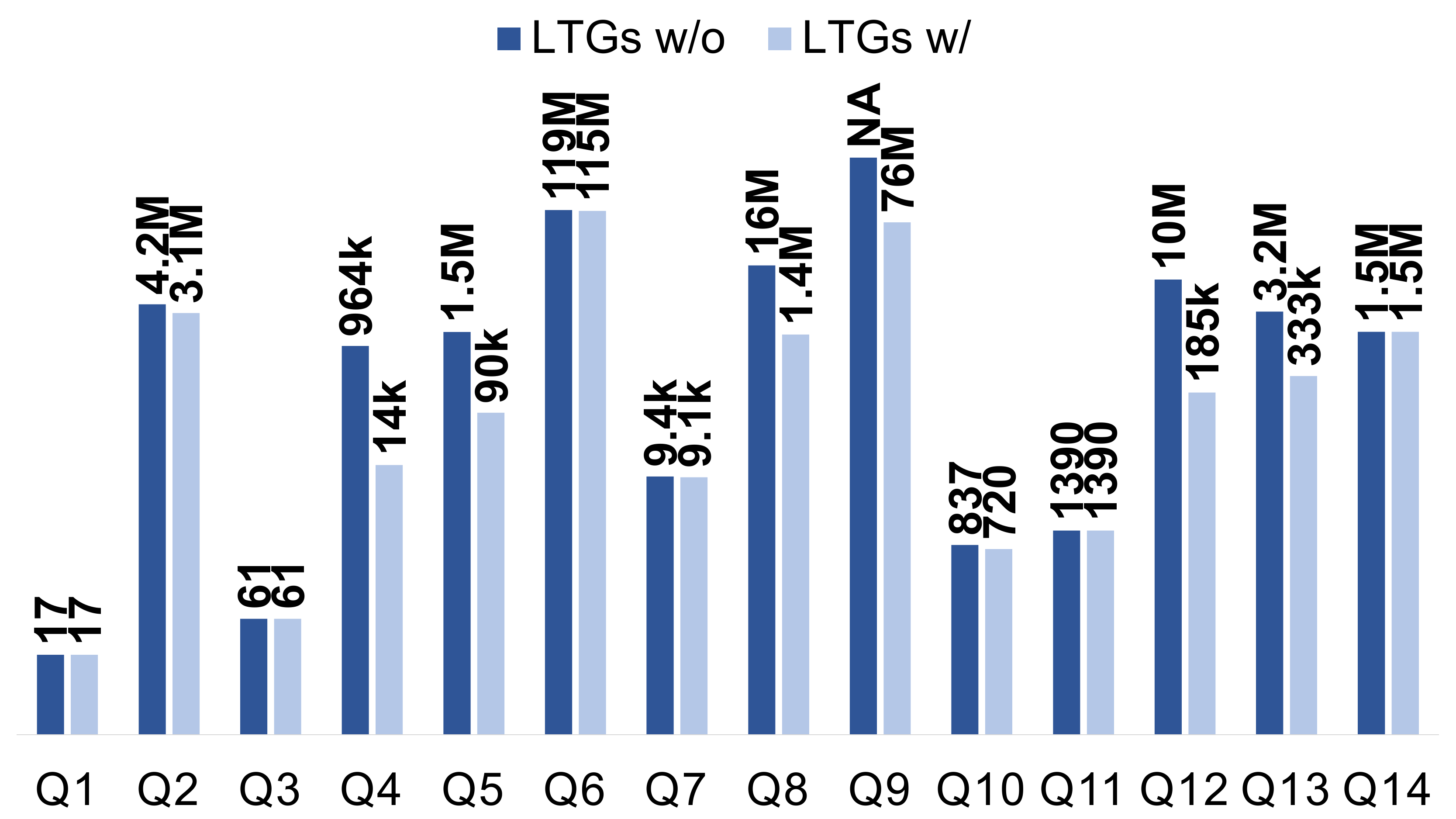}} \\
        \end{tabular}
    \end{adjustbox}
    \caption{Number of derivations for the $\lubm$ queries.}
    \label{fig:derivations:lubm}
\end{figure}

\subsection{QA methodology} \label{section:experiments:methodology}

To align with the evaluation of
$\problog$ and $\vp$, we applied
the \textit{magic sets} (MS) transformation \cite{DBLP:journals/jlp/BeeriR91,bancilhon:pods86,BenediktMT18}.
MS is a database technique that, given a query and a non-probabilistic program
$\Pp$, rewrites the rules in $\Rp$ so that the bottom-up evaluation of the
rewritten rules mimics the top-down evaluation of the query using $\Rp$.
\citeauthor{DBLP:journals/corr/abs-1911-07750}~\shortcite{DBLP:journals/corr/abs-1911-07750} have shown that MS also supports probabilistic programs.

Our experimental methodology for all scenarios other than the $\vqar$ ones proceeds as follows. For each scenario with program ${\Pp = (\Rp,\Fp, \pi)}$ and query $Q$, $\Pp$ is transformed into a new program $\Pp^{Q}$ using
MS. In $\problog$, $\vp$, and $\scallop$, query answering first computes the least parameterized model $\mathcal{M}$ of $\Pp^{Q}$
(\emph{reasoning} step). Then, for each query fact $Q(\vec{c})$
in $\mathcal{M}$, we compute the probability of its associated formula (\emph{probability
computation} step). Query
answering in \ours{} firstly computes the lineage TG $G$ for $\Pp^{Q}$
(\emph{reasoning} step), then the lineage of each $Q(\vec{c})$ in $G(\Fp)$
(\emph{lineage collection} step) and, finally, the probability of the lineage
(\emph{probability computation} step). With VQAR, we did not apply MS but used directly the queries proposed by \scallop{}'s authors \cite{scallop}.

\begin{table}[t]
    \footnotesize
    \caption{Absolute (ms) and relative runtime overhead for collapsing
    the lineage during reasoning.}

        \begin{tabular}{C{0.3cm}R{1.4cm}R{1.4cm}||C{0.5cm}R{1.4cm}R{1.25cm}}
                & $\lubmM$ & $\lubmL$ & & $\lubmM$ & $\lubmL$ \\
             Q1 &  0.009 (0.01\%) & 0.007 (0.001\%) & Q8 & 28 (1.7\%) & 27 (0.6\%)\\
             Q2 & 1.6 (0.5\%) & 54 (0.1\%) & Q9 & 115 (0.7\%) & 1165 (0.7\%) \\
             Q3 & 0.02 (0.05\%) & 0.01 (0.002\%) & Q10 & 151 (0.1\%) & 0.2 (0.01\%)\\
             Q4 & 0.8 (0.4\%) & 0.7 (0.04\%) & Q11 & 0.03 (0.6\%) & 0.04 (0.1\%) \\
             Q5 & 2.8 (1\%) & 3 (0.1\%) & Q12 & 6.3 (1.5\%) & 6 (0.3\%)\\
             Q6 & 204 (3.4\%) & 2239 (2.8\%) & Q13 & 1.6 (0.8\%) & 17 (0.8\%)\\
             Q7 & 183 (0.1\%) & 0.3 (0.01\%) & Q14 & 0.9 (6.9\%) & 10 (6.5\%) \\
    \end{tabular}
    \label{table:overheadcompression}
\end{table}

\begin{table}[t]
    \footnotesize
    \caption{Average runtime (ms) and standard deviation for computing probability per query answer for $\lubmM$.}
    \label{tab:differentsolvers}
      \begin{tabular}{cr || rrr}
          & vProbLog  & \multicolumn{3}{c}{$\ours$ w/} \\
          \cline{3-5}
          & + PySDD & + PySDD & + d-tree & + c2d \\
          Q1 & 7.5 $\pm$0.1 & 0.001 $\pm$0.0008         & 0.001 $\pm$0.0007
             & 0.002 $\pm$0.0007\\
          Q2 & 7.3 $\pm$ 0.2 & 3.6 $\pm$1.7             & 13 $\pm$27
             & 1.5s $\pm$389 \\
          Q3 & 7.4 $\pm$0.08 & 0.06/0.1                 & 0.2 $\pm$0.08
             &  78 $\pm$9\\
          Q4 & 7.4 $\pm$ 0.07 & 0.5 $\pm$0.7            & 8.9 $\pm$49
             & 146 $\pm$182\\
          Q5 & 7 $\pm$0.08 & 0.1$\pm$0.1                & 0.4 $\pm$2
             & 119 $\pm$80\\
          Q6 & NA & NA                                  & NA
             & NA \\
          Q7 & 7.1 $\pm$0.07 & 1.9 $\pm$1.3             & 70 $\pm$163
             & 443 $\pm$164\\
          Q8 & NA & 2 $\pm$1.3                          & 84 $\pm$110
             & 521 $\pm$147\\
          Q9 & NA & 33 $\pm$39                          & 6 $\pm$10
             & TO \\
          Q10 & 7.5 $\pm$0.09  & 6 $\pm$2               & 358 $\pm$200
              & 614 $\pm$163\\
          Q11 & 6.9 $\pm$ 0.08 & 0.0009 $\pm$0.0003     & 0.0008 $\pm$0.0007
              & 0.002 $\pm$0.0005\\
          Q12 & 7.5 $\pm$ 0.09 & 2.6 $\pm$1             & 44 $\pm$36
              & 980 $\pm$59 \\
          Q13 & 6.8 $\pm$ 0.04 & 0.2 $\pm$0.2           & 1 $\pm$1
              & 234 $\pm$75\\
          Q14 & 6.9 $\pm$0.3 & 0.0008 $\pm$0.0005       & 0.0007 $\pm$0.001
              & 0.002 $\pm$0.0008\\
    \end{tabular}
\end{table}

\sys{}, \problog{}, and \vp{} are exact probabilistic reasoning engines, while \scallop{} is an approximate one that keeps
only the top-$k$ explanations for each derived fact. To ensure a fair comparison, we configured \scallop{} so that the number of computed explanations is as close as
possible to the ones computed by other exact engines. To this end, we set $k=30$ as
the default value, the highest possible value for which \scallop{} can
answer most queries (for higher values, the computation goes out of memory most of the
time). Notice that even with $k=30$, \scallop{} still
approximates in some cases, having an advantage.
We use  \scallop{}($k$) to indicate  \scallop{} applied for a specific $k$.
We observed that when query answering terminates, it does so
within a few minutes. Therefore, we set a 30 minutes timeout to let as
many queries to be answered as possible without waiting for too long.

\subsection{Results}

For \lubm{}, we present a comparison between \sys{} and all the
other engines. For \dbpedia, \claros, \smokers, \yago, and \wnrr{}, the
comparison does not consider $\problog$ and $\scallop$. Regarding $\problog$,
its performance in \lubmM turned out to be too low to be further considered, as also
observed by \cite{DBLP:journals/corr/abs-1911-07750}. Regarding $\scallop$,
either it did not support some rules in the benchmarks or
the data was too large to be loaded (\scallop's authors'
highlight scalability as a direction for future work~\cite{scallop}).

For \vqar, we show a comparison only between \sys{} w/ and $\scallop$: neither \sys{} w/o nor $\vp$ were able to compute the least parameterized model due to the
combinatorial explosion of the derivations.
$\ours$ is the only technique that can compute the full least
parameterized model despite this explosion.

Notice that computing the full lineage
does not necessarily mean that we can always compute its
exact probability, as the problem is \#P-hard~\cite{scaled-dissociations}.
To deal with such cases,
approximations can be employed either upfront, by reducing the size of the lineage
(\scallop) or after the full lineage has been collected like~\cite{dissociations,scaled-dissociations,Olteanu-approximate}.
Approximating the probability of the lineage is an orthogonal problem. Hence, we leave the
integration of such techniques with $\ours$ as future work,
focusing on the queries for which the answers' probabilities can be computed exactly using PySDD. In $\vqar$, these are 417/1000 queries.
For the remaining ones, either PySDD fails or the lineage is too
large ($>1M$ disjuncts) that fully computing it, although possible in some
cases, goes beyond the timeout.

\leanparagraph{Overview of the experimental results}
Table~\ref{tab:total-lubm10} reports the \emph{total runtime} to answer the
queries in $\lubmM$ (left-hand side) and $\lubmL$ (right-hand side)
using all engines. In \sys{}, the
total runtime is the sum of the reasoning, lineage collection, and probability
computation times. Figure~\ref{fig:performance:lubm} shows a breakdown of the above steps for $\vp$ and $\ours$.
Lineage collection is not relevant to $\vp$, since $\Delta \Tcp$
does not require this step.
Figure \ref{fig:derivations:lubm}
shows the number of derivations produced when answering the $\lubm$ queries using
$\ours$. The number of derivations is a rough estimator of the
difficulty of each query and is independent of the implementation.
``NA'' in Table~\ref{tab:total-lubm10} and Figures~\ref{fig:performance:lubm} and
\ref{fig:derivations:lubm}
denotes either timeout or out of memory (a detailed breakdown is shown later).
Table~\ref{table:overheadcompression} shows the overhead (both
    absolute in ms and relative to the total reasoning time) to
    collapse the lineage on $\lubm$, while
    Table~\ref{tab:differentsolvers} reports the average runtime in ms to compute
    the probabilities of the query answers using different techniques.

\begin{figure*}[tb]
    \begin{adjustbox}{max width=\textwidth}
        \begin{tabular}{ccccc}
             & \textbf{Reasoning (ms)} & \textbf{Probability (ms)} &  \textbf{Total (ms)} & \textbf{Derivations}  \\
            \rotatebox{90}{\hspace{3mm}{$\boldsymbol{\dbpedia}$}} &
            \x\x {\includegraphics[width = 0.25 \textwidth]{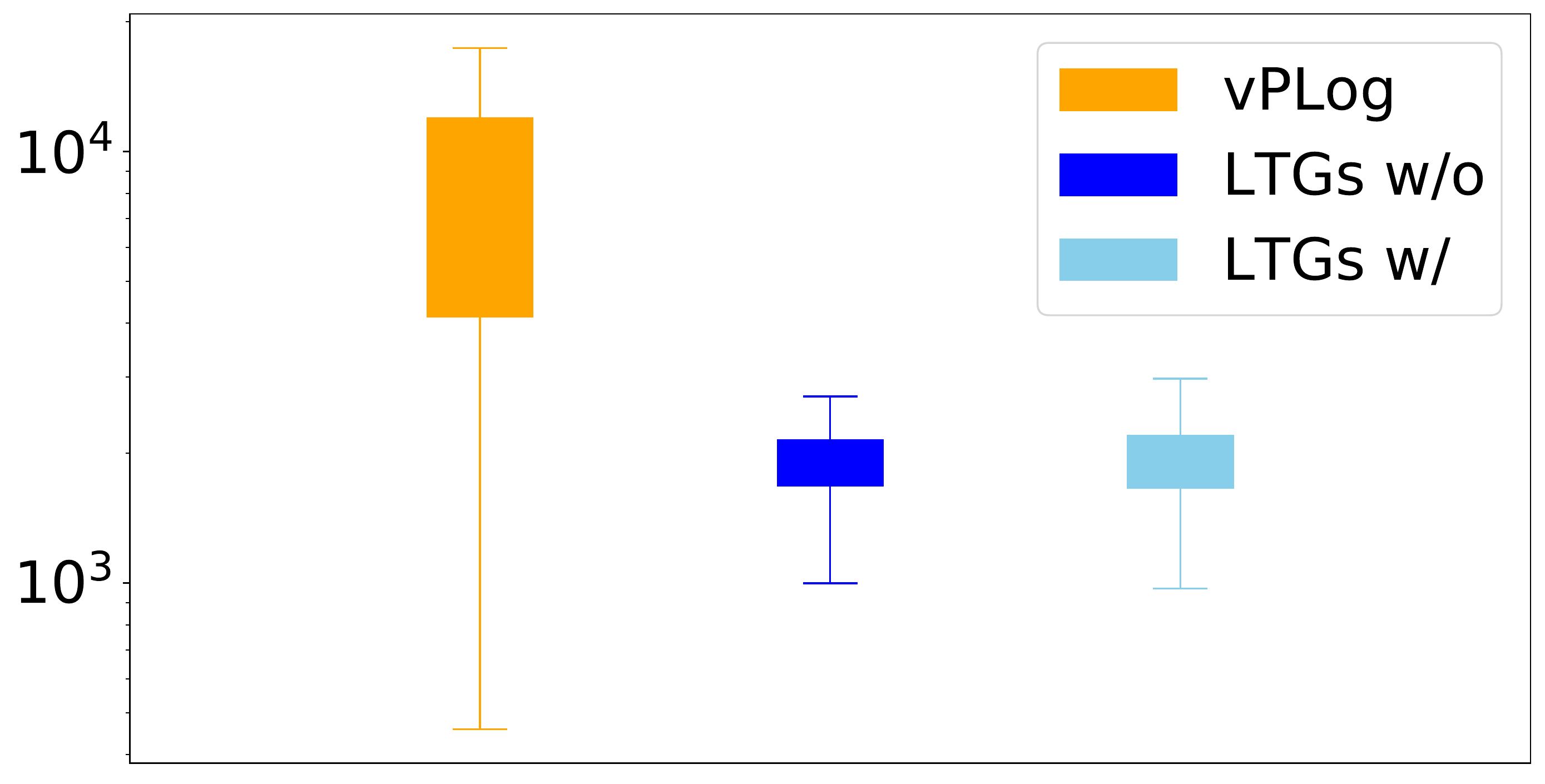}} &
            \x\x {\includegraphics[width = 0.25 \textwidth]{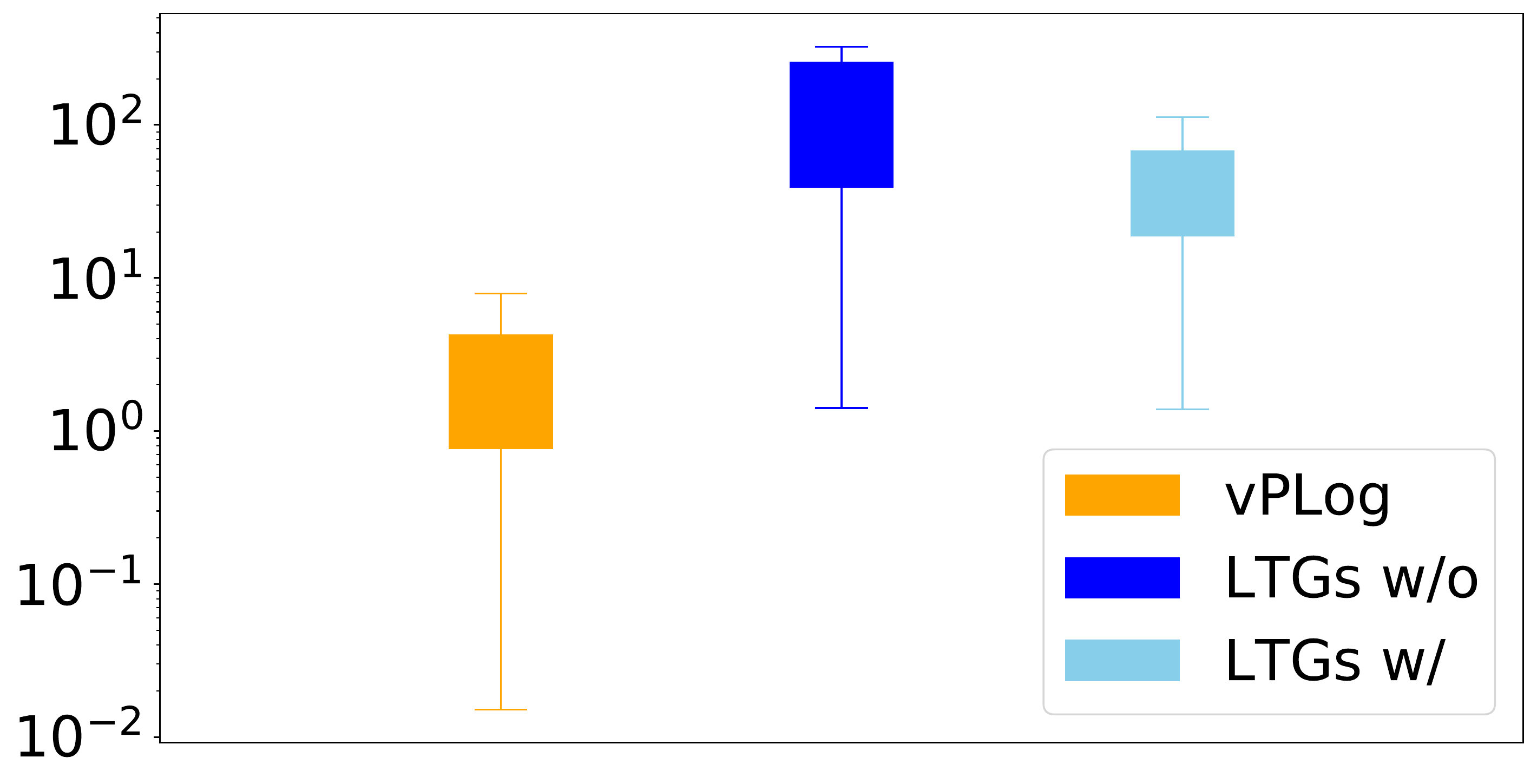}} &
            \x\x {\includegraphics[width = 0.25 \textwidth]{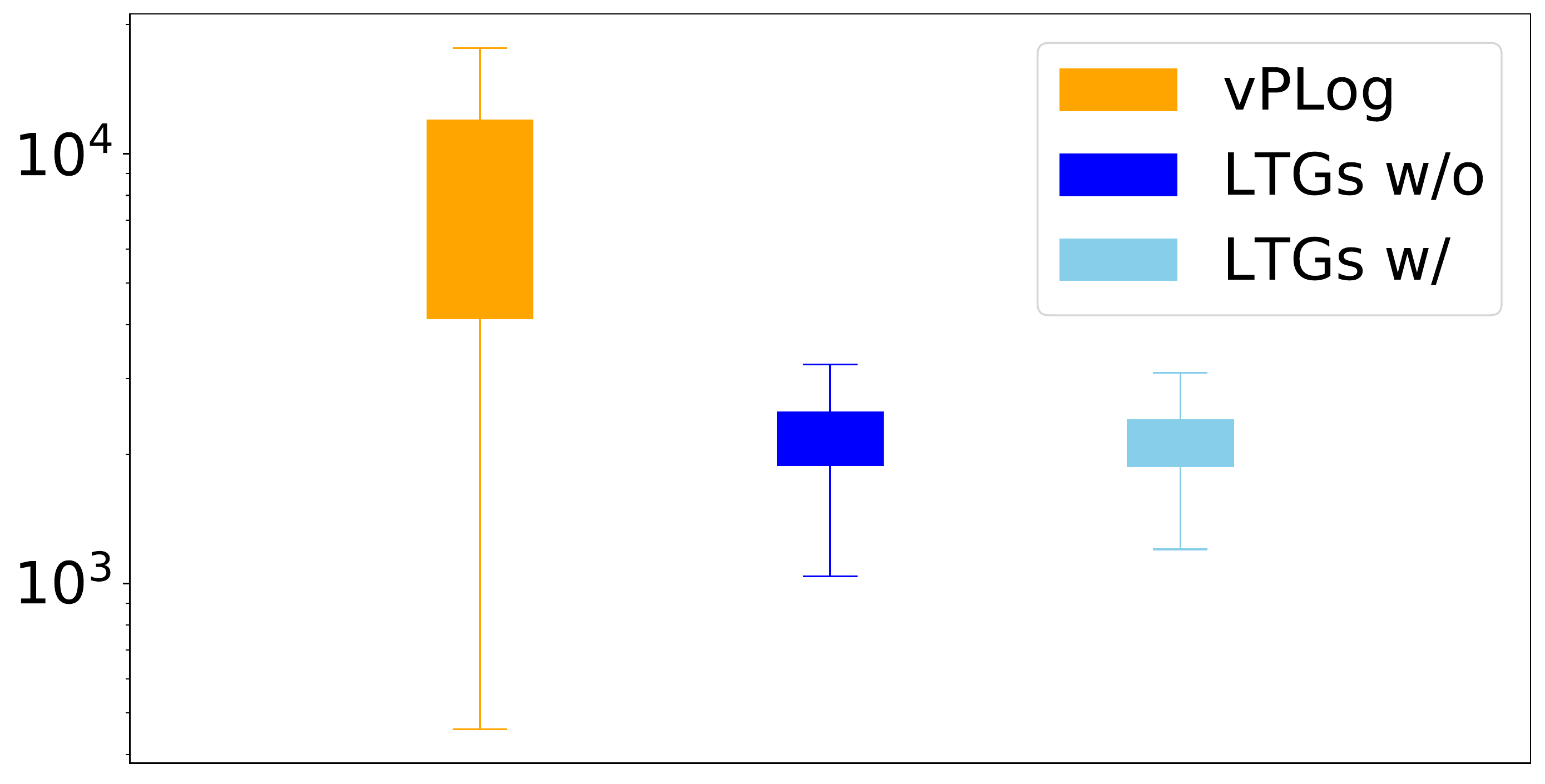}} &
            \x\x {\includegraphics[width = 0.25 \textwidth]{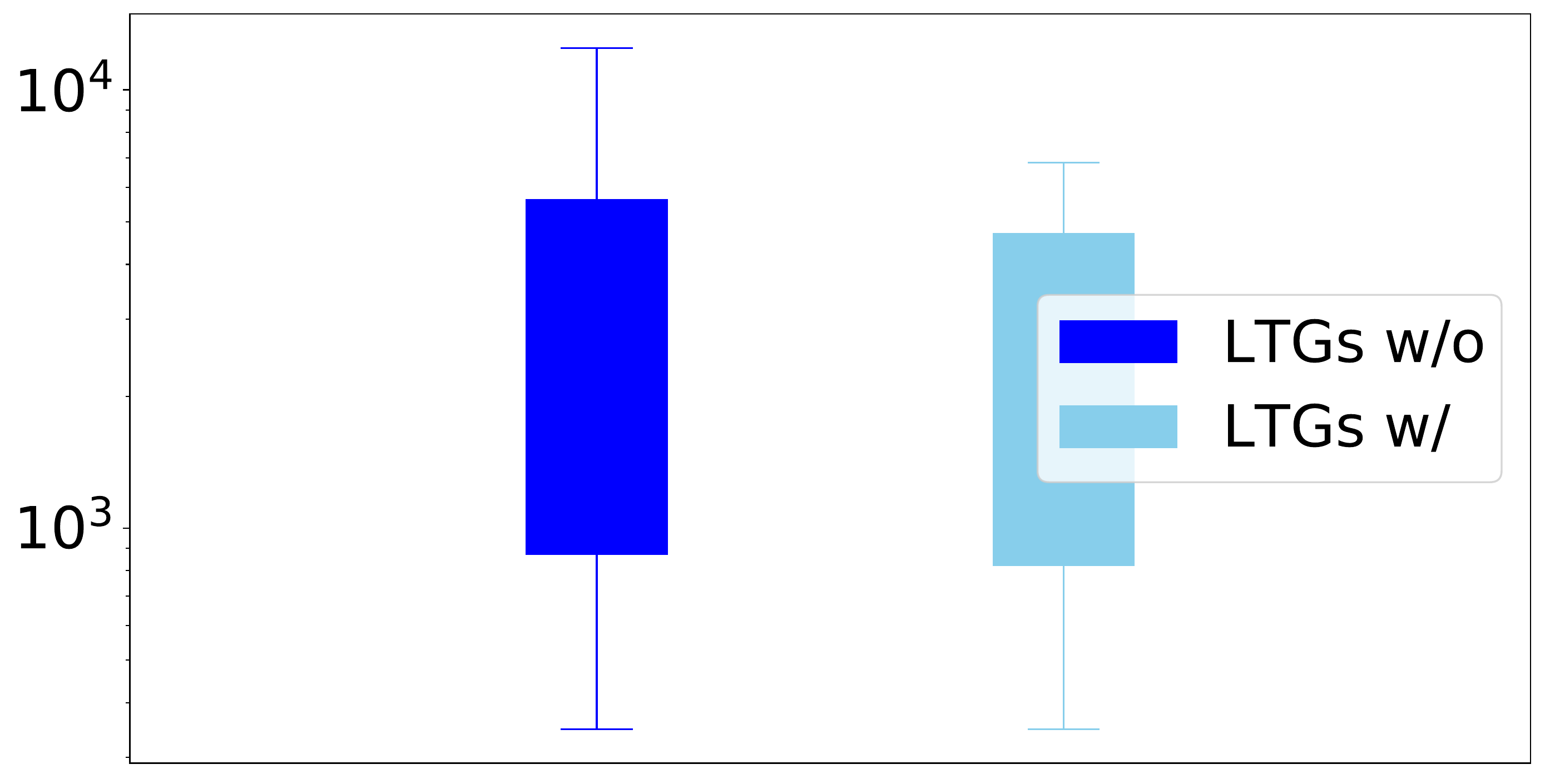}} \\

            \rotatebox{90}{\hspace{3mm}{$\boldsymbol{\claros}$}} &
            \x\x {\includegraphics[width = 0.25 \textwidth]{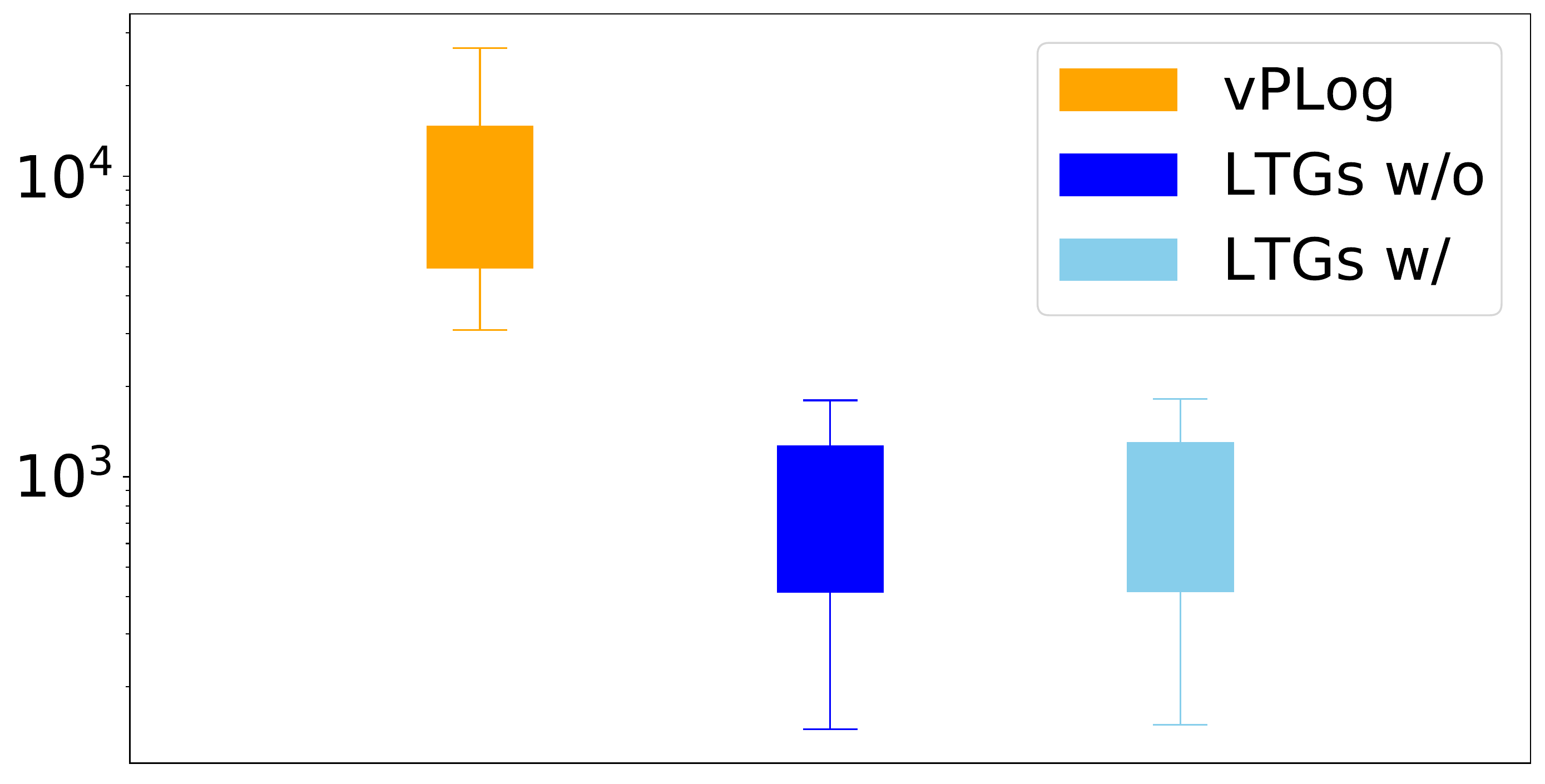}} &
            \x\x {\includegraphics[width = 0.25 \textwidth]{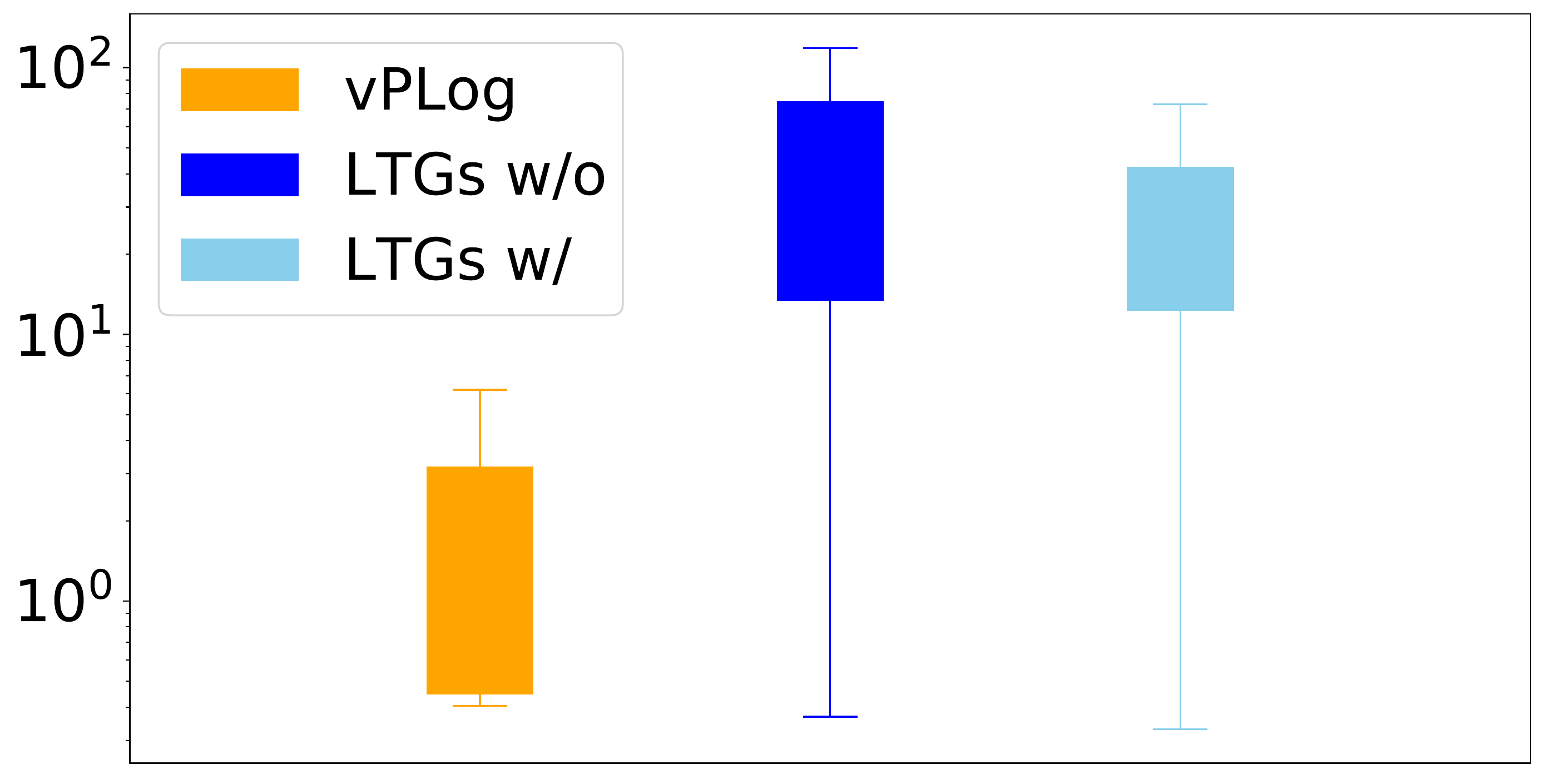}} &
            \x\x {\includegraphics[width = 0.25 \textwidth]{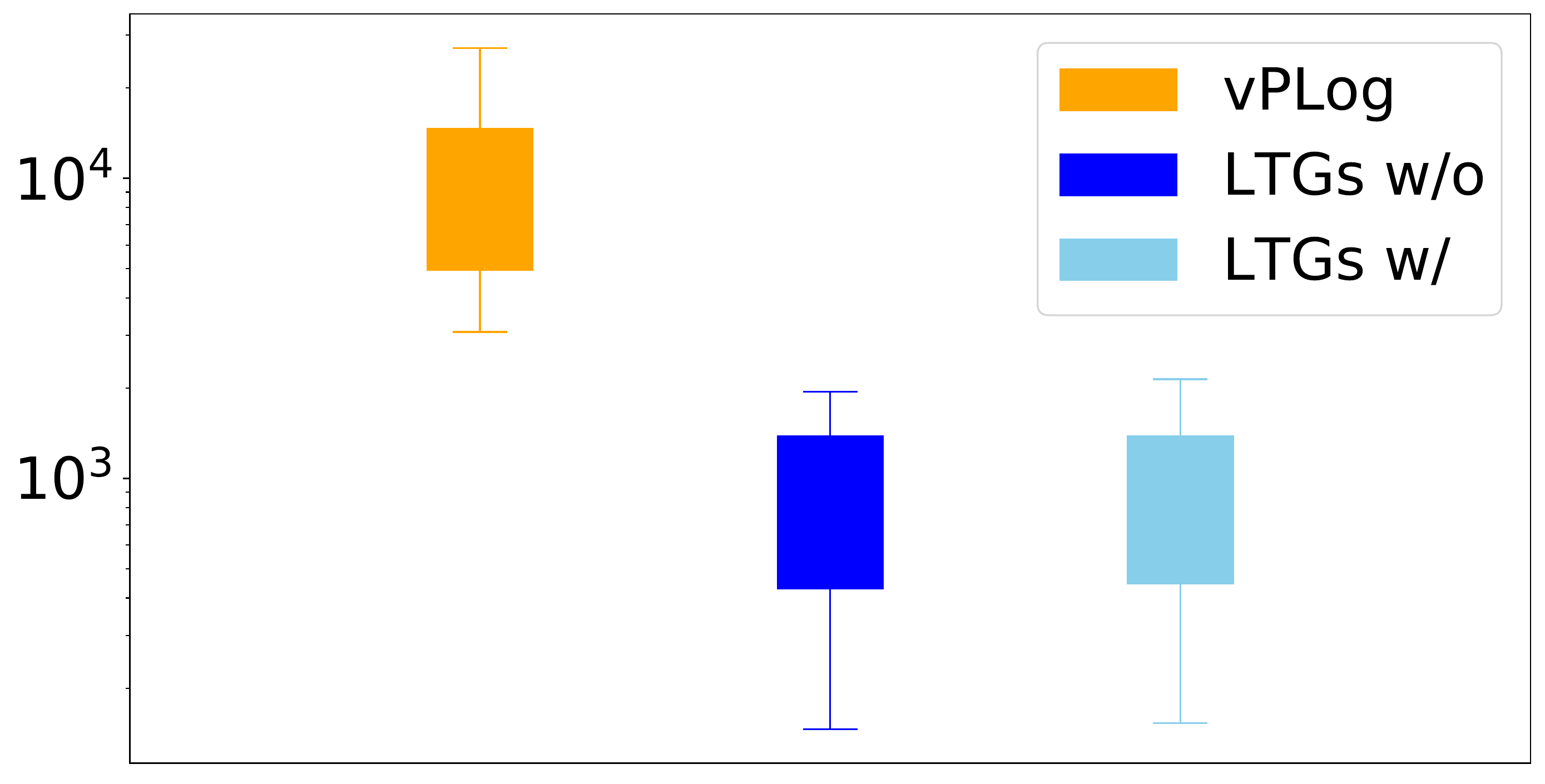}} &
            \x\x {\includegraphics[width = 0.25 \textwidth]{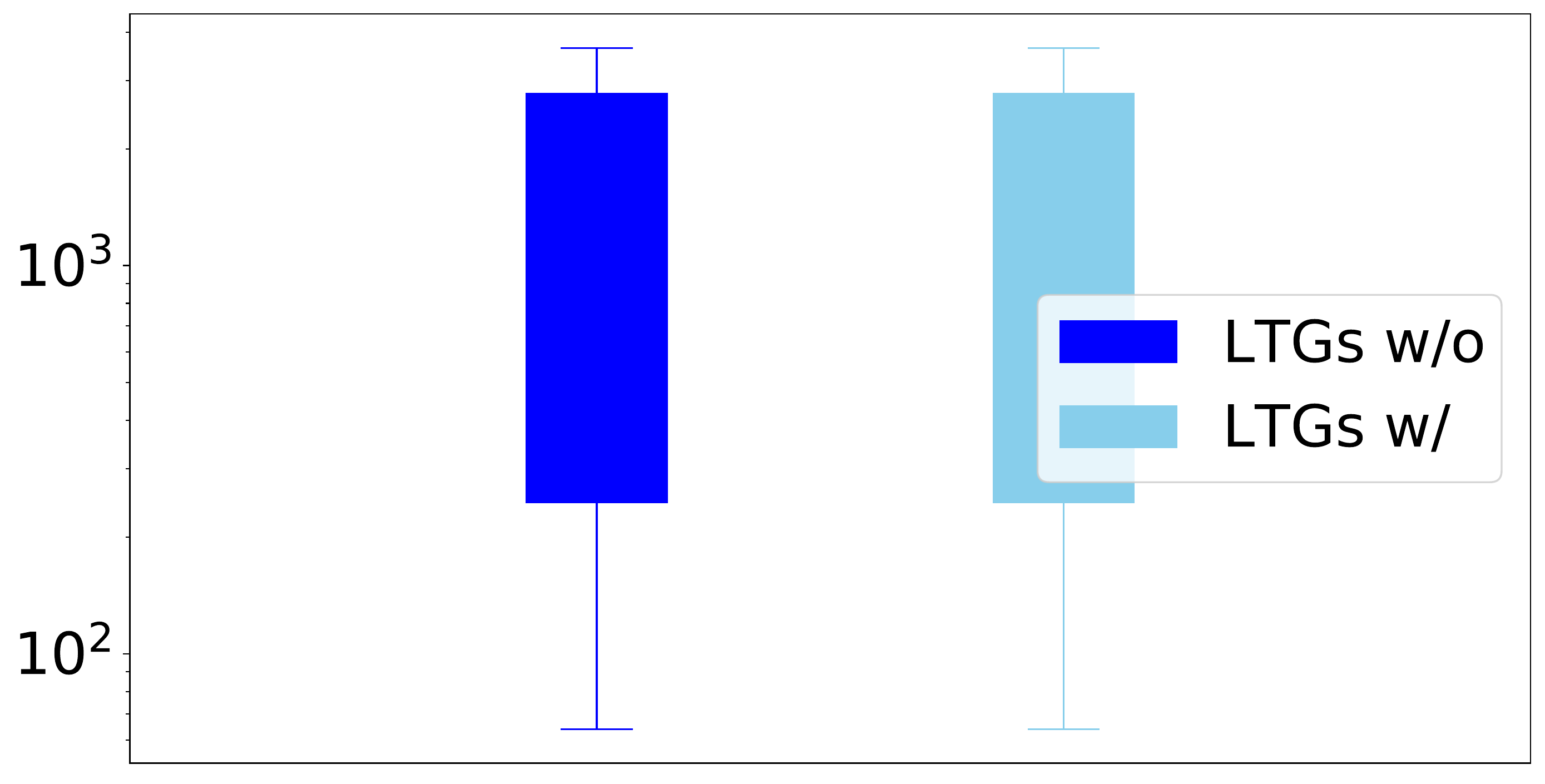}} \\

            \rotatebox{90}{\hspace{9mm}{$\boldsymbol{\yago}$}} &
            \x\x {\includegraphics[width = 0.25 \textwidth]{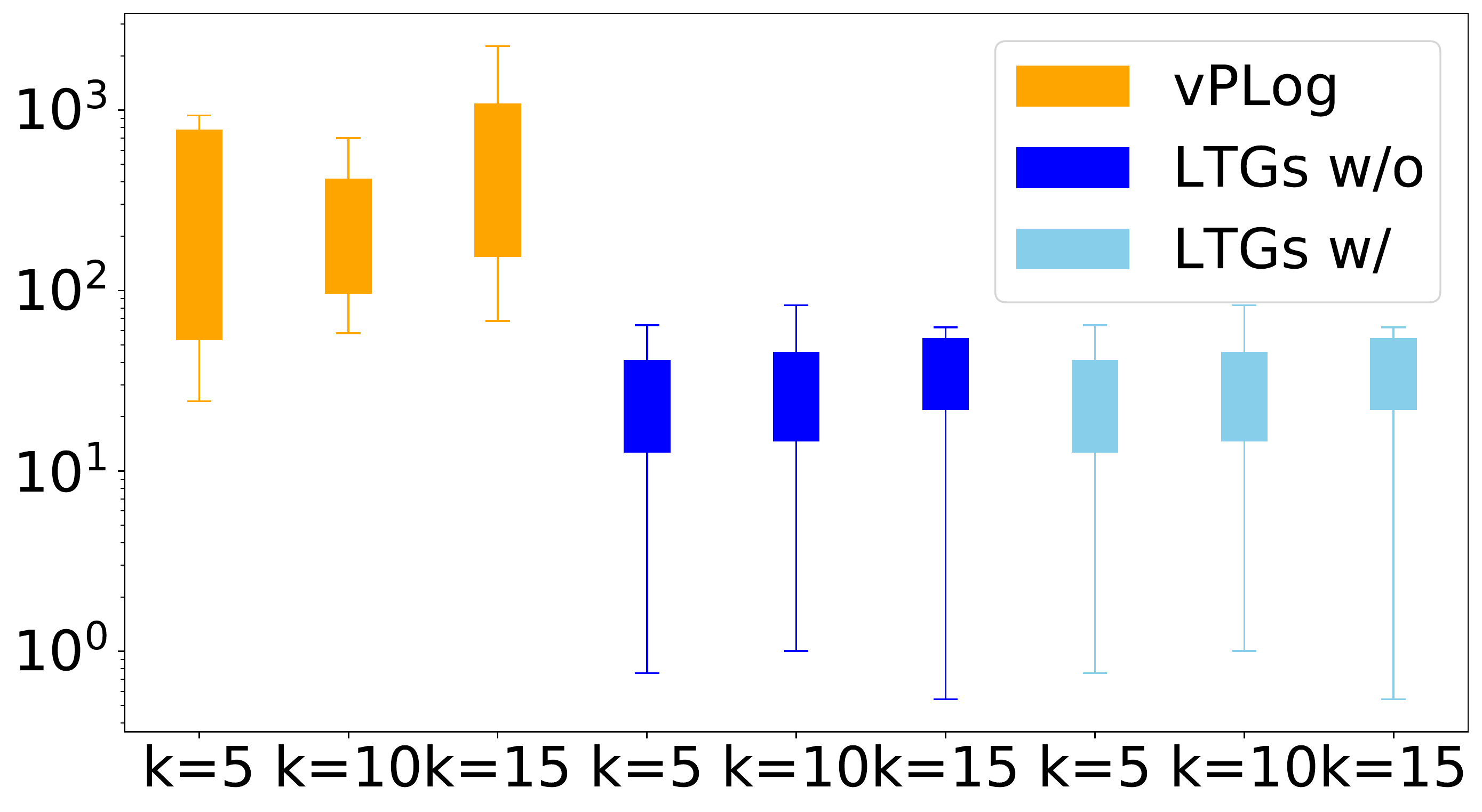}} &
            \x\x {\includegraphics[width = 0.25 \textwidth]{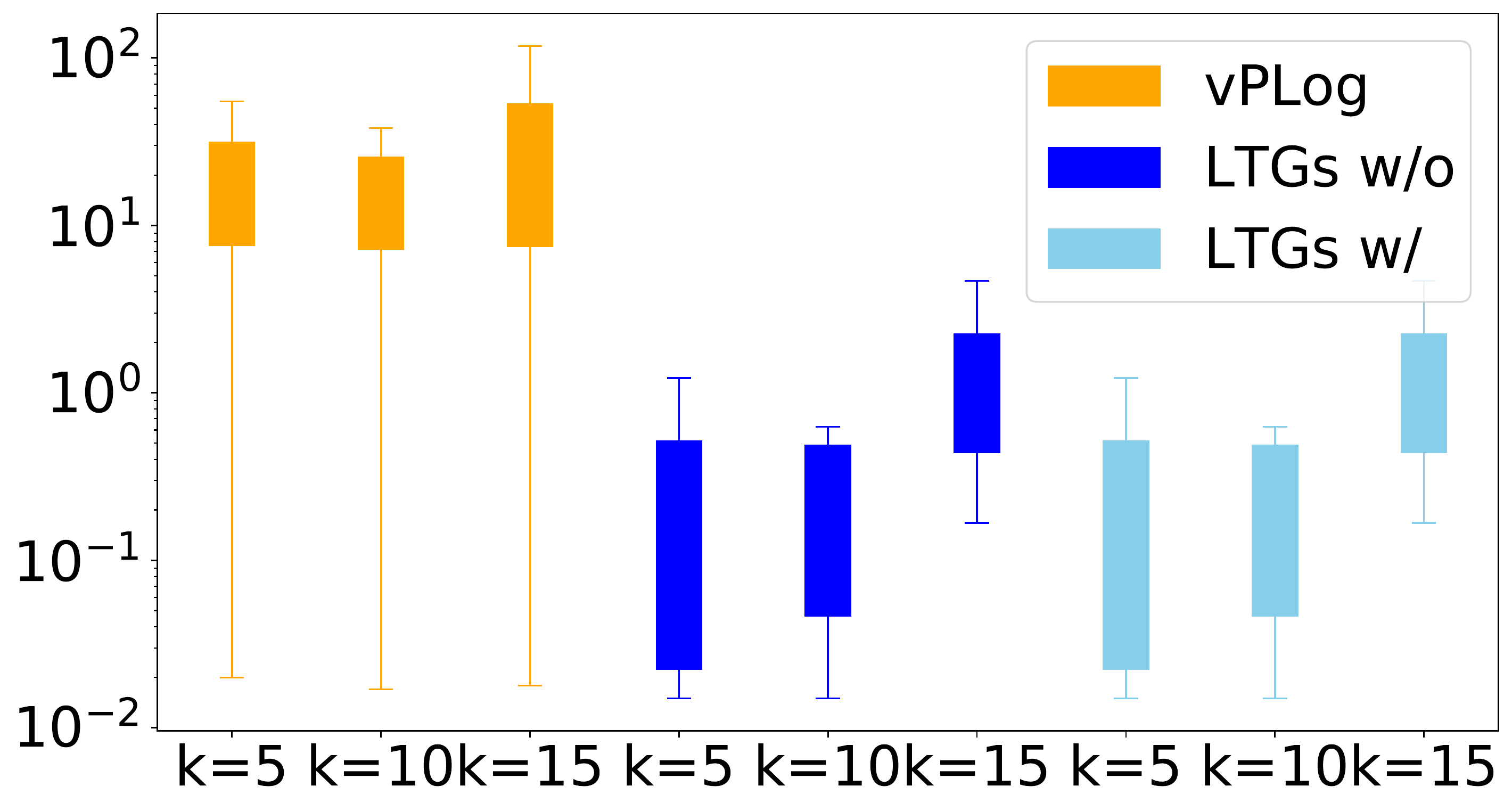}} &
            \x\x {\includegraphics[width = 0.25 \textwidth]{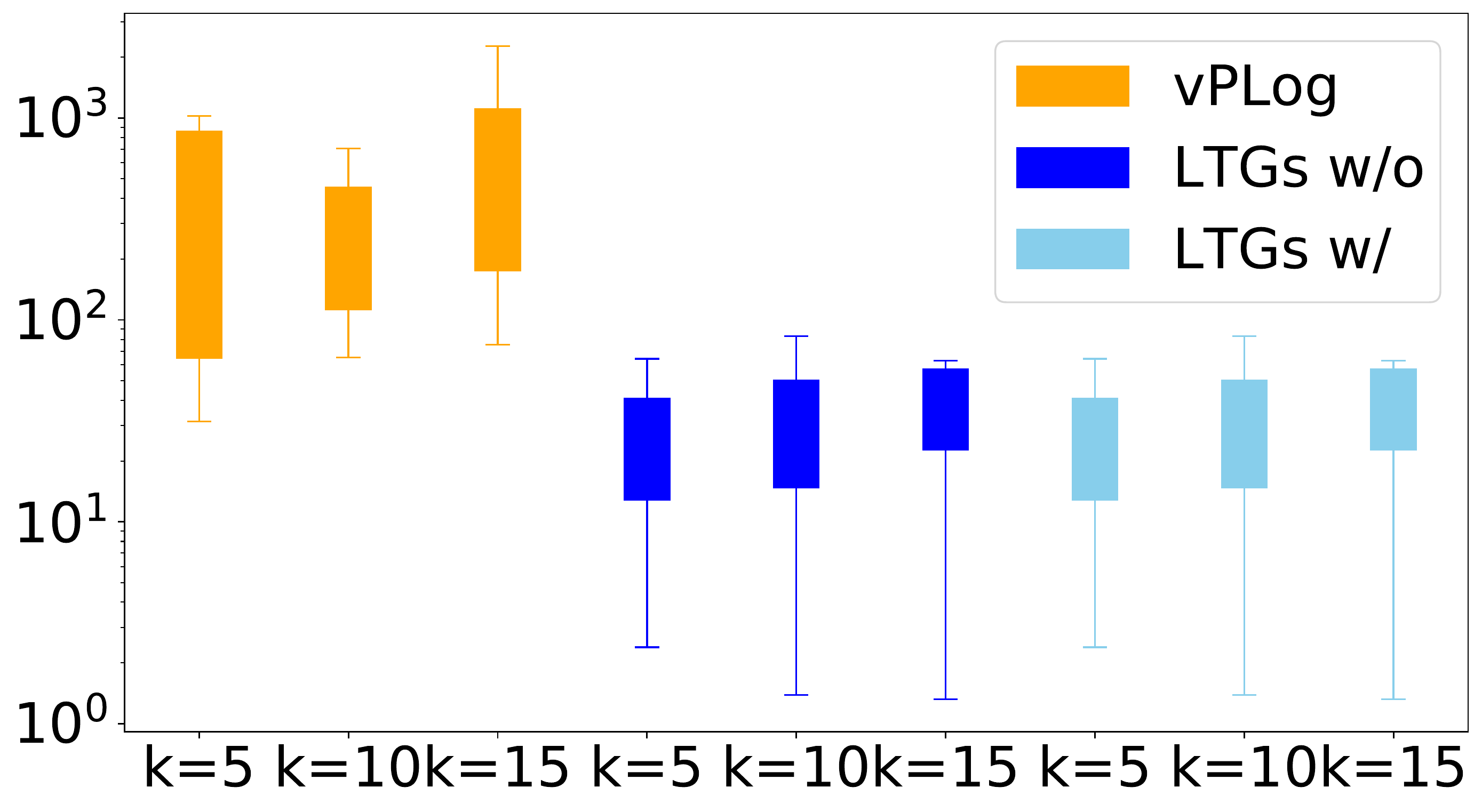}} &
            \x\x {\includegraphics[width = 0.25 \textwidth]{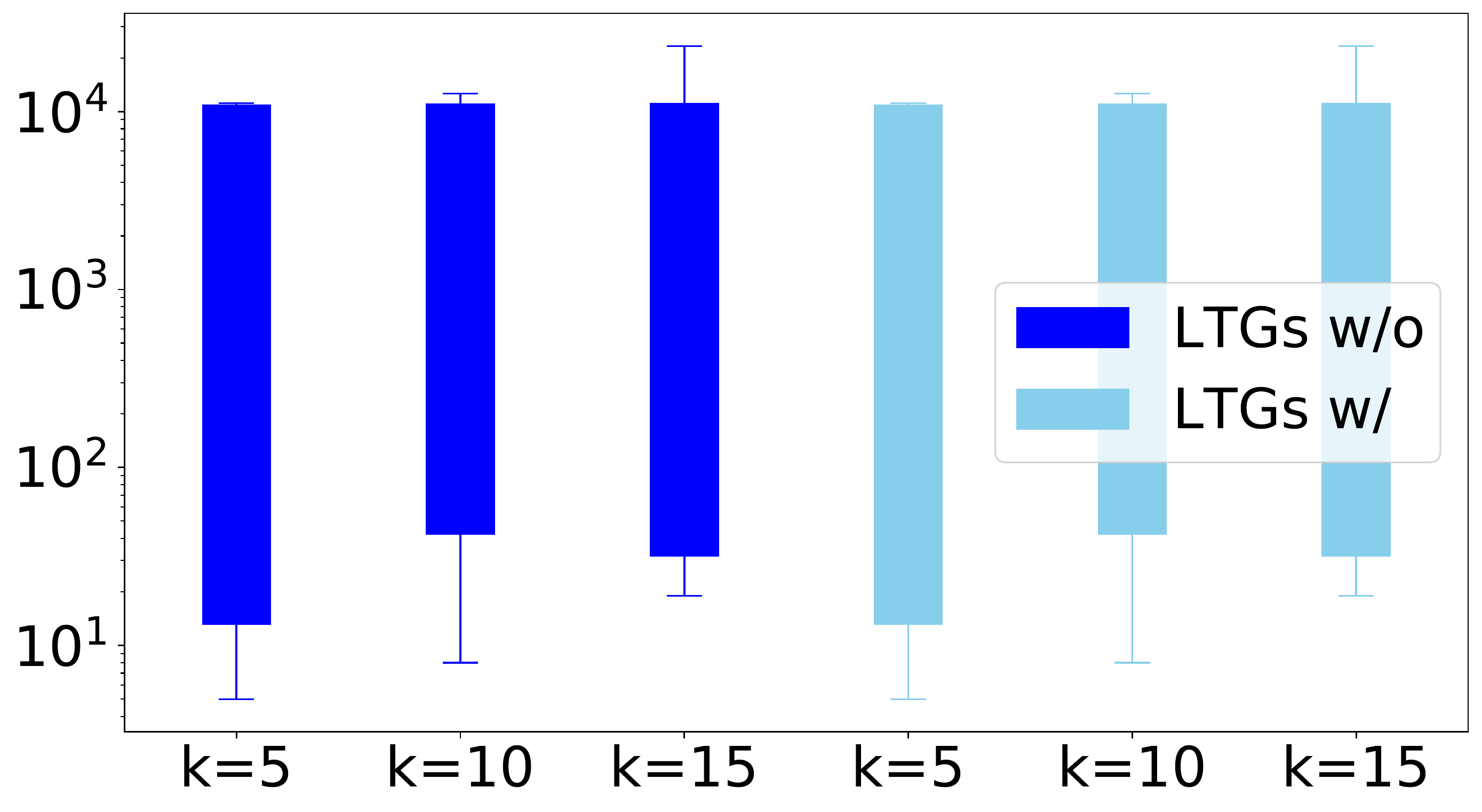}} \\

            \rotatebox{90}{\hspace{9mm}{$\boldsymbol{\wnrr}$}} &
            \x\x {\includegraphics[width = 0.25 \textwidth]{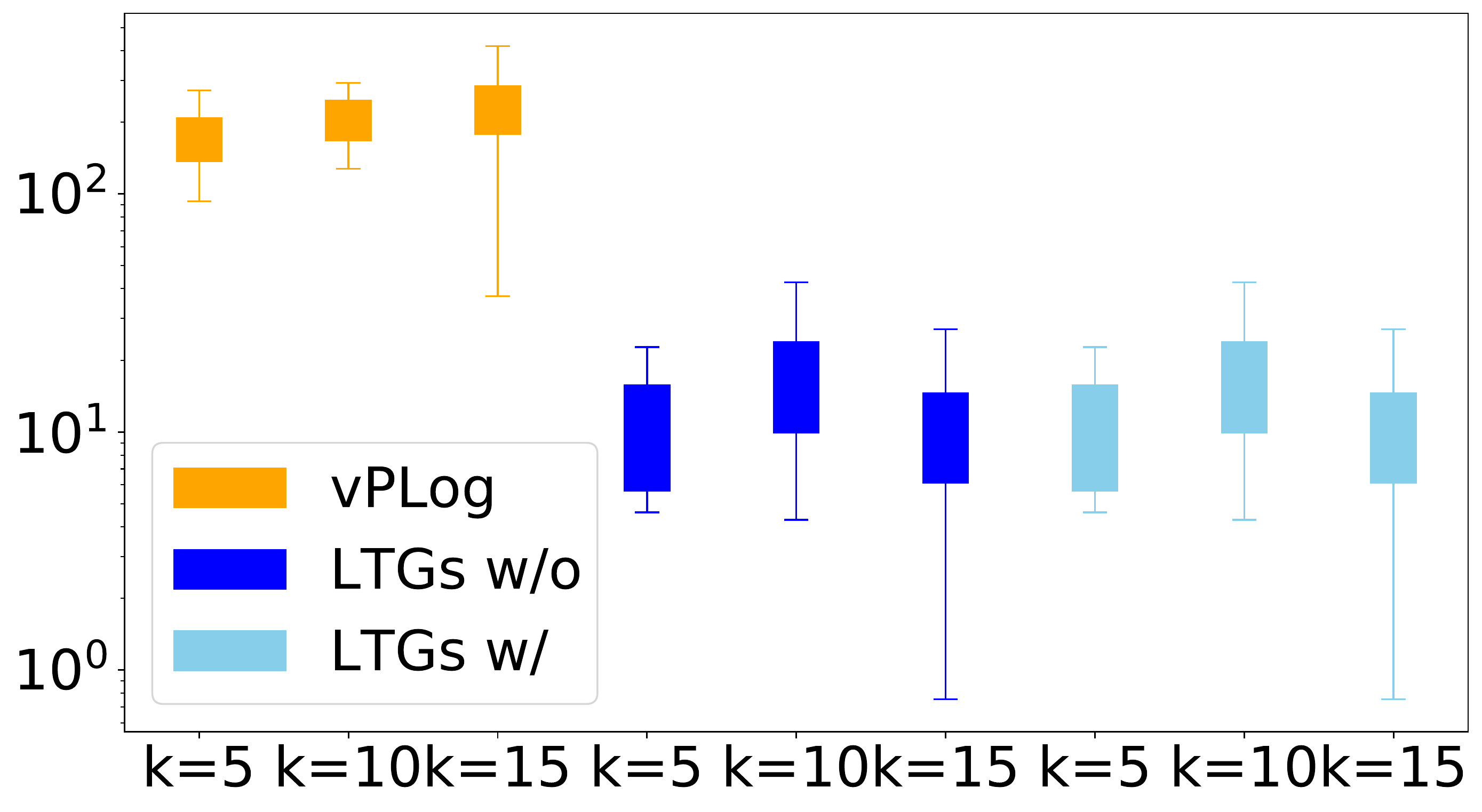}} &
            \x\x {\includegraphics[width = 0.25 \textwidth]{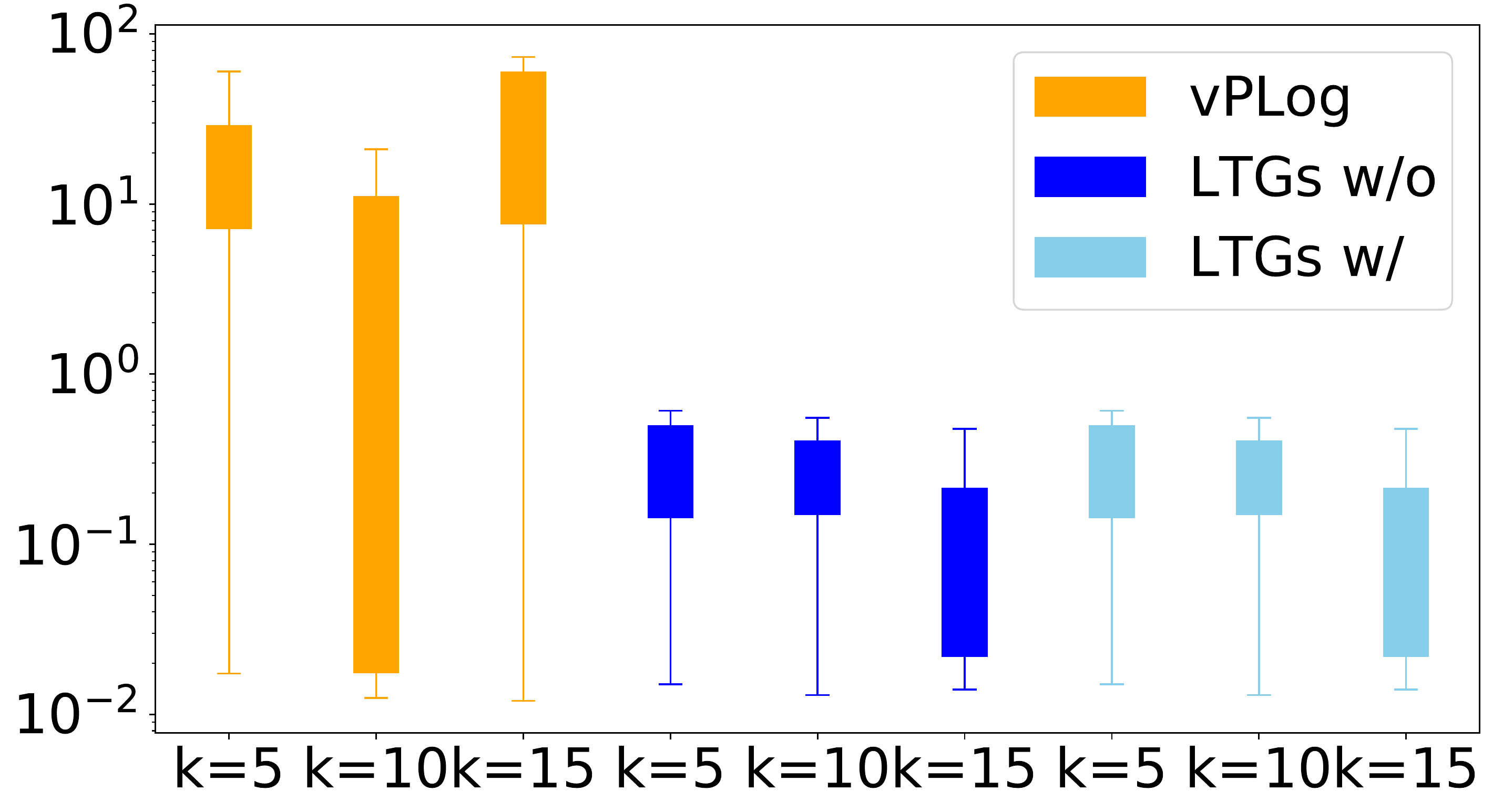}} &
            \x\x {\includegraphics[width = 0.25 \textwidth]{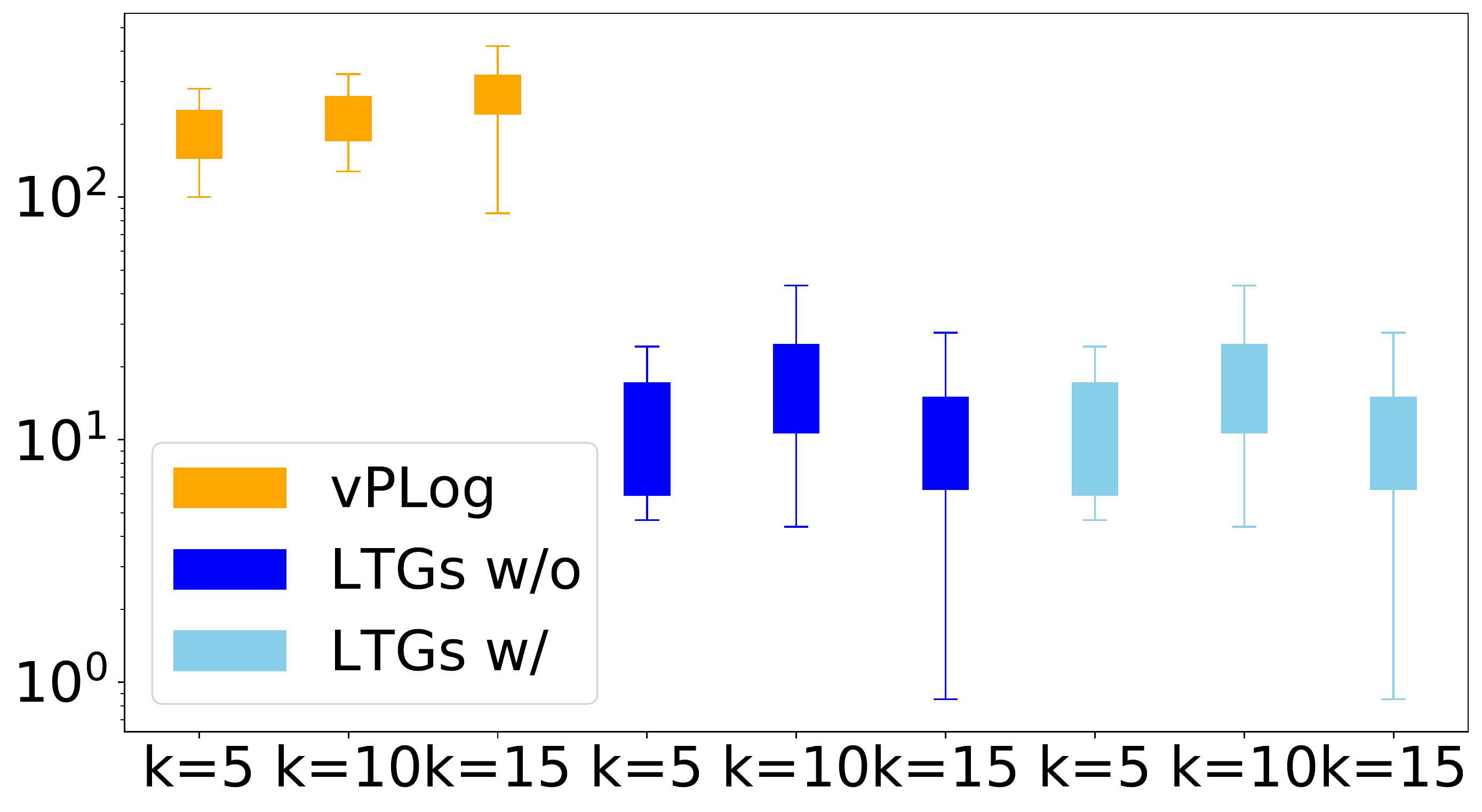}} &
            \x\x {\includegraphics[width = 0.25 \textwidth]{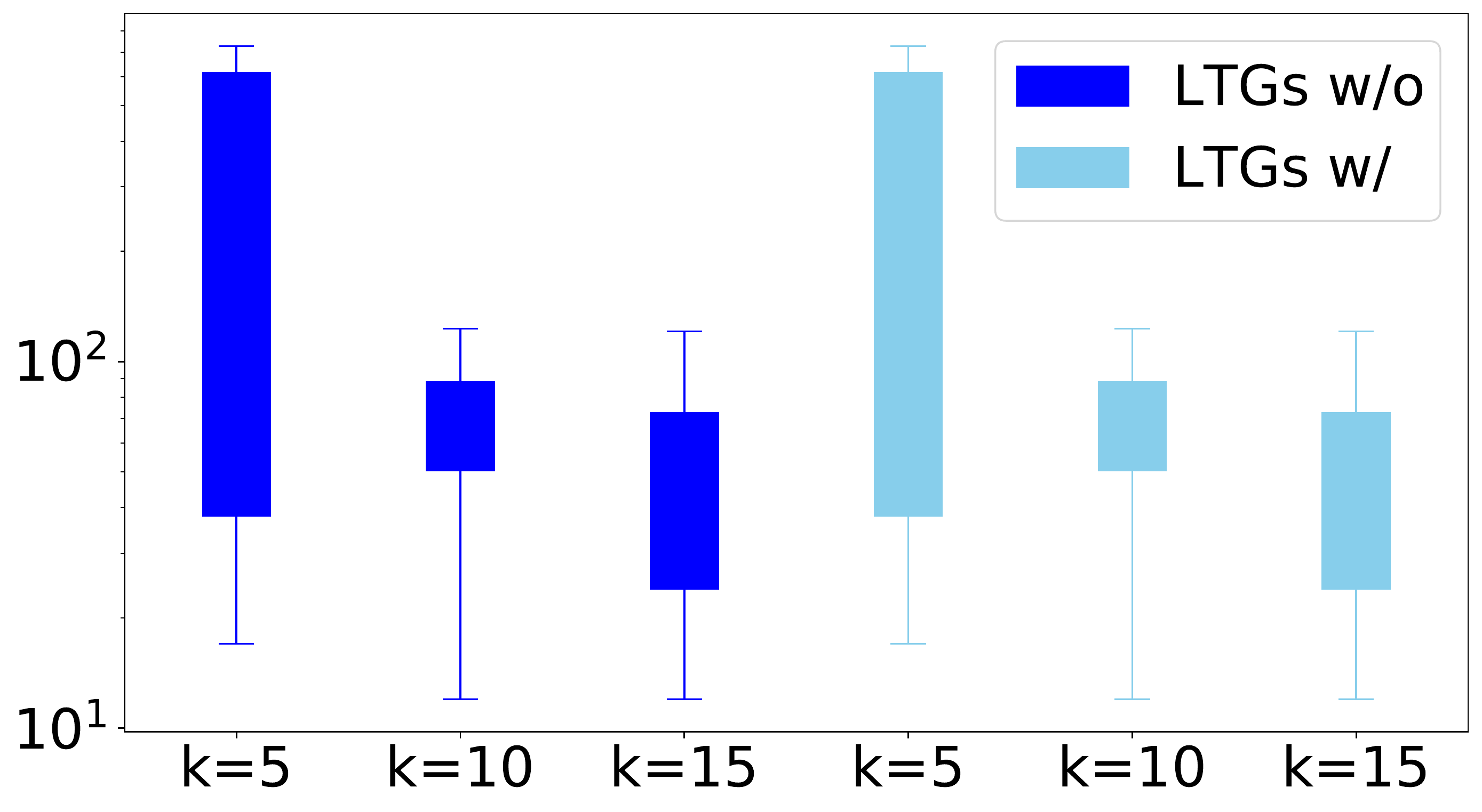}} \\

            \rotatebox{90}{\hspace{9mm}{$\boldsymbol{\smokers}$}} &
            \x\x {\includegraphics[width = 0.25 \textwidth]{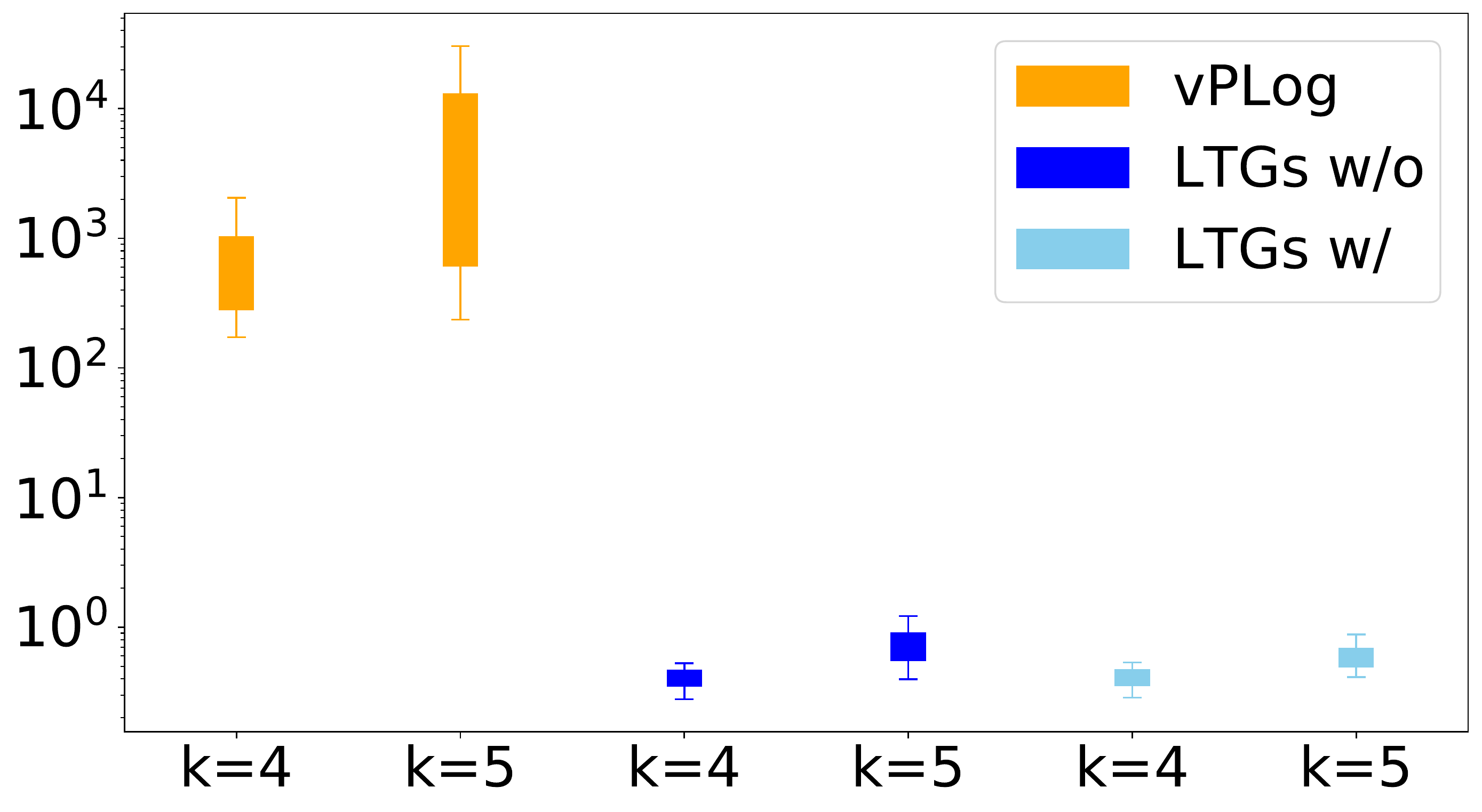}} &
            \x\x {\includegraphics[width = 0.25 \textwidth]{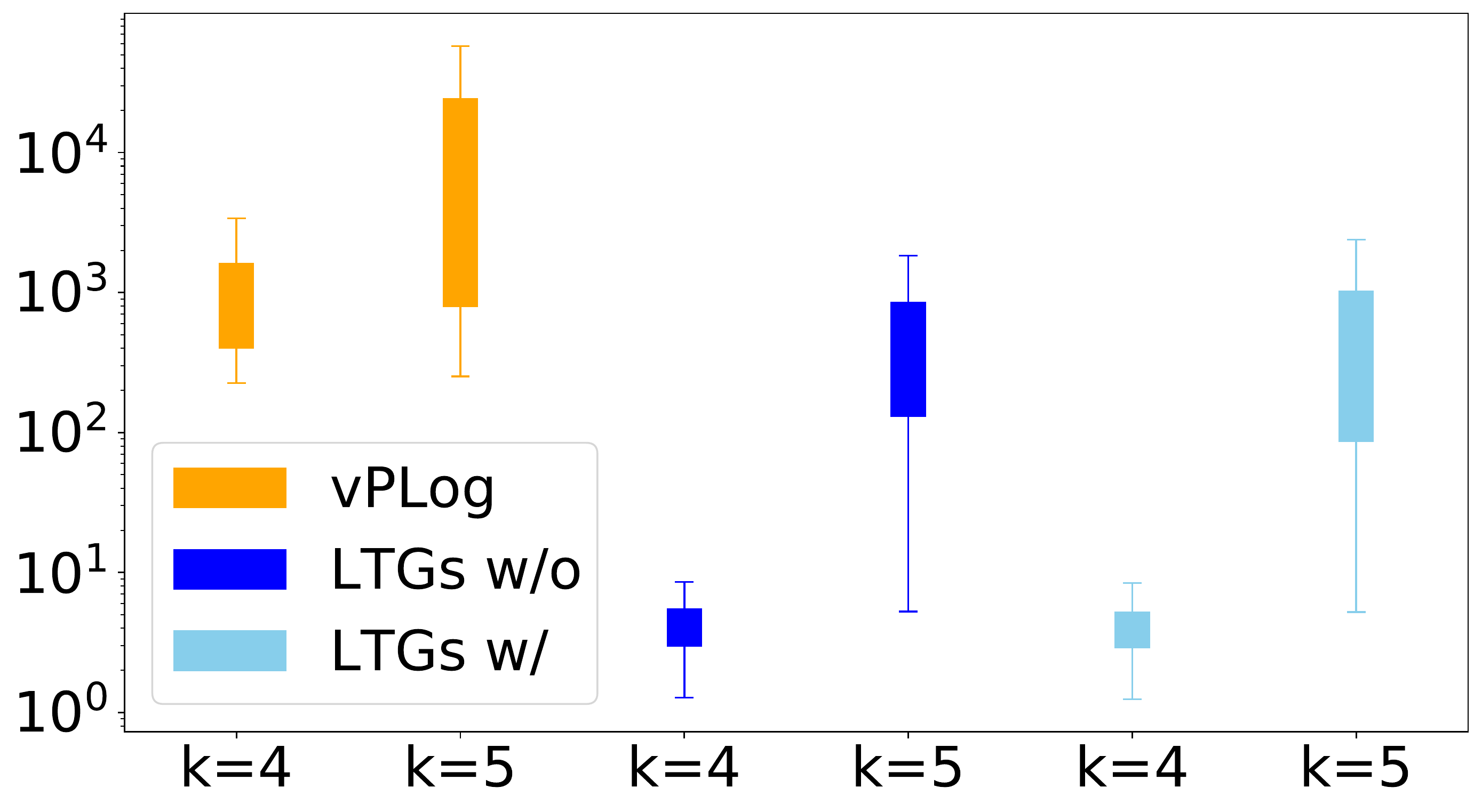}} &
            \x\x {\includegraphics[width = 0.25 \textwidth]{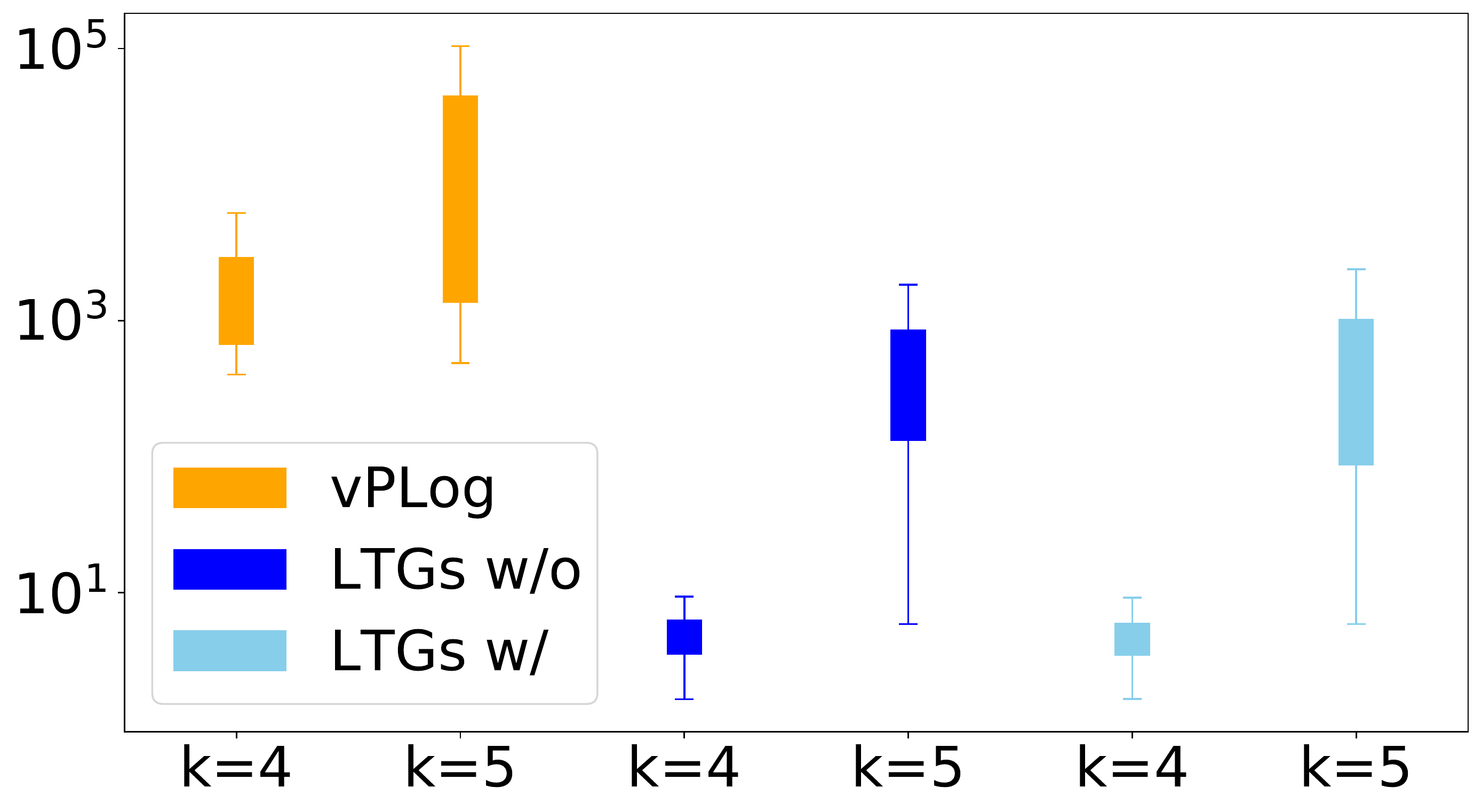}} &
            \x\x {\includegraphics[width = 0.25 \textwidth]{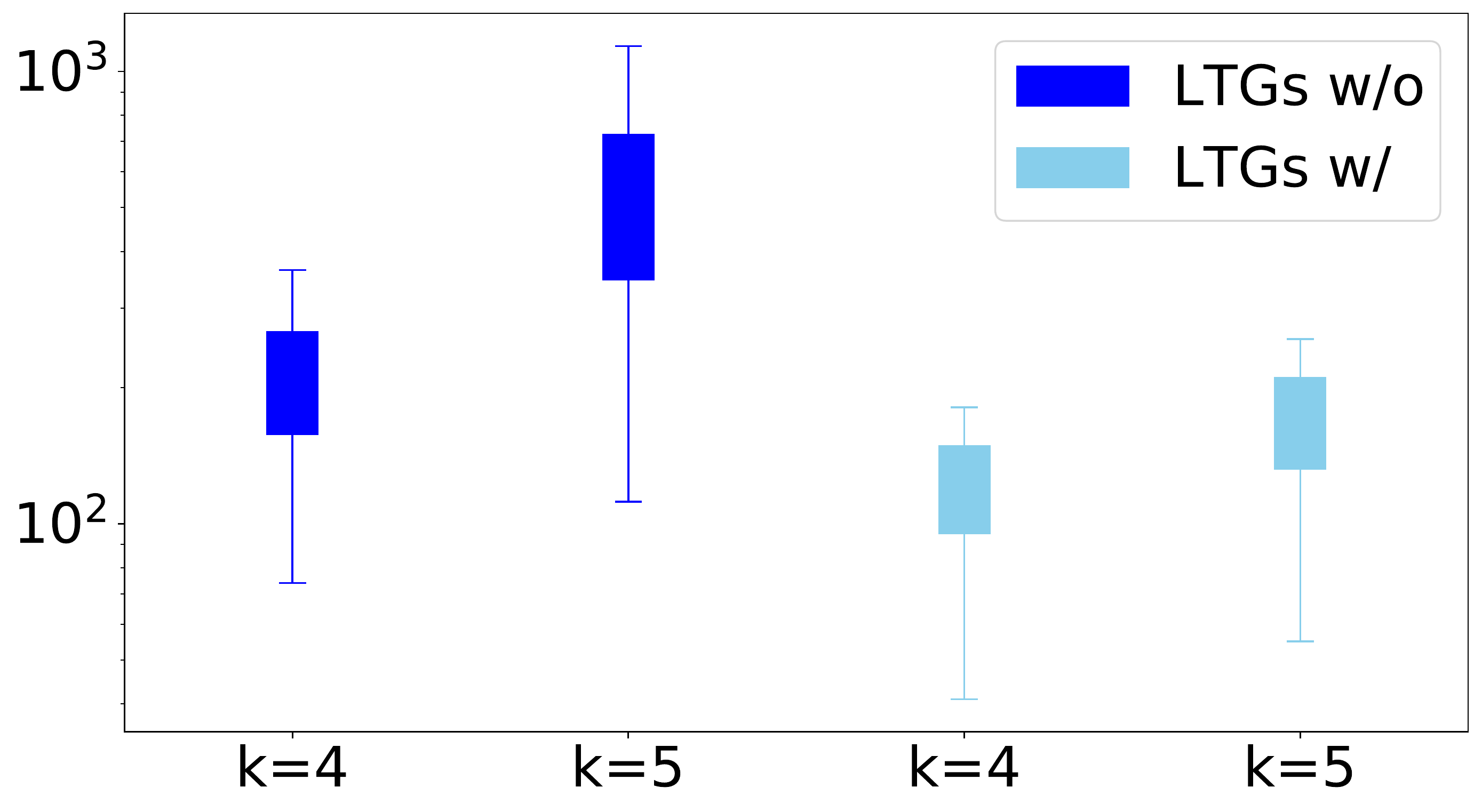}} \\
        \end{tabular}
    \end{adjustbox}
    \caption{Time in ms to answer queries and total number of derivations for different scenarios with $\vp$ and \ours{}.}
    \label{figure:experiments}
\end{figure*}

Figure~\ref{figure:experiments} reports the reasoning, probability computation,
and the total query answering time for the $\dbpedia$, $\claros$, $\smokers$,
$\yago$, and $\wnrr$ scenarios. The figures under ``Derivations'' show the
number of derivations for $\sys$.  We used boxplots since the number of queries
per scenario is large. The boxplots aggregate the times of all queries whose
evaluation is completed within 30 minutes. For $\smokers$, $k$ indicates the
maximum reasoning depth which can be either four or five. In contrast, the
different $k$'s in the $\yago$ and $\wnrr$ scenarios denote the number of
highest confidence rules kept per predicate. Table~\ref{tab:memory} reports the
number of queries whose evaluation was not completed within the timeout
(``TO" column) or which ran out of memory (``OOM" column) (the
\# queries per scenario is in Table~\ref{tab:benchmarks}). For
the queries that were successfully answered, we recorded the peak RAM
used by the engine. Table~\ref{tab:memory} reports the min and max values
obtained in each scenario (denoted by its initial, e.g., ``L" denotes
$\lubm$) to show the memory requirements in the best and worst case.

Figures~\ref{figure:experiments:vqar:a},~\ref{figure:experiments:vqar:b},
    and~\ref{tab:hardest-scallop-loss} report results collected from the 417
    \vqar{} queries for which \sys{} w/ can compute exact answers. To show the impact of
    the approximation on runtime, Figure~\ref{figure:experiments:vqar:a}
    reports a comparison of the total runtime needed by \scallop(1) (S(1)),
    \scallop(20) (S(20)), and \sys{} w/ (Total \sys{}). For \sys{} w/, the figure also reports a breakdown of the runtime needed for reasoning (Reas.), lineage collection
    (Lin.), and probability computation (Prob.).
    All times are in ms. Since our engine allows exact query answering, we also evaluate the impact of
    approximations on the answers' probabilities, i.e., we assess how close the approximate probabilities to the actual ones are.
    To this end, Figure~\ref{figure:experiments:vqar:b} reports the relative probability errors of the answers computed by \scallop(1) and \scallop(20).
    The relative probability error of an answer $\alpha$ is computed by
    $(\epsilon_{\alpha} - \hat{\epsilon}_{\alpha})/\epsilon_{\alpha}$, where
    $\epsilon_{\alpha}$ denotes the exact probability (as computed by $\ours{}$)
    and $\hat{\epsilon}_{\alpha}$ the approximation. In this experiment, the 417 queries produced
    5949 answers.
    Figure~\ref{figure:experiments:vqar:b} groups the answers based on their relative errors and reports the total number of answers within each group, i.e., regarding S(1), there are 168
    answers for which the error falls in $[0,10\%)$.
    Finally, to provide some anecdotal evidence, we chose the 5/417 queries
    that $\scallop$ takes the most time to answer for different $k$'s.
    Table~\ref{tab:hardest-scallop-loss} presents the total runtime of those queries with \scallop{} and \sys{} w/, and the highest probabilities of their answers.

\begin{table}[t]
    \caption{Min and max peak RAM usage (GB) to answer the benchmark
    queries and \# of OOM and TO queries.}

\footnotesize
\begin{tabular}{p{0.5cm}
        c
	p{0.8cm}  ||
    c
	p{0.8cm}
	||
    c
	p{0.8cm}
	}
       & \multicolumn{2}{c}{vProbLog} & \multicolumn{2}{c}{$\ours$ w/o} & \multicolumn{2}{c}{$\ours$ w/}\\
       & {Min/Max} & {OOM/TO} & {Min/Max} & {OOM/TO} & {Min/Max} & {OOM/TO} \\
	    \hline
    L10         &		11/11   &   1/2             &	1.7/19	& 	    1/0             & 1.9/11    & 1/0\\
    L100        &		13/14	&   2/12            &	1.8/5.7	&	    2/0             & 1.9/4.8   & 2/0\\
    D           &		30/38	&   3/0             & 	2/4.9	&	    3/0             & 1.9/2.5   & 3/0\\
    C           &		19/20	&   3/0             &	2.5/6.1	&       3/0	            & 2.4/5.2   & 3/0\\
    Y5     	    &  		11/11   &   12/8            &   1.8/1.8	&       12/0            & 1.8/1.8   & 11/0\\
    Y10     	&  		11/11	&   30/19           &   1.8/1.8	&       25/0            & 1.8/1.8   & 25/0\\
    Y15     	&  		11/11	&   12/18           &   2.4/2.4	&       12/8            & 1.7/1.7   & 12/8\\
    W5          &  		11/11	&    0/0            &   1.8/1.8	&       0/0             & 1.8/1.8 & 0/0\\
    W10         &  		11/11	&    0/0            &   1.8/1.9	&       0/0             & 1.8/1.8 & 0/0\\
    W15    	    &  		11/11	&    0/0            &   1.9/1.9	&       0/0             & 1.8/1.8 & 0/0\\
    S4  	    &  		11/11	&    0/0            &  	1.7/1.7	&       0/0	            & 1.7/17 & 0/0\\
    S5   	    &  		11/11	&    0/0            &  	1.8/17	&       0/0	            & 1.7/19 & 0/0\\
    V   	    &  		        &1000/0	        &  	   		&  1000/0	 	& 1.4/24 &  560/23\\
\end{tabular}
\label{tab:memory}
\end{table}

\subsection{Key conclusions}

\begin{figure*}[tb]
        \subfloat[]{\label{figure:experiments:vqar:a}\includegraphics[width =
        0.3\textwidth]{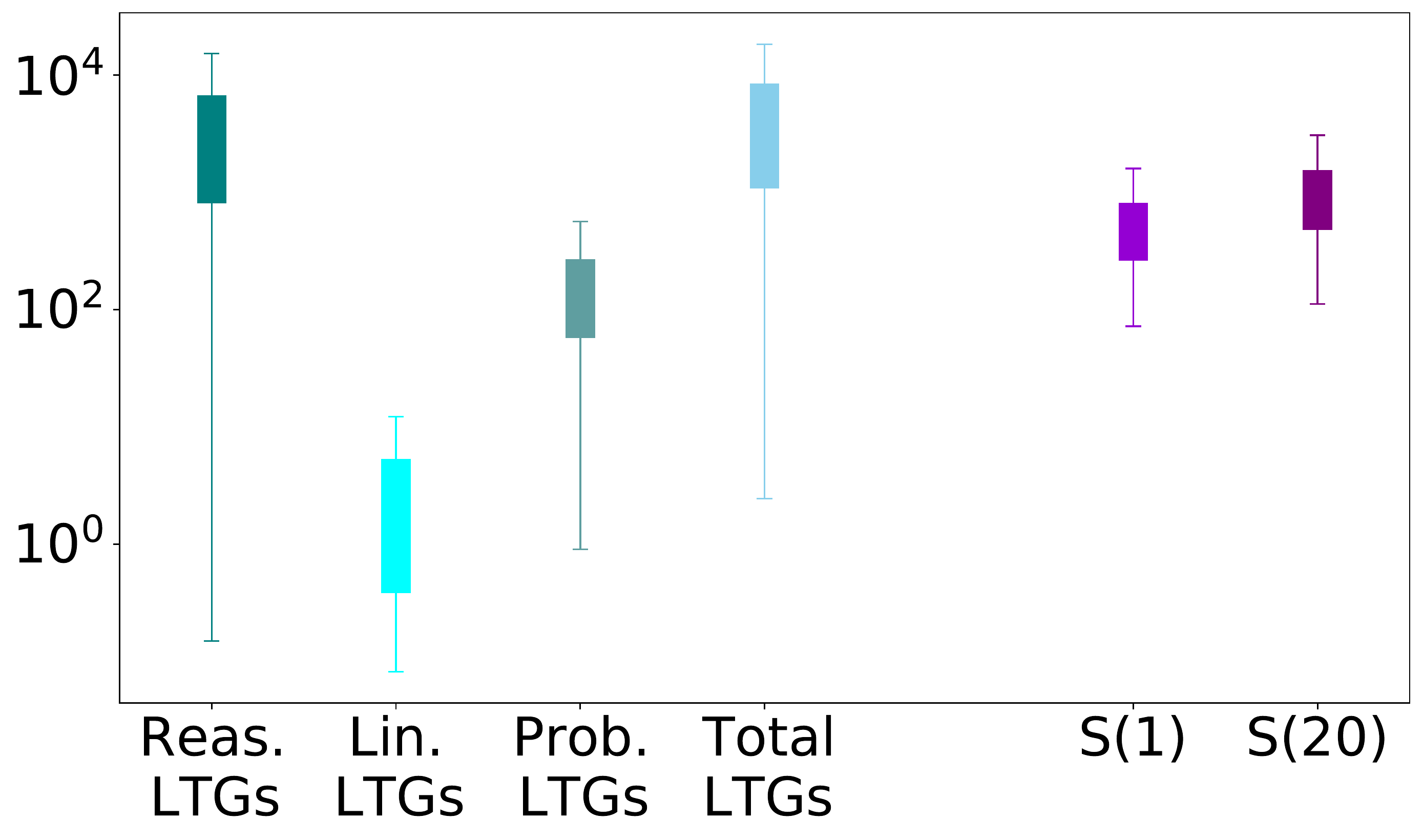}}
        \subfloat[]{\label{figure:experiments:vqar:b}\includegraphics[width = 0.3
    \textwidth]{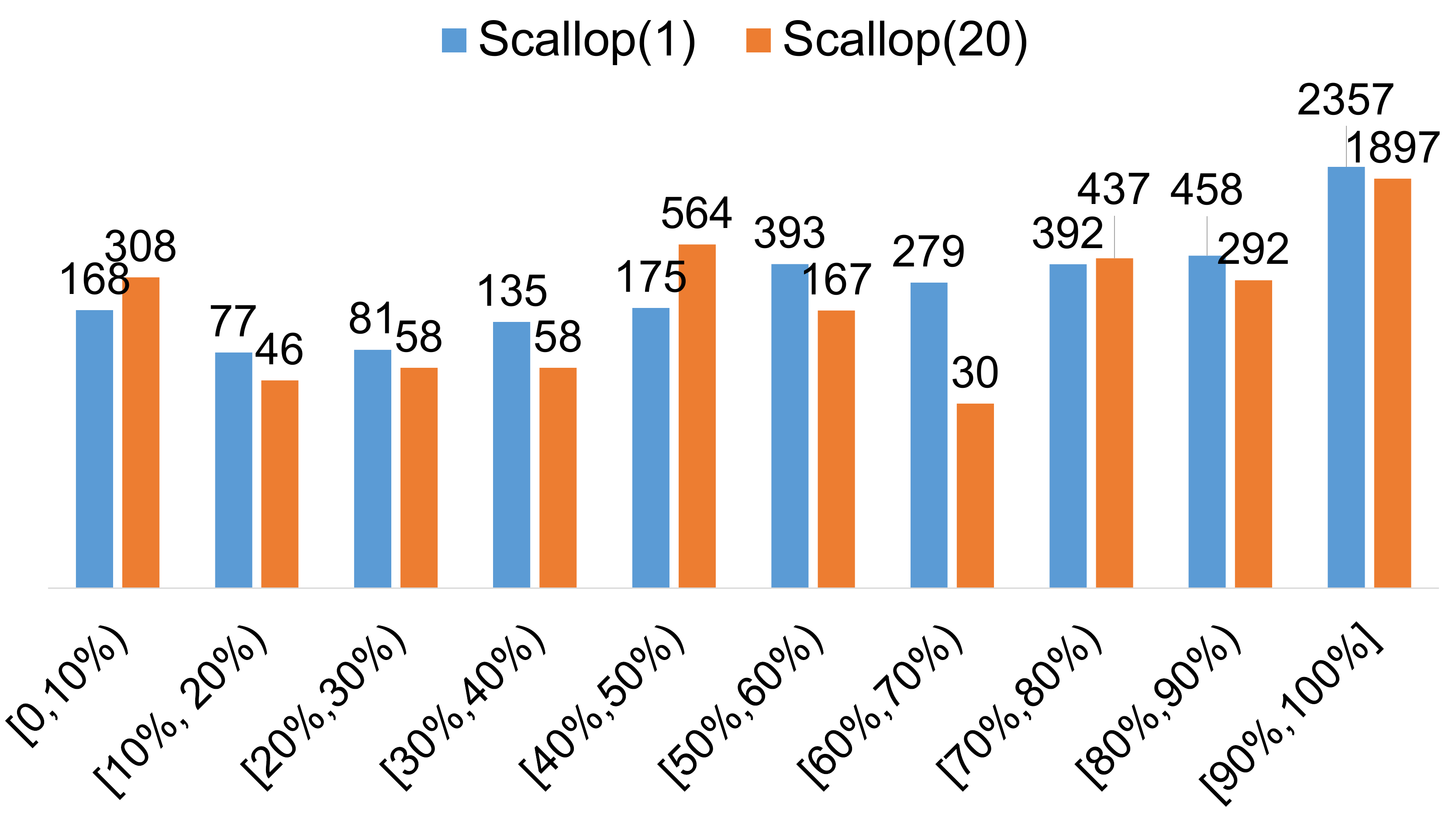}}
    \subfloat[]{\begin{minipage}{8cm}\vspace{-10em}    \label{tab:hardest-scallop-loss}
    \footnotesize
    \begin{tabular}{ll ccccc}
        & Query ID & \rotatebox{60}{2343894\_40} & \rotatebox{60}{2327997\_45} &
        \rotatebox{60}{2322829\_40} & \rotatebox{60}{2416754\_49} &
        \rotatebox{60}{2346575\_46}\\
        \multirow{4}*{\rotatebox{90}{Runtime}}  & \scal{1} & 1.5s & 800ms &
        721ms & 793ms & 1.1s \\
                                                & \scal{20} & 1311s & 148s & 88s
                                                & 45s & 40s \\
                                                & \scal{30} & TO & 1415s & 89s &
                                                42s & 41s \\
                                                & $\ours$ w/& 353s & 7.3s & 6.1s
                                                & 20s & 17.6s \\
                                                \hline
        \multirow{4}*{\rotatebox{90}{Probability}}  & \scal{1} & 0.03 & 0.003 &
                                                    0.04 & 0.006 & 0.68 \\
                                                & \scal{20} & 0.12 & 0.02 & 0.05
                                                & 0.007 & 0.97 \\
                                                & \scal{30} & TO & 0.02 & 0.05 &
                                                0.007 & 0.97 \\
                                                & $\ours$ w/& 0.13 & 0.02 & 0.11
                                                & 0.015 & 0.97 \\

    \end{tabular}
\end{minipage}}
        \caption{Results related to the 417 $\vqar$ queries that $\ours$ can answer exactly. (a)
        Runtime breakdown for $\ours$ w/ and total runtime in ms for $\scallop$; (b)
        relative probability errors obtained when approximating; (c) anecdotal
evidence with five queries.}
\end{figure*}

\conclude{$\ours$ outperforms prior art in terms of runtime}
Table~\ref{tab:total-lubm10} and Figure~\ref{figure:experiments} indicate that
query answering with $\ours$ is faster than with the other engines. For instance,
the maximum total runtime drops from 195s in $\dbpedia$
and 26s in $\claros$ with $\vp$, to 129s and 11s, respectively, with
$\ours$. The average total runtime drops from 14s ($\dbpedia$) and 8s
($\claros$) with $\vp$, to 6.6s and 1.4s respectively, with $\ours$. The improvements are
even larger for $\yago$ and $\wnrr$: the mean runtime drops from 0.7s to 0.1s in $\yagoL$ and from 0.2s to 0.01s in $\wnrrL$.

More importantly, $\ours$ can mean the difference between answering and not
answering the query at all. For instance, $\ours$ can successfully answer 13/14
queries in $\lubmM$ (most of them in the order of seconds) and 12/14 queries in
$\lubmL$. Regarding $Q_6$ in $\lubmM$, \ours{} completed reasoning and lineage collection successfully, but PySDD ran out of memory.
Regarding $Q_6$ and $Q_9$ in $\lubmL$ using $\ours$, it is lineage collection
that ran out of memory: reasoning finished successfully.
In comparison to $\vp$, which is the second-best exact engine, $\sys{}$ is faster
in all queries except for the ones in which both systems time out, with the improvements brought by $\sys{}$ being more than one
order of magnitude, see $Q_1$ and $Q_5$. Table~\ref{tab:total-lubm10} also
shows that $\ours$ is often faster than \scallop{}(30), outperforming it in all
cases except $Q_6$ in $\lubmM$, where \scallop{} returns approximate answers in
817s, and $Q_{14}$ in $\lubmL$. Regarding $Q_6$, the probability computation
done by $\ours$ is intrinsically expensive and hence $\scallop$, which does
approximations, runs faster.
Regarding $Q_{14}$, the query requires almost no reasoning and $\vp$ had a lower overhead.

\conclude{Collapsing the lineage can significantly improve the performance}
Consider, for instance, query $Q_{12}$ in $\lubmM$. It takes 10.6s to answer
$Q_{12}$ with $\ours$ w/o and
only 387ms with $\ours$ w/. This is because
reasoning for $Q_{12}$ in $\lubmM$ takes 10s with $\ours$ w/o and only 341ms with $\ours$ w/, see Figure~\ref{fig:performance:lubm}. Significant performance improvements are observed in other queries as well, e.g., $Q_4$ in $\lubmM$ and $Q_2$ in $\lubmM$.
Overall, $\ours$ w/ is at least 25\% faster than $\ours$ w/o in most of the cases. The biggest difference is observed in the $\vqar$ queries, where lineage collapsing allows us to compute the full least parameterized model for all queries.
The cause behind the reasoning time improvements is the drastic decrease in the
number of derivations. For instance, $Q_{12}$ in $\lubmM$ involves 10M
derivations with $\ours$ w/o, see
Figure~\ref{fig:derivations:lubm}. Instead, the same query involves 185k
derivations with $\ours$ w/. The number of derivations significantly decreases
also in $\dbpedia$ and $\smokers$, while it remains roughly the same in the other cases, see Figure~\ref{figure:experiments}.

Often, the overhead introduced by the operation of collapsing the lineage is
negligible (i.e., less than 1\%, see Table~\ref{table:overheadcompression}). However,
there are a few cases where it is not, like with $Q_{12}$ and $Q_{14}$.
Regarding $Q_{12}$, even though the overhead is non-negligible, it brings an
improvement in terms of runtime which outweighs the cost of collapsing. With $Q_{14}$, however, this is not the case because the query does not
trigger enough reasoning to justify the operation of collapsing, rendering collapsing no longer beneficial.

\conclude{The runtime overhead to collect the lineage is small}
In most scenarios, the overhead of computing the lineage is
relatively small in comparison to the time needed for reasoning. For instance, in
$\dbpedia$ and $\claros$, the maximum to collect the lineage is {1581
ms} and {1533} ms, respectively. Since $\ours$ can reason more efficiently,
this overhead is a fair price to pay to obtain a much lower total
runtime. In $\lubm$, we observed two cases where the cost of lineage collection
is prohibitively high: $Q_6$ and $Q_9$ in $\lubmL$. This is
because the associated number of answers is so large that
the cost of lineage collection outweighs the reduction in reasoning runtime.

\conclude{$\ours$ outperforms prior art in terms of memory} As we can see in
Table~\ref{tab:memory}, $\ours$ is up to four times more memory efficient than
\vp{} in the $\lubm$ scenarios; the improvements exceed the six times in the
$\yago$ and $\wnrr$ scenarios. This is because \sys{} does not fully materialize
the trees but stores instead pointers to the parent trees (structure sharing).
The operation of collapsing the derivation trees further reduces the memory
consumption. Looking again at Table~\ref{tab:memory}, we notice that in the best
case the max RAM usage is almost reduced by half (\dbpedia{}, 4953MB vs.
2550MB). The only scenario where $\ours$ requires more RAM is with $\smokers$5 (S5), which is a case where the column-based data structures used by $\vp$ take less space.
Regarding $\lubmM$ and \ours{} w/o, the higher maximum RAM consumption (19GM in \ours{} w/o vs 11GB in \vp{}) is due to $Q_9$, a query that \vp{} cannot answer.
Although $\ours$ is overall more
memory-efficient, there are still queries for which the RAM is not enough.
The reason lies in the worst-case intractability of the problem at hand.
One such example is $Q_6$ from $\lubmM$ where the reasoning and lineage collection step are computed successfully, but probability computation done by PySDD runs out of memory.
A major contribution of our work is that it significantly reduces the number of such cases, see $\vqar$.
Furthermore, the fact that in the
most challenging cases \sys{} reasons using 24GB of RAM hints that \sys{} does not require expensive hardware for supporting complex scenarios.

\conclude{$\ours$ can be used in combination with different probability
computation techniques} Table~\ref{tab:total-lubm10}
shows that even when using different probability computation tools
than PySDD, $\ours$ can have state-of-the-art performance:
when combined with c2d, $\ours$ w/ outperforms \vp{} in 5/14 queries in $\lubmM$
and 9/14 queries in $\lubmL$; when using d-tree, $\ours$ w/ is almost always faster than \vp{}.
Table~\ref{tab:differentsolvers} shows that PySDD tends to be the most efficient
library in terms of runtime, while c2d is the one
with the slowest runtime per answer. This is because the translation from DNF
into CNF
via the Tseitin transformation creates inter-dependencies among different disjunctions of the lineage
formulas that make the decomposition of the formula required by c2d harder.
It is also interesting to
point out that PySDD performs much better when it is coupled with $\ours$ than
with $\vp$. The reason is that PySDD translates the lineage into an
internal form called vtree \cite{vtree}. The cost of that translation depends on the structures
of the formulas, i.e., two formulas (one returned by $\ours$ and one returned by
$\vp$) may be logically equivalent, but the cost of translating them into
vtrees may be different. This explains the discrepancy of runtimes, which was
also observed in~\cite{DBLP:journals/corr/abs-1911-07750}.

\conclude{\ours{} is competitive to approximate techniques}
    Unsurprisingly, approximating reasoning by keeping only the top $k$ proofs
    returns lower runtimes (Figure~\ref{figure:experiments:vqar:a}). However, by
    doing so, the returned probabilities may be far from the actual ones. The
    difference can be substantial if we do aggressive approximations, like with
    $\scallop$(1). For instance, we can see from
    Figure~\ref{figure:experiments:vqar:b} that the relative error of 2357
    answers (out of 5949) is greater than $90\%$.
    Figure~\ref{tab:hardest-scallop-loss} gives a couple of illustrative
    examples: the answers of queries 2322829\_40 and 2416754\_49 have
    probabilities that are at least two times lower than the actual ones.  To
    reduce the approximation error, one would need to increase $k$, e.g., by
    running \scallop{}(20) and \scallop{}(30). However, by doing so, the runtime
    of \scallop{} increases to the point where it is no longer beneficial to
    approximate.

\section{Related work}

Several approaches perform reasoning under uncertainty including
ICL~\cite{poole:pilp08}, PRISM~\cite{sato:iclp95}, MLNs
\cite{DBLP:journals/ml/RichardsonD06}, and PSL
\cite{DBLP:journals/jmlr/BachBHG17}. ICL and PRISM support only rules where the
same predicate cannot occur both in their premise and in their conclusion. MLNs
and PSL are not based on logic programming but first-order logic. Hence, they
do not support non-ground recursive rules, as first-order logic cannot specify
the closure of a transitive relation \cite{transitive-closure-logic}.
Stochastic Logic Programs \cite{SLPs}, TensorLog \cite{tensorlog}, and
Probabilistic Datalog \cite{Fuhr00,DBLP:conf/sigir/Fuhr95} do not support the
possible world semantics.  \citeauthor{BaranyCKOV17}~\citeyear{BaranyCKOV17}
proposed another probabilistic version of Datalog, called PPDL
\cite{BaranyCKOV17}. Different from our work, the semantics of PPDL is
defined using Markov chains. We are not aware of any PPDL engine.

Several approaches aim to reduce the cost of computing the
full lineage by computing a subset of it.
The first ProbLog engine
\cite{deraedt:ijcai07} implemented \textit{iterative deepening}; ProbLog2
implements $k$-best \cite{k-best} and \emph{$k$-optimal} \cite{k-optimal}
approximations; \scallop{} keeps the $k$ most likely explanations per fact achieving state-of-the-art
performance \cite{scallop}. The main difference between these approaches and \sys{} is that the
latter performs exact reasoning. Our evaluation
shows that even though \sys{} must often do more work than
approximate methods, there are cases where $\sys{}$ is still significantly more efficient.

A multitude of approximations techniques tackle the DNF probability computation problem
\cite{DBLP:journals/vldb/DalviS07,Olteanu-approximate,Re-approximate}.
\citeauthor{scaled-dissociations} in \cite{scaled-dissociations} have recently
proposed an anytime approximation technique that builds upon
``dissociation”-based bounds \cite{dissociations}. Integrating such techniques
into $\ours$ is an interesting direction as it can extend applicability when the lineage is too large to be
further processed.

Research of query answering over PDBs \cite{pdbs} has provided us with
a wealth of results, especially on tractability of complex queries, e.g., \cite{dichotomy1,dichotomy2},
and approximations, e.g., \cite{SlimShot}. Recent work includes finding explanations for queries, e.g., \cite{GribkoffBUDA14},
and querying subject to constraints \cite{DBLP:conf/ijcai/FriedmanB19}.
Systems like MystiQ \cite{Mystiq} and MayBMS \cite{maybms} propose
extensions to DBMSs like PostgreSQL for supporting the semantics of PDBs,
while PrDB \cite{prdb} introduces techniques extending databases with graphical models.
In \cite{10.14778/2556549.2556564} and \cite{6544819}, \citeauthor{10.14778/2556549.2556564}
study the problem of answering queries over temporal PDBs and introduce techniques
for top-k query answering over PDBs, respectively. In contrast to our work, the above line of research focuses on
supporting SQL queries and not rule-based reasoning beyond view reformulation.

$\ours$ relates to high-performance Datalog engines including VLog \cite{vlog},
RDFox \cite{rdfox} and Vadalog \cite{vadalog}. It has been shown that
(non-probabilistic) TG-based reasoning outperforms the above engines in
terms of runtime and memory consumption~\cite{tsamoura2021materializing}.
Furthermore, TG-based reasoning provides the means to naturally maintain the
derivation provenance without extra overhead due to the induced TG.
None of the aforementioned engines can be easily extended to that fashion, as
they all implement the chase \cite{chasebench}, which ``disconnects" the facts
from the rules that derived them.

\problog{}, \vp{}, and \scallop{} closely relate to provenance semirings.
\citeauthor{provenance-semirings} have defined provenance for
Datalog using \emph{semirings} that supports the
possible world semantics \cite{provenance-semirings}.
The difference between~\cite{provenance-semirings} and the aforementioned
engines is that the latter improve the runtimes exploiting ideas from
bottom-up Datalog evaluation. %
The authors in \cite{senellart-provenance} provide an SNE, bottom-up method for
approximating the provenance for a specific class of semirings for finding
best-weight derivations. With the same spirit, \cite{deutch-provenance} proposes
approximate provenance computation techniques, while
\cite{fixed-point-provenance} develops semiring provenance for very general
logical languages involving negation and fixed-point operations.

\section{Conclusion}
\label{section:conclusion}

We presented a new scalable technique for probabilistic rule-based reasoning
over PDBs which
computes a compact probabilistic model, leveraging the topology of the TG and
structure sharing. Our experiments
show that our
engine outperforms prior art both in terms of runtime and memory consumption,
often significantly, and sometimes it can make the difference
between answering a query in a few seconds and not answering it at all.
Future research includes extending $\ours$ for reasoning over KG embedding models (e.g., \cite{DBLP:conf/uai/FriedmanB20}).
A promising direction is \ours{}' integration with
approximate techniques that compute only part of the
lineage
\cite{DBLP:conf/fgcs/Poole92,deraedt:ijcai07,k-best,k-optimal,approximate1,approximate2}, or with techniques that guide the computation of the proofs via machine learning i.e., reinforcement learning \cite{DBLP:conf/nips/KaliszykUMO18}.


\bibliographystyle{ACM-Reference-Format}
\bibliography{references}

\onecolumn
\appendix

\section{Computing execution graphs incrementally}
\label{appendix:increment}

We elaborate on the step of incremental EG computation
(line~\ref{algorithm:online:inc} of Algorithm~\ref{alg:online} and line~\ref{algorithm:compressed:incremental} of Algorithm~\ref{alg:online:compressed}).
EGs can be computed incrementally, as it was shown by
\citeauthor{tsamoura2021materializing}~\shortcite{tsamoura2021materializing}.
Suppose that we have already built an EG of depth ${k-1}$ and we want to add new
nodes of depth $k$ so that we do not miss any rule execution.  As the edges
dictate the nodes over which we are instantiating the premises of the rules, we
need to make sure that the extended EG exhaustively adds all possible edges.  We
refer to each combination of nodes of depth ${< k}$ whose facts may instantiate
the premise of a rule $r$ when reasoning over an EG, as $k$-compatible nodes for
$r$:

\begin{definition}\label{definition:compatible-nodes} [From \cite{tsamoura2021materializing}]
	A combination of $n$ nodes ${(u_{1},\dots,u_{n})}$ from $G$ is $k$-compatible with
	$r$, where ${k \geq 2}$ is an integer, if:
	\begin{compactitem}
		\item the predicate in the head of $u_i$ is $p_i$;
		\item the depth of each $u_i$ is less than $k$; and
		\item at least one node in ${(u_{1},\dots,u_{n})}$ is of depth ${k-1}$.
	\end{compactitem}
\end{definition}

The above ideas are summarized in an iterative procedure, which builds at each step $k$ an EG $G^k$ of depth $k$:
\begin{compactitem}
	\item (\textbf{Base step}) if ${k=1}$, then for each base\footnote{Base rules reference only database relations in their premises; non-base ones reference only derived relations instead.
	 Every ruleset can be rewritten to that form \cite{tsamoura2021materializing}.} rule $r$ add to $G^k$ a node $v$ associated with $r$.

	\item (\textbf{Inductive step}) otherwise, for each non-base rule $r$ and each combination of nodes ${(u_{1},\dots,u_{n})}$ from $G^{k-1}$ that is $k$-compatible with $r$,
	add to $G^k$: (i) a fresh node $v$ associated with $r$ and (ii) an edge ${u_{i} \rightarrow_i v}$, for each ${1 \leq i \leq n}$.
\end{compactitem}

\section{Proofs for Section 4}

\begin{algorithm}[tb]
    	\caption{$\step(\Ht)$, where $\Pp = (\Rp, \Fp, \pi)$, $\Ht = \{(\alpha_i, \lambda_i)\}$, and $\lambda_i$ is a Boolean formula over elements in $\Fp$} \label{alg:tcp}
	\begin{algorithmic}[1]
		\State $\mathcal{I}' \defeq \emptyset$\; $\Delta \defeq \emptyset$
		\For{\textbf{each} rule $r$ in $\Pp$ and each instantiation $\alpha \leftarrow \alpha_1 \wedge \dots \wedge \alpha_n$ of $r$, s.t. each ${(\alpha_i, \lambda_i)}$ is in $\Ht$ for some
		$\lambda_i$}	 														\label{alg:tcp:inst:begin}
			\State \textbf{add} ${(\alpha, \bigwedge_{i}^n \lambda_i)}$ to $\Delta$					\label{alg:tcp:inst:lambda}
		\EndFor																\label{alg:tcp:inst:end}

		\For{\textbf{each} $\alpha$, s.t. a  pair of the form ${(\alpha, \cdot)}$ is in $\Delta \cup \Ht$ }
			\State \textbf{add} ${(\alpha, \bigvee_{(\alpha, \varphi)} \varphi) \in \Delta}$ to $\mathcal{I}'$
		\EndFor
		\State \textbf{return} $\mathcal{I}'$
	\end{algorithmic}
\end{algorithm}

As stated in Section~\ref{sec:correctness}, Lemma~\ref{lemma:tree-formula-correspondence} concerns
the simplified versions of Algorithm~\ref{alg:online} and $\Tcp$ in which no termination condition is employed.
In the case of Algorithm~\ref{alg:online} that simplification means that the check in line~\ref{algorithm:filtering}
is avoided, so that each tree $\tau$ is added to $\treeset{v}{\Fp}$.
We recapitulate in Algorithm~\ref{alg:tcp} the steps taking place
during each iteration of simplified $\Tcp$, when an instance $\Ht$ is provided in its input.
Throughout, we fix a probabilistic logic program $\Pp = (\Rp, \Fp, \pi)$ and denote by $G^i$, the
EG computed at the end of the $i$-th iteration of Algorithm~\ref{alg:online}
and by $G^i(\Fp)$ the derivation trees that are stored within the nodes of $G^i$.
We also denote by $\Ht^i$ the instance computed at the end of the $i$-th iteration of $\Tcp$, where $\Ht^0 = \{(f,f) \mid p::f \in \Fp \}$.
The proof of Lemma~\ref{lemma:tree-formula-correspondence} relies on Claim~\ref{claim:tree-combination},
which again concerns the simplified Algorithm~\ref{alg:online}.

\begin{claim} \label{claim:tree-combination}
	Let $\tau_1,\dots,\tau_n$ be derivation trees in $G^i(\Fp)$, for $i \geq 0$, so that each $\tau_j$
	has root $\alpha_j$ and is stored in a node of depth $<i$.
	Let also $r$ be a rule and ${\alpha \leftarrow \alpha_1 \wedge \dots \wedge \alpha_n}$ be an instantiation of it.
	Then, the following IH holds for each $i \geq 0$:
	\begin{compactitem}
		\item $\rho$. There is a tree $\tau$ in $G^{i}(\Fp)$ so that $\tau$ has root $\alpha$ and there is an edge from the root of $\tau_j$ to the root of $\tau$.
	\end{compactitem}
\end{claim}
\begin{proof}
	For $i=0$, the IH $\rho$ trivially holds as $G^0$ is the empty graph by definition.
	For $i+1$ and assuming that $\rho$ holds for ${i \geq 0}$, we have the following.
	Let ${(u_1,\dots,u_n)}$ be the tuple of nodes in $G^i$, such that
	$\tau_j \in \treeset{u_j}{\Fp}$, for ${1 \leq j \leq n}$.
	If each $u_j$ is of depth $<i$, then $\rho$ trivially holds.
	Hence, we consider the case in which at least one $u_j$ is of depth $i$.

	As the root of each $\tau_j$ is $\alpha_j$ and due to the instantiation ${\alpha \leftarrow \alpha_1 \wedge \dots \wedge \alpha_n}$,
	it follows from Definition~\ref{definition:compatible-nodes} that
	${(u_1,\dots,u_n)}$ is $i$-compatible with $r$.
	As ${(u_1,\dots,u_n)}$ is $i$-compatible with $r$, it follows from the step in line~\ref{algorithm:online:inc}
	of Algorithm~\ref{alg:online} that $G^{i+1}$ includes a node $v$, so that the edge
	${u_j \rightarrow_j v}$ is in $G^{i+1}$, for ${1 \leq j \leq n}$.
	Furthermore, due to ${\alpha \leftarrow \alpha_1 \wedge \dots \wedge \alpha_n}$,
	due to Definition~\ref{def:instrule} and since $\tau_j \in \treeset{u_j}{\Fp}$, for ${1 \leq j \leq n}$,
	${\trees(\alpha,v,\Fp)}$ includes a tree $\tau$ with root atom $\alpha$
	and edges from the root of each $\tau_j$ to the root of $\tau$, where ${1 \leq j \leq n}$.
 	Tree $\tau$ will be stored in $\treeset{v}{\Fp}$.
 	As  $\tau \in \treeset{v}{\Fp}$, it follows that $\rho$ holds for $i+1$ concluding the proof of Claim~\ref{claim:tree-combination}.
\end{proof}

\lemmacorrespondence*
\begin{proof}
	For $i=0$, the IH holds in both directions as ${\Ht^0 = \{(f,f) \mid p::f \in \Fp \}}$ and ${G^0(\Fp) = \Fp}$ by definition
	and the lineage of a fact in $\Fp$ is the fact itself.

	($\Rightarrow$) For $i+1$ and assuming that $\omega$ holds for $i$, the proof proceeds as follows.
	Suppose that a rule $r$ is instantiated under ${\alpha \leftarrow \alpha_1 \wedge \dots \wedge \alpha_n}$ at step ${i+1}$
	of $\Tcp$
	and let ${(\alpha_j, \lambda^i_{\alpha_j}) \in \Ht^i}$, for ${1 \leq j \leq n}$.
	As the IH holds for $i$, we know that for each
	$\alpha_j$, there is a set of trees $\mathcal{T}_j$ in $G^i(\Fp)$ all having the same root $\alpha_j$,
	so that $\bigvee \nolimits_{\tau \in \mathcal{T}_j} \phi(\tau) \equiv \lambda^i_{\alpha_j}$ holds.
	Furthermore, due to the IH, each tree in $\mathcal{T}_j$ is stored in a node of depth $\leq i$, for ${1 \leq j \leq n}$.
	Due to the above, the formula computed out of the above rule instantiation in the step in line~\ref{alg:tcp:inst:lambda}
	of Algorithm~\ref{alg:tcp} is given by
	\begin{align}
		\bigwedge_{j=1}^n \lambda^i_{\alpha_j} \equiv \bigwedge_{j=1}^n  \bigvee \nolimits_{\tau \in \mathcal{T}_j} \phi(\tau) \equiv \bigvee_{(\tau_{1}, \dots, \tau_{n}) \in \mathcal{T}_1 \times \dots \times \mathcal{T}_n} \phi(\tau_{1}) \wedge \dots \wedge \phi(\tau_{n}) \label{eq:lemma:1}
	\end{align}
	From Claim~\ref{claim:tree-combination}, we know that
	for each tuple ${(\tau_1,\dots,\tau_n) \in \mathcal{T}_1 \dots \times \dots \mathcal{T}_n}$,
	there is a tree $\tau$ in $G^{i+1}(\Fp)$ so that
	$\tau$ has root fact $\alpha$ and there is an edge from the root of $\tau_j$ to the root of $\tau$.
	From the above and the definition of $\phi$, ${\phi(\tau) = \phi(\tau_1) \wedge \dots \phi(\tau_n)}$ holds.
	Consequently, the lineage of $\alpha$ in $G^{i+1}(\Fp)$ is given by
	\begin{align}
		\bigvee_{(\tau_{1}, \dots, \tau_{n}) \in \mathcal{T}_1 \times \dots \times \mathcal{T}_n} \phi(\tau_{1}) \wedge \dots \wedge \phi(\tau_{n})
	\end{align}
	Due to the above, $\omega$ holds for $i+1$: firstly, the formula $\lambda^{i+1}_{\alpha}$
	associated with atom $\alpha$ at the end of the $(i+1)$-th iteration of the steps shown in Algorithm~\ref{alg:tcp} is given by
	\begin{align}
		\lambda^{i+1}_{\alpha} = \lambda^{i}_{\alpha} \vee \bigvee \limits_{\alpha \leftarrow \alpha_1 \wedge \dots \wedge \alpha_n, s.t. (\alpha_j,\lambda_{\alpha_j}^i) \in \Ht^i}     \bigwedge_{j=1}^n \lambda^i_{\alpha_j}
	\end{align}
	secondly, due to $\omega$, for all trees ${\tau_1,\dots,\tau_m}$ in $G^i(\Fp)$ with root atom $\alpha$, we have
	\begin{align}
		\bigvee \nolimits_{j=1}^m \phi(\tau_j) \equiv \lambda^{i}_{\alpha}
	\end{align}
	and finally, the lineage of $\alpha$ in $G^{i+1}(\Fp)$ is the disjunction the formula $\phi(\tau)$ of each $\tau$ in
	$G^{i+1}(\Fp)$ with root fact $\alpha$.

	($\Leftarrow$) The proof is analogous to the other direction and relies on Claim~\ref{claim:tree-combination} and \eqref{eq:lemma:1}.
\end{proof}

\propequiv*
\begin{proof}
	If $\tau'$ is a subtree of $\tau$, then
	$\phi(\tau')$ is a subconjunct of $\phi(\tau)$. As such, ${\phi(\tau) \vee \phi(\tau') \equiv
\phi(\tau')}$ holds.
\end{proof}

\thmtermination*
\begin{proof}
	When a (probabilistic) logic program admits a finite Herbrand base, then there is
    a finite number of different rule instantiations. Due to the above, if we
    organize the rule instantiations in a graph $\Gamma$ so that $\Gamma$
    includes an edge from atoms ${\alpha_1, \dots, \alpha_n}$ to atom $\alpha$,
    for each rule instantiation ${\alpha \leftarrow \alpha_1, \dots, \alpha_n}$,
    there will be a depth ${k > 0}$ so that either there is no rule
    instantiation in which the atoms in the premise of a rule are of depth $<k$,
    or there are such instantiations, but there is a repetition in the atom
    derivations, in the sense, that an atom $\alpha$ of derivation depth $k$
    includes $\alpha$ in its set of ancestor nodes. The above indicates that
    Algorithm~\ref{alg:online} does terminate with programs admitting a finite
    Herbrand base. Notice that (probabilistic) Datalog programs always admit a
    finite Herbrand base, therefore Algorithm~\ref{alg:online} always
    terminates.
\end{proof}

\thmparameterizedTG*
\begin{proof}
	Recall that according to Definition~\ref{definition:trigger-graph},
	an EG $G$ for a probabilistic program $\Pp$ s a \emph{lineage TG} for $\Pp$, if
	for each atom ${\alpha \in \Ap \setminus \Fp}$, the lineage of $\alpha$ in $G(\Fp)$
	is logically equivalent to the lineage of $\alpha$ in $\Pp$.
	Lemma~\ref{lemma:tree-formula-correspondence} guarantees
	an equivalence between {the formulas computed out of the derivation trees and the disjuncts of the $\lambda$
	formulas in the least parameterized model of $\Pp$ at each step of $\Tcp$} assuming that both techniques employ no termination checks.
	Furthermore, Lemma~\ref{theorem:termination} suggests termination of Algorithm~\ref{alg:online}
	and consequently, due to Lemma~\ref{lemma:tree-formula-correspondence}, equivalence with
	least parameterized model of $\Pp$.
\end{proof}

\thmbounds*
\begin{proof}
	This follows straightforwardly from the fact that the lineage of an atom $\alpha$ in $G^k(\Fp)$ is a disjunct of its full lineage in $G(\Fp)$.
\end{proof}

\section{Proofs for Section 5}

\thmparameterizedTGcompressed*
\begin{proof}
	The proof of Theorem~\ref{theorem:parameterized-TG-compressed} follows from Lemma~\ref{lemma:xi1}, \ref{lemma:xi2} and \ref{lemma:collapsing-termination} which we show below.
	Throughout, we fix a probabilistic logic program $\Pp = (\Rp, \Fp, \pi)$ and denote by $G^i_1$ and $G^i_2$, the 
	EGs computed at the end of the $i$-th iteration of Algorithm~\ref{alg:online} and Algorithm~\ref{alg:online:compressed}.

	\begin{lemma} \label{lemma:xi1}
		For each $i \geq 0$, the following IH holds:
		\begin{compactitem}
			\item $\xi_1$. for each derivation tree $\tau$ with root fact $\alpha$ that is stored in a node of depth $i$ in $G_1^i$, 
			there exists a derivation tree $\varepsilon$ that is stored in a node of depth $i$ in $G_2^i$, such that ${\tau \in \unfold{\varepsilon}}$.  
		\end{compactitem}
	\end{lemma}
	\begin{proof}
		For $i=0$, $\xi_1$ trivially holds, as both $G_1^0$ and $G_2^0$ are empty. 
		For $i+1$ and assuming that $\xi_1$ holds for $i \geq 0$, the proof proceeds as follows. 
		Let $\tau$ be a derivation tree that has been added to a node $v$ by the end of the $i+1$-th iteration of Algorithm~\ref{alg:online}. 
		Let us assume $v$ is associated with rule $r$ and that the root of $\tau$ is a fact $\alpha$.
		According to the definition of redundancy from Section~\ref{sec:compression} and the steps in lines \ref{algorithm:filtering} and \ref{algorithm:storetrees} of Algorithm~\ref{alg:online}, 
		it follows that $\tau$ is not redundant w.r.t. $\alpha$.  
		Furthermore, from Definition~\ref{def:instrule}, 
		it follows that there is an instantiation $\alpha \leftarrow \alpha_1 \land \ldots \land \alpha_n$ of $r$
		so that for each ${1 \leq j \leq n}$, (i) there is a node $u_j$ of depth $\leq i$, 
		(ii) a tree $\tau_j$ is stored in $\treeset{u_j}{\Fp}$ and (iii) the edge $u_j \rightarrow v$ is in $G_1^{i+1}$. 
		Due to the step in line~\ref{algorithm:online:inc} of Algorithm~\ref{alg:online}, 
		at least one node from $u_1,\dots,u_n$ is of depth $i$.
		Since $\xi_1$ holds for $\leq i$, we have that for each ${\tau_j}$, there exists a derivation tree $\varepsilon_j$ 
		stored in a node $u'_j$ in $G_2^{i}$ having the same depth with $u_j$, so that $\tau_j \in \unfold{\varepsilon_j}$.  
		Due to the step in line~\ref{algorithm:compressed:incremental} of Algorithm~\ref{alg:online:compressed}, 
		since at least one node in $u_1,\dots,u_n$ is of depth $i$ and since the combination of nodes  
		${(u_{1},\dots,u_{n})}$ from $G_1^{i}$ is ($i+1$)-compatible with $r$ (see Definition~\ref{definition:compatible-nodes}), then 
		the combination of nodes ${(u'_1,\dots,u'_n)}$ from $G_2^{i}$ is ($i+1$)-compatible with $r$. 
		Hence, ${(u'_1,\dots,u'_n)}$ will be considered in line~\ref{algorithm:compressed:incremental} at the beginning of 
		the ($i+1$)-th iteration of Algorithm~\ref{alg:online:compressed}. Consequently, 
		$G_2^{i+1}$ will include a node $v'$ associated with rule $r$ and an edge from $u'_j \rightarrow v'$, for ${1 \leq j \leq n}$.   
		Furthermore, since for each ${1 \leq j \leq n}$, $\varepsilon_j$ is stored in $u'_j$, so that $\tau_j \in \unfold{\varepsilon_j}$, 
		since the root of $\varepsilon_j$ is $\alpha_i$, since  
		$\alpha \leftarrow \alpha_1 \land \ldots \land \alpha_n$ is an instantiation of $r$
		and due to Definition~\ref{def:instrule}, it follows that for each ${1 \leq j \leq n}$, $\trees(\alpha,v',\Fp)$ will include a tree $\varepsilon$
		and $G_2^{i+1}$ will include an edge $u'_j \rightarrow v'$.
		We distinguish the following cases.
		
		\begin{compactitem}
			\item No collapsing. Hence ${Z = \trees(\alpha,v',\Fp)}$ in line~\ref{algorithm:donotcompr} of Algorithm~\ref{alg:online:compressed}. 
			Then Algorithm~\ref{alg:online:compressed} will iterate $\varepsilon$ in line \ref{algorithm:beginuncompr}. 
			From Definition~\ref{def:unfolding}, we know that for each 
			${(\delta_1,\dots,\delta_n) \in \unfold{\varepsilon_1} \times \dots \times \unfold{\varepsilon_n}}$, 
			$\unfold{\varepsilon}$ includes a tree $\delta$, such that $\roottree{\delta}$ is $\alpha$ and 
		         there is an edge from each $\roottree{\delta_i}$ to $\roottree{\delta}$. Due to the above, 
		         and since $\tau_j$ is in $\unfold{\varepsilon_j}$, for ${1 \leq j \leq n}$, it follows that $\tau$ is in $\unfold{\varepsilon}$.  
		         As $\tau \in \unfold{\varepsilon}$, since $\tau$ is not redundant w.r.t. $\alpha$ and
		         due to the definition of redundancy from Section~\ref{sec:compression}, 
		         $\varepsilon$ is not redundant w.r.t. $\alpha$ and hence, $\varepsilon$ is added to $\treeset{v'}{\Fp}$ in line~\ref{algorithm:redundancypass} 
		         of Algorithm~\ref{alg:online:compressed}.  
		         
		         \item Collapsing. Hence, $\varepsilon$ is collapsed with other trees into a new tree $\varepsilon'$. 
		         From the proof of the previous case, we know that $\tau$ is in $\unfold{\varepsilon}$ and $\varepsilon$ is not redundant w.r.t. $\alpha$.   
		         Since $\unfold{\varepsilon'} \supseteq \unfold{\varepsilon} \supseteq  \{\tau\}$, 
		         $\tau$ will be included in the unfolding of $\varepsilon'$. 
		         Since $\tau \in \unfold{\varepsilon'}$ and since $\tau$ is not redundant w.r.t. $\alpha$, 
		         it follows that $\varepsilon'$ is not redundant w.r.t. $\alpha$. Hence, $\varepsilon'$
		         is added to $\treeset{v'}{\Fp}$ in line~\ref{algorithm:redundancypass} of Algorithm~\ref{alg:online:compressed}.
		\end{compactitem}
		The above shows that $\xi_1$ holds for $i+1$ concluding the proof of Lemma~\ref{lemma:xi1}. 
	\end{proof}

	\begin{lemma}  \label{lemma:xi2}
		For each $i \geq 0$, the following IH holds:
		\begin{compactitem}  
			\item $\xi_2$. for each derivation tree $\varepsilon$ with root fact $\alpha$ that is stored in a node of depth $i$ in $G_2^i$, 
			each tree in ${\unfold{\varepsilon}}$ that is not redundant w.r.t. $\alpha$ is stored in a node of depth $i$ in $G_1^i$.    
		\end{compactitem}
	\end{lemma}	
	\begin{proof}
		Similarly to the proof of IH $\xi_1$, $\xi_2$ trivially holds for $i=0$, as both $G_1^0$ and $G_2^0$ are empty. 
		For $i+1$ and assuming that $\xi_2$ holds for $i$, the proof proceeds as follows. 
		Let $\varepsilon$ be a derivation tree that has been added to a node $v$ by the end of the ($i+1$)-th iteration of Algorithm~\ref{alg:online:compressed}.
		Let us assume that node $v$ is associated with rule $r$ and that the root of $\varepsilon$ is fact $\alpha$. 
		Since $\varepsilon$ is added to node $v$, 
		it follows that $\epsilon$ is not redundant w.r.t. $\alpha$.  
		Furthermore, from Definition~\ref{def:instrule}, 
		it follows that there is an instantiation $\alpha \leftarrow \alpha_1 \land \ldots \land \alpha_n$ of $r$
		so that for each $\alpha_j$, there is a $\epsilon_j$ stored in some node $u_j$ of depth $\leq i$
		and $u_j \rightarrow v$ is in $G_2^{i+1}$. 
		Due to the step in line \ref{algorithm:compressed:incremental} of Algorithm~\ref{alg:online:compressed},  
		at least one node from $u_1,\dots,u_n$ is of depth $i$.
		Since $\xi_2$ holds for $\leq i$, for each $\varepsilon_j$, where ${1 \leq j \leq n}$, each tree in 
		$\unfold{\varepsilon_j}$ that is not redundant w.r.t. $\alpha$ is stored in a node $u'_j$ in $G^i_1$.   
		Furthermore, due to $\xi_2$, $u'_j$ has the same depth with $u_j$. 
		
		To prove that $\xi_2$ holds for $i+1$, we need to show that each tree in 
		$\unfold{\varepsilon}$ that is not redundant w.r.t. $\alpha$ is stored in a node of depth $i$ in $G_1^{i+1}$.    
		According to Definition~\ref{def:unfolding}, $\unfold{\varepsilon}$ 
		includes for each ${(\delta_1,\dots,\delta_n) \in \unfold{\varepsilon_1} \times \dots \times \unfold{\varepsilon_n}}$,
		a tree $\delta$ whose root is $\alpha$ and that has
		an edge from each $\roottree{\delta_i}$ to $\roottree{\delta}$. 
		From the above, due to Definition~\ref{def:instrule}, 
		due to the instantiation $\alpha \leftarrow \alpha_1 \land \ldots \land \alpha_n$ of $r$ and 
		since each $u_j$ has the same depth with $u'_j$, for ${1 \leq j \leq n}$, 
		it follows that each combination of nodes $(u'_1,\dots,u'_n)$ is $i+1$-compatible with
		$r$. From the above and due to the step in line~\ref{algorithm:online:inc} of Algorithm~\ref{alg:online}, 
		we have that for each combination $(\tau'_1,\dots,\tau'_n)$, where $\tau'_j \in \unfold{\varepsilon_j}$, for $1 \leq j \leq n$,
		there will be a node $v'$ of depth $i+1$ in $G_1^{i+1}$ and a tree $\tau'$, so that  
		${\tau' \in \trees(\alpha,v',\Fp)}$ and there is an edge from the root of $\tau'$ to the root of each 
		$\tau'_j$. Finally, each tree in $\trees(\alpha,v',\Fp)$ that is not redundant w.r.t. $\alpha$ will be stored in 
		$\treeset{v'}{\Fp}$ in line~\ref{algorithm:storetrees} of Algorithm~\ref{alg:online}, proving $\xi_2$ for $i+1$. 
	\end{proof}
	
	\begin{lemma} \label{lemma:collapsing-termination}
		For a probabilistic program $\Pp=(\Rp,\Fp, \pi)$, $\mat(\Pp)$ terminates at step $i$ if-f $\matcompr(\Pp)$ terminates at step $i$. 
	\end{lemma}	
	\begin{proof}
	($\Rightarrow$) (By contradiction) Suppose that $\mat(\Pp)$ terminates at step $i$.  
	Then, for each node $v$, all the derivation trees that are computed in line~\ref{algorithm:tree:start} of Algorithm~\ref{alg:online} are redundant w.r.t. $\alpha$.
	Hence, $\treeset{v}{\Fp} = \emptyset$, for each $v$ in $G^i_1$.  
	To reach a contradiction, suppose that $\matcompr(\Pp)$ does not terminate at step $i$. That means that 
	some $\varepsilon$ in line~\ref{algorithm:redundancypass} of Algorithm~\ref{alg:online:compressed} is not redundant w.r.t. $\alpha$. 
	Due to the above, there exists at least one tree $\tau' \in \unfold{\varepsilon}$ in $G_1^i(\Fp)$ that is not redundant w.r.t. $\alpha$.  
	However, if such a $\tau'$ exists, then, according to Lemma~\ref{lemma:xi2}, Algorithm~\ref{alg:online} should have iterated over it in line~\ref{algorithm:tree:start}, eventually adding 
	it to the corresponding node in line~\ref{algorithm:storetrees}, reaching a contradiction.   
	
	($\Leftarrow$) (By contradiction) Suppose that $\matcompr(\Pp)$ terminates at step $i$.  
	Then, for each node $v$ in $G^i_2$, we have that $\treeset{v}{\Fp} =\emptyset$.  
	Suppose by contradiction, that $\mat(\Pp)$ does not terminate at step $i$. 
	Then, some tree $\tau$ that is visited in line~\ref{algorithm:tree:start} of Algorithm~\ref{alg:online} is not redundant w.r.t. $\alpha$. 
	However, from Lemma~\ref{lemma:xi1}, we know that tree $\tau$ will be in the unfolding of some 
	derivation tree $\varepsilon'$ in $G_2^i(\Fp)$. 
	Furthermore, since $\tau$ is not redundant w.r.t. $\alpha$, 
	it follows that $\varepsilon'$ is not redundant w.r.t. $\alpha$. 
	Hence, $\varepsilon'$ should be stored in the corresponding node, see line~\ref{algorithm:redundancypass}
	of Algorithm~\ref{alg:online:compressed}, reaching a contradiction.
	\end{proof}

	After proving Lemma~\ref{lemma:xi1}, \ref{lemma:xi2} and \ref{lemma:collapsing-termination}, 
	we are ready to return back to the proof of Theorem~\ref{theorem:parameterized-TG-compressed}. 
	Due to Lemma~\ref{lemma:collapsing-termination}, we know that both Algorithm~\ref{alg:online} and \ref{alg:online:compressed} 
	terminate at the same iteration. Let $i$ be that iteration. 
	For an atom $\alpha$, let ${\tau_1,\dots,\tau_n}$ be all trees in $G^i_1(\Fp)$ 
	with root $\alpha$. Due to Lemma~\ref{lemma:xi1}, we have that 
	for each $\tau_j$, for ${1 \leq j \leq n}$, there exists a derivation tree ${\varepsilon_j \in G^i_2(\Fp)}$, such that 
	${\tau_j \in \unfold{\varepsilon_j}}$. Furthermore, due to Lemma~\ref{lemma:xi2}, 
	each tree $\delta \in \unfold{\varepsilon_j}$ is either one of  
	${\tau_1,\dots,\tau_n}$, or is redundant w.r.t. $\alpha$. 
	Let $\tau'_1,\dots,\tau'_m$ be all trees in $\bigcup \nolimits_{j=1}^n \unfold{\varepsilon_j}$
	that are redundant w.r.t. $\alpha$. 
	Due to the above, we have that the lineage of $\alpha$ in $G^i_2(\Fp)$ is given by:
	\begin{align}
		\bigvee \nolimits_{j=1}^n \bigvee \nolimits_{\delta \in \unfold{\varepsilon_j}} \phi(\delta) =  \bigvee \nolimits_{j=1}^n \phi(\tau_j) \vee \bigvee \nolimits_{j=1}^m \phi(\tau'_j)
	\end{align}
	
	Due to the correctness of Algorithm~\ref{alg:online}, for each such tree $\tau'_{\kappa}$, for ${1 \leq \kappa \leq m}$, there is a tree $\tau_{\ell}$, where ${1 \leq \ell \leq n}$, 
	such that $\tau_{\ell}$ is a subtree of $\tau'_{\kappa}$-- otherwise, the lineage for $\alpha$ in $G^i_1(\Fp)$ would not be incomplete. 
	From Proposition~\ref{proposition:equiv} and the above, it follows that 	
	\begin{align}
		\bigvee \nolimits_{j=1}^n \phi(\tau_j) \vee \bigvee \nolimits_{j=1}^m \phi(\tau'_j) \equiv \bigvee \nolimits_{j=1}^n \phi(\tau_j)		\label{eq:compressed:equivalence}
	\end{align}
	Equation \eqref{eq:compressed:equivalence} indicates that the lineage of an atom $\alpha$ in $G^i_1(\Fp)$ is logically equivalent to the lineage of $\alpha$ in $G^i_2(\Fp)$. 
	From the above and since for each atom ${\alpha \in \Ap \setminus \Fp}$,
	the lineage of $\alpha$ in $G^i_1(\Fp)$ is logically equivalent to the lineage of $\alpha$ in $\Pp$ (see Theorem~\ref{theorem:parameterized-TG}),
	it follows that the lineage of $\alpha$ in $G^i_2(\Fp)$ will be logically equivalent to the lineage of $\alpha$ in $\Pp$. Hence, 
	Algorithm~\ref{alg:online:compressed} will also compute a lineage TG for probabilistic program $\Pp$, completing the proof of 
	Theorem~\ref{theorem:parameterized-TG-compressed}.
\end{proof}
\section{Additional experimental results}
\label{app:eval}

\leanparagraph{Query generation}
We applied a technique, called $\queryGen$, that generates synthetic queries of increasing complexity. This technique is very similar to what has been
applied by~\cite{DBLP:conf/ecai/JoshiJU20}. It is also similar to the technique
used in~\cite{chasebench}, extending it
with a better control on the level of reasoning involved in answering the
synthetic queries.

$\queryGen$ takes as input a set of rules $\Rp$ and a set of facts $\Fp$ and returns a set of queries over the derived relations.
The technique starts by computing a graph $O$. Graph $O$ includes a node for each column occurring in a derived relation and an undirected edge between each pair of columns whose data overlaps. As $O$ encodes overlapping columns in the derived relations, its construction is based on the model $M$ of the non-probabilistic program ${(\Rp, \Fp)}$, i.e., $\queryGen$ first computes the model of ${(\Rp, \Fp)}$ and then spots overlaps in the derived data.

After computing $O$, $\queryGen$ performs random walks on ${O}$ to compute an initial set of queries $\mathcal{Q}$,
where each query involves up to $P$ derived predicates and up to $E$ free variables, which are randomly chosen.
In step two, the technique computes for each query in $\mathcal{Q}$ (i)
the number of its recursive predicates, (ii) the number of rules defining each of its predicates
and (iii) the maximum distance between any of its predicates to an extensional predicate.
By predicates, we refer to predicates occurring in the body of each query.
Criteria (i) and (iii) are computed based on the dependency graph $\Delta$ of $\Rp$,
where $\Delta$ includes a node for each predicate occurring in $\Rp$ and an edge from a predicate
$b$ to a predicate $h$, if there exists a rule in $\Rp$ whose conclusion includes an $h$-atom
and whose premise includes a $b$-atom. In particular, a predicate $h$ is recursive if $h$ occurs
in a cyclic path in $\Delta$; and the distance between a predicate $h$ and an extensional predicate $b$ is defined as the length of the longest path between $b$ and $h$.
To create a challenging benchmark, we eliminate the queries from $\mathcal{Q}$ with the lowest values
for criteria (i)--(iii). In step three, $\queryGen$ executes the remaining queries over $M$ -- recall that $M$ is the model of the program
${(\Rp, \Fp)}$-- and discards the empty ones. In the fourth and final step, $\queryGen$ randomly chooses a constant occurring
in the set of answers of each one of non-empty queries in $\mathcal{Q}$ and uses this constant to bind the corresponding free variable.

Let us elaborate on each step taking place in $\queryGen$. The first and the second step serve as a pivot to create non-empty queries over the derived predicates
in $\Rp$. The ranking of the queries based on criteria (i)--(iii) aims at selecting the most difficult ones in terms of reasoning: the higher the values of
(i)--(iii) become, the more reasoning is required to answer those queries. Step three ensures that the queries are indeed none-empty, while
step four aims at reducing the number of answers per query.
We restricted to queries including $\{1,\dots,4\}$ atoms and up to three free variables.

\leanparagraph{AnyBurl scenarios} We provide more details on how we created a probabilistic database from AnyBurl
rules. AnyBurl is a rule mining technique and as such it annotates the mined
rules with confidence values and not the KB facts. In our setting, it is the
facts that are associated with probabilities (and not rules).  However, this is
not an issue. As mentioned in Section~\ref{section:preliminaries}, there is a
simple trick that allows us to transform such rulesets into a probabilistic
program. The trick consists of adding an extra ``dummy'' fact to the premise of
the rule and setting its probability with the confidence value of the rule. The
KB facts created out of the training and validation triples are assigned
probability equal to one.

\leanparagraph{Scallop}
In the $\lubm$ scenarios, we used Scallop's Github release\footnote{\url{https://github.com/scallop-lang/scallop-v1}.}
as the at the NeurIPS website\footnote{\url{https://proceedings.neurips.cc/paper/2021/hash/d367eef13f90793bd8121e2f675f0dc2-Abstract.html}.}
does not provide an interface for specifying rules and facts.
We did not employ Scallop in $\dbpedia$ and $\claros$,
as the GitHub release does not support programs as large as the ones, throwing an exception at data loading time.
Scaling Scallop to very large data sizes was left as future work by the authors \cite{scallop}.
Furthermore, for the $\smokers$ scenarios, Scallop may miss some proofs within the $k$ threshold.
All the above issues are communicated to the authors.
For the $\vqar$ scenarios, we used the engine available at the NeurIPS website, since the one in GitHub does not support ternary predicates.

\begin{table*}[h]
    \caption{Min and max reasoning depth (DP), \# derivations (DR) and
    \# rules (R) relevant to the benchmark queries. The statistics consider only the queries that did not timeout or ran out of memory.
    For $\vqar$, column DR shows the number of derivations after collapsing the lineage. 
}
\begin{adjustbox}{max width=\textwidth}
\begin{tabular}{p{2cm} | c  c  c  c  c  c  c  c  c  c  c  c c}
	    & $\lubmM$ & $\lubmL$ & $\dbpedia$ & $\claros$ & $\yagoS$ & $\yagoM$ & $\yagoL$ & $\wnrrS$ & $\wnrrM$ & $\wnrrL$ & $\smokers 4$ & $\smokers 5$ & $\vqar$ \\
	    \hline
	     			   Min/Max DP      		& 3/22			& 3/22				& 4/19		& 5/25		& 6/12		& 6/12		& 6/14			& 8/12			& 7/12			& 4/14		& 4			& 5 & 15/15 \\
	     			   Min/Max DR      		& 17/107M			& 17/117M				& 26/691k		& 64/502k		& 5/91k		& 8/48k		& 19/116k		& 17/727		& 12/105		& 12/2k		& 74/364	& 112/1138 & 370k/390k\\
	     			   Min/Max R   	     	& 2/208			& 2/208				& 611/5552		& 6613/6672		& 35/275	& 86/1217	& 95/2104		& 147/216		& 257/440		& 95/720	& 3/3		& 3/3 & 6/6\\
\end{tabular}
\end{adjustbox}
            \label{tab:queries}
\end{table*}

\begin{figure*}[tb]
	\begin{adjustbox}{max width=\textwidth}
		\begin{tabular}{ccccc}
			 $\boldsymbol{\dbpedia}$ & $\boldsymbol{\claros}$ & $\boldsymbol{\yago}$ & $\boldsymbol{\wnrr}$ & $\boldsymbol{\smokers}$\\
			\x\x {\includegraphics[width = 0.2 \textwidth]{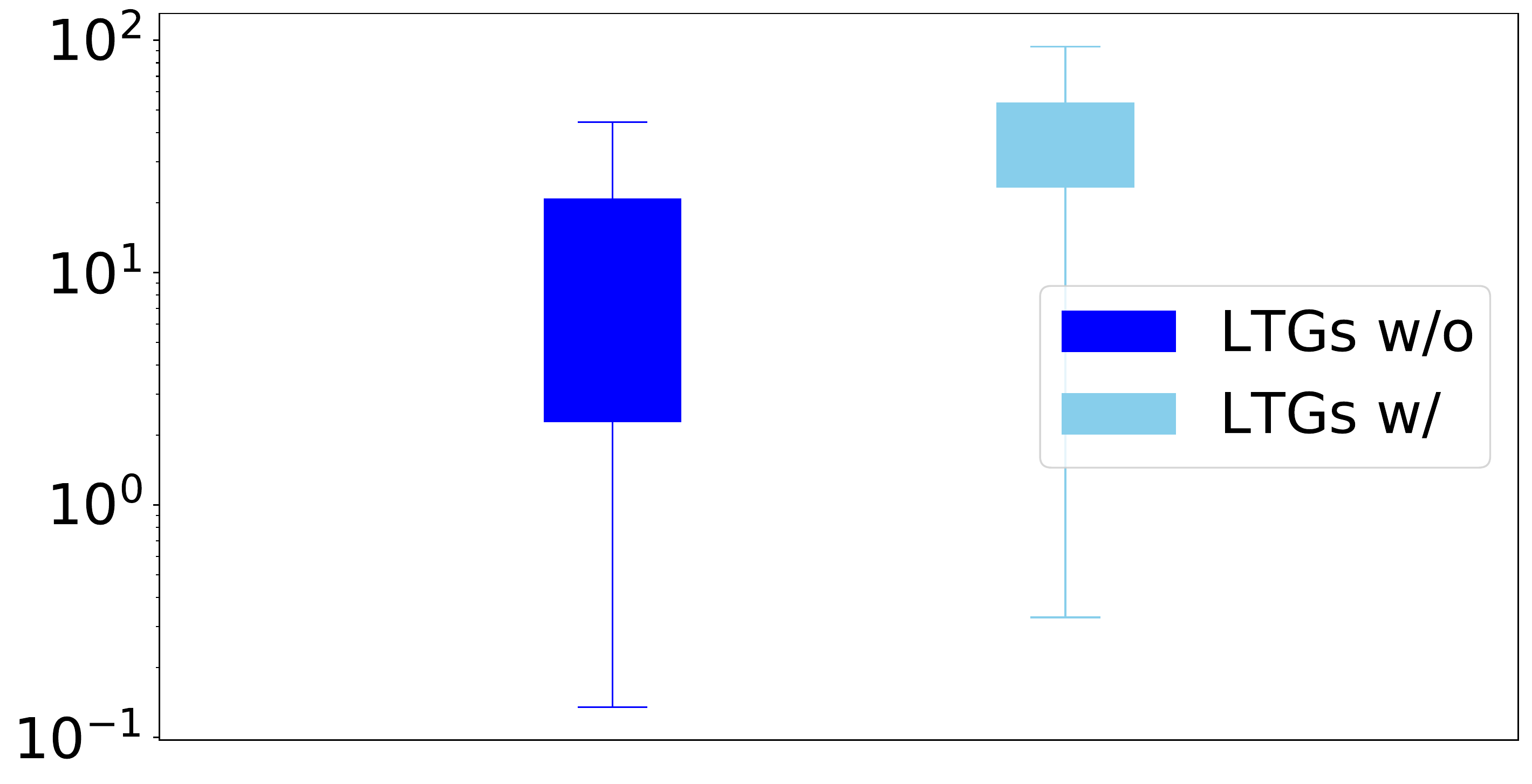}}  &
			\x\x {\includegraphics[width = 0.2 \textwidth]{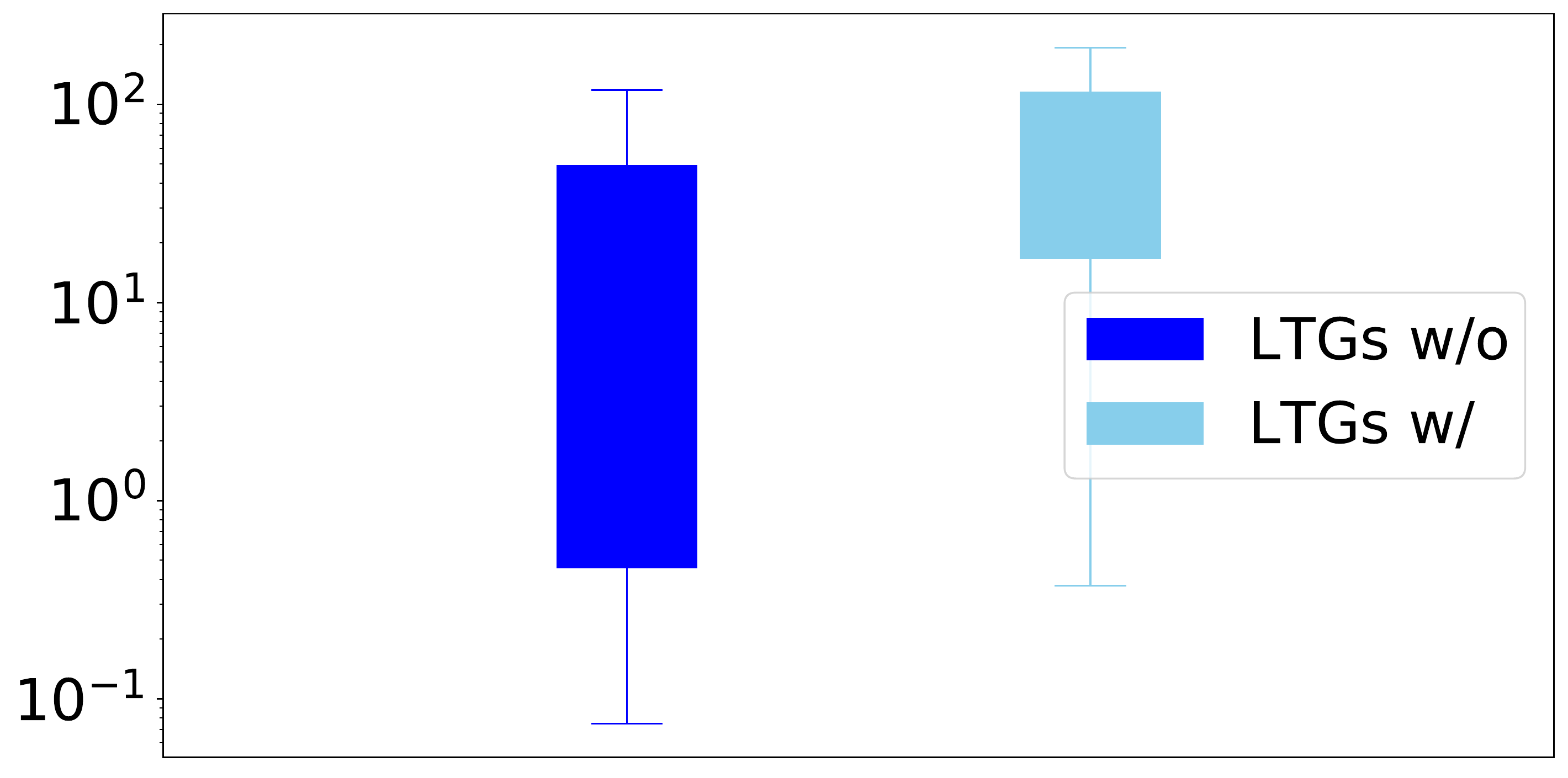}}  &
			\x\x {\includegraphics[width = 0.2 \textwidth]{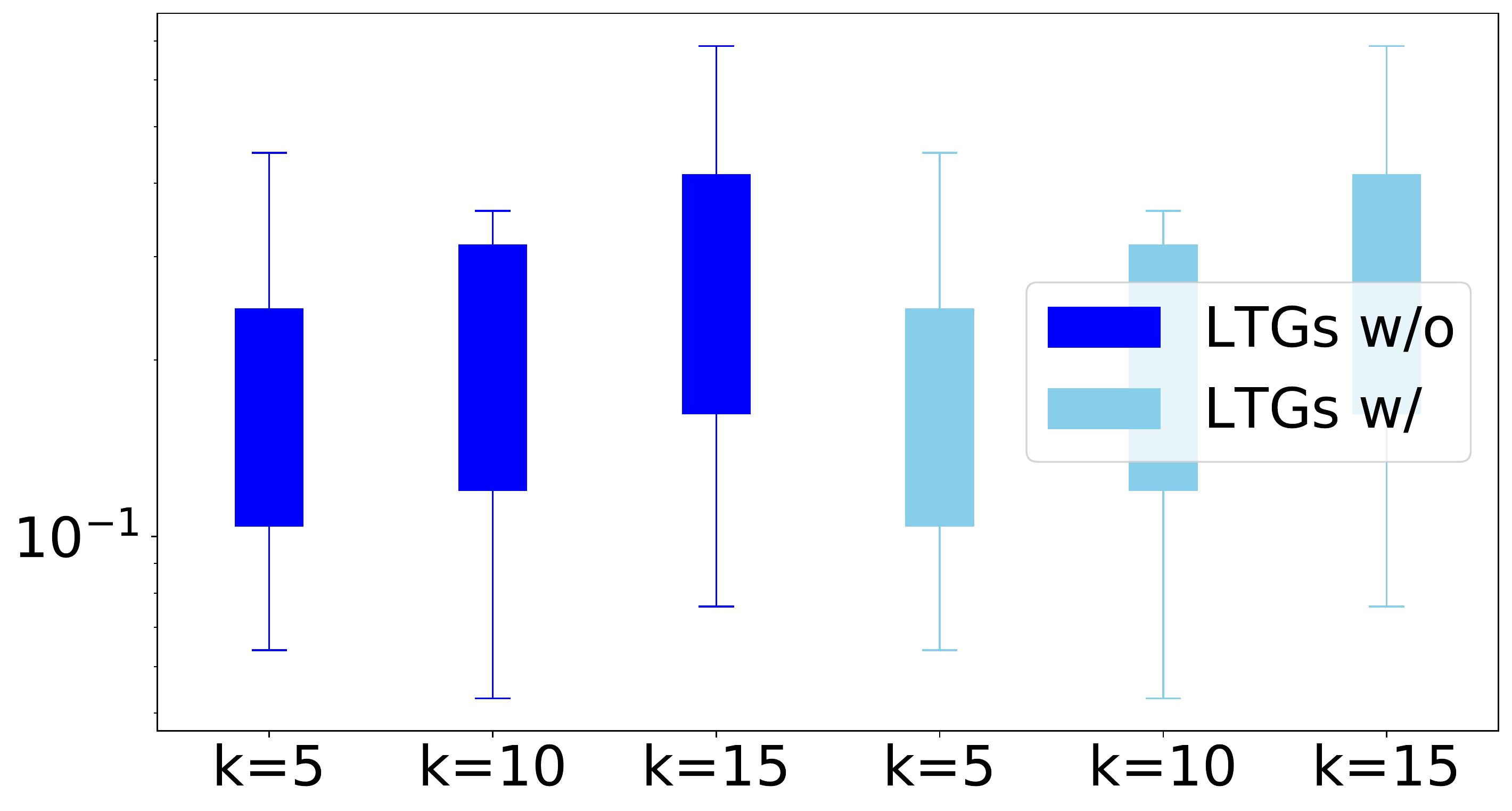}}  &
			\x\x {\includegraphics[width = 0.2 \textwidth]{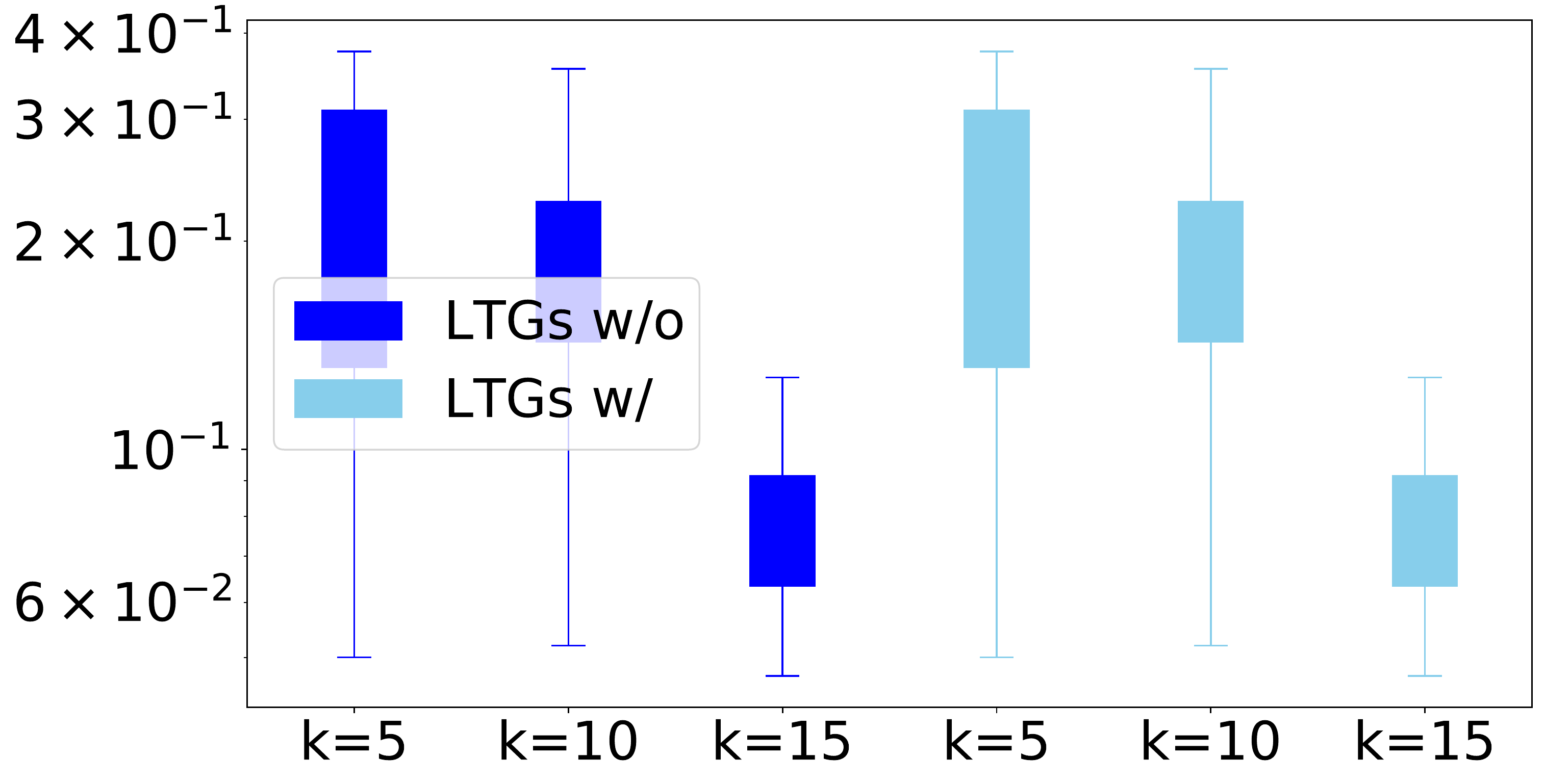}} &
			\x\x {\includegraphics[width = 0.2 \textwidth]{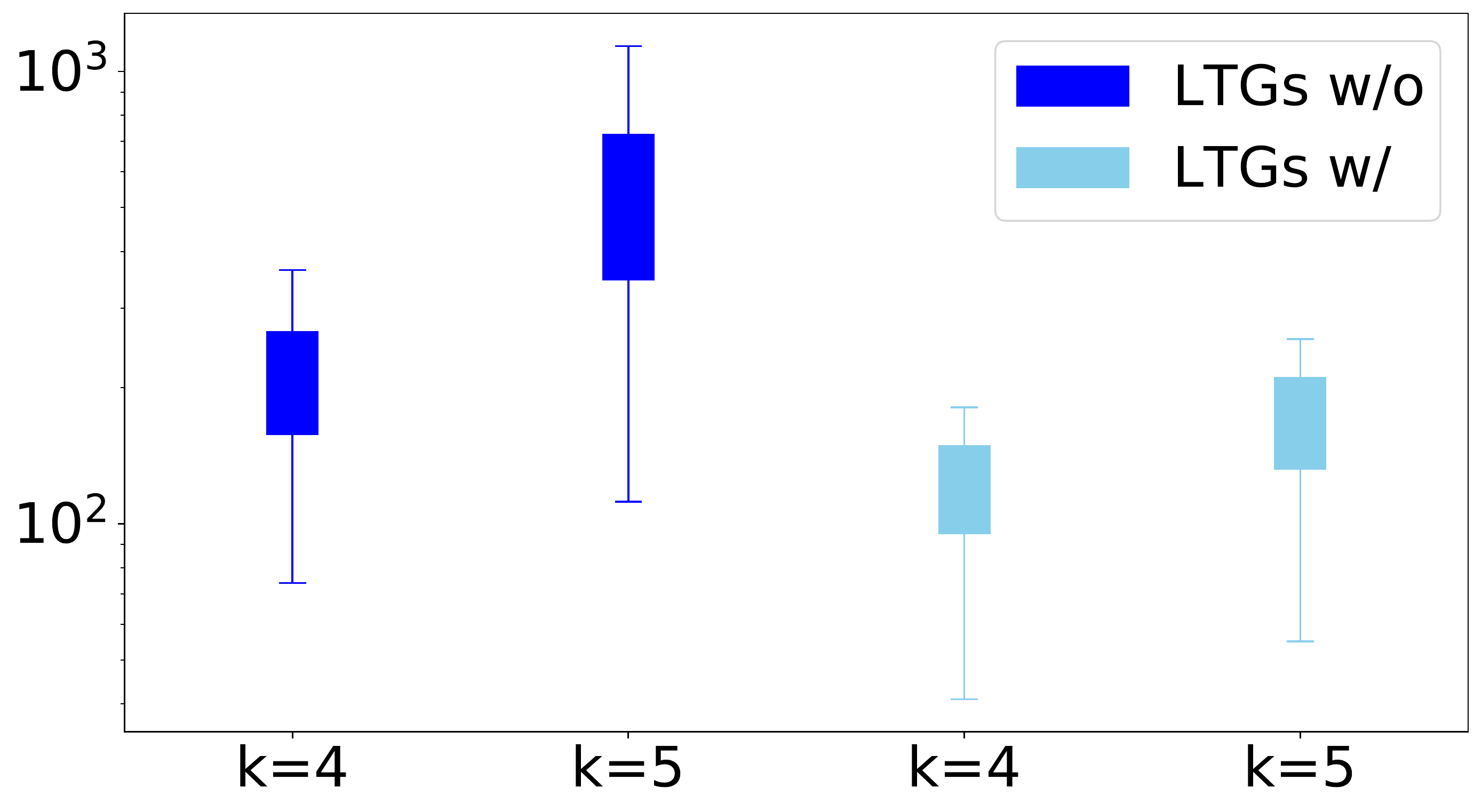}}
		\end{tabular}
	\end{adjustbox}
	\caption{Time to perform lineage collection for different scenarios using $\ours$.}
	\label{figure:appendix:lineage}
\end{figure*}

\end{document}